\documentclass{aa}  

\usepackage{xcolor}
\usepackage{graphicx}
\usepackage{txfonts}
\usepackage[]{hyperref}
\begin{document}

   \title{Statistical view of orbital circularisation with 14\,000 characterised TESS eclipsing binaries}
    
   \author{L. W. IJspeert\inst{1}
          \and
          A. Tkachenko\inst{1}
          \and
          C. Johnston\inst{2, 1}
          \and
          C. Aerts\inst{1, 3, 4}
          }

   \institute{Institute of Astronomy, KU Leuven,
              Celestijnenlaan 200D, 3001 Leuven, Belgium\\
              \email{luc.ijspeert@kuleuven.be}
        \and 
              Max Planck Institute for Astrophysics, Karl-Schwarzschild-Straße 1, 85741 Garching, Germany\\
        \and
              Department of Astrophysics, IMAPP, Radboud University Nijmegen, 
              P. O. Box 9010, 6500 GL Nijmegen, the Netherlands\\
        \and
              Max Planck Institute for Astronomy, 
              K\"onigstuhl 17, 69117 Heidelberg, Germany
             }

   \date{Received [month] dd, 2022; accepted [month] dd, 2022}

 
  \abstract
   {Eclipsing binaries are crucial for understanding stellar physics, allowing detailed studies of stellar masses, radii, and orbital dynamics. Recent space missions like the Transiting Exoplanet Survey Satellite (TESS) have significantly expanded the catalogue of observed eclipsing binaries with uninterrupted time series photometry, providing an opportunity for large-scale ensemble studies.}
   {This study aims to analyse the statistical properties of circularisation in a large sample of intermediate-to-high mass eclipsing binaries observed by TESS. We explore the dependence of orbital circularisation on stellar properties and orbital parameters to improve our understanding of the physical processes affecting these systems. We further aim to assess the role of stellar pulsations in circularisation.}
   {We compiled a catalogue of O- to F-type stars to search for eclipsing binary signals in the data available from the first 4 years of the TESS mission. Using automated classification and data analysis methodologies, we arrive at a well-characterised sample of 14,573 eclipsing binaries. We supplement our catalogue with \textit{Gaia} effective temperatures, and investigate the statistical characteristics of the sample as a function of temperature, orbital period, and scaled orbital separation.}
   {The orbital circularisation was measured with statistical methods to obtain three distinct measurements of the critical period and separation in four temperature ranges. These measurements cover a range of orbital periods and separations where both circularised and eccentric systems exist. Pulsations were identified in the g- and p-mode regimes and a reduced fraction of eccentric systems was found among them.}
   {Our analysis revealed the dependence of orbital circularisation on stellar temperatures, also seen in other studies, and confirmed previous findings that additional dissipation is needed as compared to the predictions of turbulent viscosity and non-resonant radiative damping. We speculate that pulsations may play a role in the circularisation of close binaries. Our study highlights the need for dissipative mechanisms that can produce a wide range of critical periods from a range of initial conditions.}

   \keywords{asteroseismology -- binaries: eclipsing -- 
             catalogues -- ephemerides -- methods: statistical -- stars: oscillations (including pulsations)
             }

   \maketitle
%

\section{Introduction}

Large numbers of stars continue to be classified and collected into catalogues for further study. Specifically in the field of eclipsing binaries (EBs) numerous samples have been published with sizes ranging from hundreds to hundreds of thousands. After observing for 9 years from the ground and across the whole sky, the All-Sky Automated Survey (ASAS) had found 11\,000 EBs \citet{Paczynski2006}. 10 years later, the Optical Gravitational Lensing Experiment (OGLE) reported on a catalogue of nearly half a million EBs in the Galactic bulge and Magellanic Clouds combined \citep{Soszynski2016, Pawlak2016}. The All-Sky Automated Survey for Supernovae compiled a list of 136\,000 EBs over about 6 years of observing \citep{Jayasinghe2019, Jayasinghe2021}. In space, Convection, Rotation and planetary Transits (CoRoT) found a total of 2269 EBs collated by \citet{Deleuil2018}. The space missions \textit{Kepler} and Transiting Exoplanet Survey Satellite (TESS) respectively, have published catalogues of 2920 \citep{kepler_eb2011, kepler_ebii2011, kepler_ebvii2016} and 4584 \citep{tess_eb2022} EBs, and later publications added 370 EBs without overlap \citep{howard2022} and 3155 EBs with overlap \citep{ijspeert2021} to the latter. Overshadowing all of the above is the recently released \textit{Gaia} catalogue of EBs containing over 2 million targets \citep{Mowlavi2023}.

Using photometric time series observations of EBs we can obtain their orbital period, eccentricity, argument of periastron, orbital inclination, fractional stellar radii, and the surface brightness ratio \citep{Southworth2020a}. If one of the stars in the system is pulsating or has rotational flux modulation, this signal can be characterised with the very same data, provided the time series is of sufficiently high quality and cadence. Such a combination can be powerful, as demonstrated in recent results \citep[][ for recent reviews, we point the reader to \citealt{Lampens2021, Guo2021}]{Liakos2017a, Liakos2017b, Johnston2019a, Miszuda2022, Johnston2023, Wagg2024}. Further adding spectroscopic observations, in which both stars can be distinguished, fully constrains the physical properties of the system, such as the individual masses and the orbital separation. This elevates EBs to a pivotal role in improving our understanding of a myriad of astrophysical phenomena \citep{Torres2010,Maxted2020}. 

Many of the EB catalogues, in particular the largest ones, constitute EBs found in relatively sparse observational data, which complicates their analysis. Space missions, sampling densely and continuously, offer unique datasets that can be used both for the detection of different types of stars and for their comprehensive analysis. One of the best examples of the power of these photometric time series is shown by in-depth asteroseismic studies of individual stars, for example: \citet{Degroote2009, Degroote2010, Briquet2011, Neiner2012, VanReeth2015, Schmid2016, Handler2019, Pedersen2021} and the review by \citet[][and references therein]{Bowman2020c}. Many studies of EBs also make good use of this high-quality data: \citet{Kjurkchieva2015, Matson2016, Guo2017, Maxted2020, Southworth2020, Southworth2021, Jennings2023, Pavlovski2023}, to give an idea.

The studies of the individual targets mentioned above are also a good example of the highly manual nature of these types of analyses. The catalogues of \textit{Kepler} and TESS EBs have not received homogeneous, comprehensive analysis to this day. An automated approach can help a great deal in reaching this goal, as much of the manual work is offloaded to an algorithm for all catalogued objects in parallel. This may not yet reach the level of depth of example studies, but it can help identify physically interesting targets and simplify a subsequent deep-dive \citep{ijspeert2024}. The enormous EB catalogues and the multiple long time-base satellite surveys covering large portions of the sky, such as TESS and the forthcoming PLAnetary Transits and Oscillations of stars \citep[PLATO; ][]{PLATO2022} mission, emphasise the need for highly automated approaches in astronomy.

Another benefit from large-scale homogeneous analysis is the potential to statistically study large samples of characterised stars. Such studies can teach us about the composition of the population of binary stars and about the distributions of their properties. Furthermore, one can learn about physical processes that express themselves as trends in the population and are hard or impossible to constrain with an individual target. The observational study of synchronisation and circularisation is such an example that benefits from large numbers. From the theory of stellar tides, we may predict timescales over which systems undergo significant orbital and rotational changes \citep{Zahn2008}. This theory relies on several physical properties that have been identified over the years, that dissipate energy within the stars. Four processes that act on the two distinguished types of tides are found to be competing for the highest efficiency. Viscous dissipation of turbulent flows acts on the equilibrium tide and is most effective in stars with convective envelopes, but does not play a role for radiative envelopes \citep{Zahn1966b, Zahn1977, Zahn1989, ZahnBouchet1989}. Radiative damping of the dynamical tide is effective in stars with an outer radiative region where gravity waves can propagate \citep{Zahn1975, Zahn1977}. Large-scale meridional flows driven by the equilibrium tidal deformation and dissipated in a thin viscous surface layer form a hydrodynamical mechanism working in stars with radiative and convective envelopes \citep{Tassoul1987, Tassoul1988, Tassoul1990}. It was later proposed that inertial waves can be excited by tides and efficiently dissipated in stars with convective envelopes \citep{Ogilvie2007}.

Observational studies have since long attempted to assess the different mechanisms and often reported discrepancies where the theories could not explain measurements, see, for instance: \citet{Savonije1983, Savonije1984, Claret1995, Claret1997} and \citet{Khaliullin2010, Khaliullin2011}. \citet{vaneylen2016} looked at the tangential components of the eccentricity of a larger sample of $\thicksim$900 \textit{Kepler} EBs and concluded that binaries with two hot (> 6250\,K) components have a higher eccentricity fraction at orbital periods shorter than 4 days. Two recent studies have provided insights from large samples of about 800 \citep{Justesen2021} and 500 \citep{Zanazzi2022} EBs in total. \citet{Justesen2021} found that binaries with a primary effective temperature under 6250\,K are in orbits circularised by tidal forces below a semi-major axis ratio of about $a/R_1\thicksim$20 in ($P_{\rm orb}$ $\thicksim$7\,days), in accordance with the theory of the equilibrium tide. For hotter primaries, the critical separation for circularisation predicted by the dynamical tide is $a/R_1\thicksim$3-4 ($P_{\rm orb}$ $\thicksim$1-2\,days), smaller than the observed separations of many of their circular systems. \citet{Zanazzi2022} introduce the measurement of the orbital period corresponding to the envelope of highest eccentricity, which traces the point where the youngest systems are still in eccentric orbits. They found an envelope period of $\thicksim$3\,days in their sample with $T_{\rm eff}$ < 7000\,K, more in accordance with lower estimates for the strength of turbulent viscosity in the equilibrium tide. \citet{Zanazzi2022} also discuss the presence of many circularised systems out to long periods of $\thicksim$10-20\,days and posit several scenarios for their existence.

We present the statistical analysis of a large new catalogue of 69\,058 EBs as observed by TESS. The aim and focus of this paper are to empirically constrain orbital circularisation with the identified sample. This embodies just one of the aspects for which this catalogue could be exploited, and other topics are deferred to follow-up papers. The EBs are selected from a catalogue of O- to F-type stars, compiled from the TESS Input Catalogue (TIC) and supplemented with \textit{Gaia} data. We use data from public releases of the first 4 years of long-cadence TESS observations to look for eclipses, with the help of methodology described in \citet[][hereafter IJ21]{ijspeert2021}. The photometry is analysed using an automated methodology by \citep[][hereafter IJ24]{ijspeert2024} and the EB classification and characterisations are published with this work in two separate catalogues. We check the characterised EBs for accuracy and then use a sample of 14\,573 to statistically explore circularisation across temperatures starting at around 6000\,K to well over 20\,000\,K. Various diagnostic measurements of circularisation are given in terms of orbital periods and scaled orbital separations. Finally, we take a look at the pulsational properties of our sample and identify four groups, among which gravity-mode (g-mode) and pressure-mode (p-mode) pulsators are found. By adding these categorisations to our catalogue, we produce the largest sample of pulsating EBs to date.


\section{Sample selection and data analysis}

We have expanded the all-sky catalogue of O, B and early A stars and the sample of EB candidates among them assembled by IJ21 towards lower temperatures. We worked analogously to IJ21 and started with the TESS Input Catalog (TIC) \citep{tic2019} to arrive at a catalogue of O, B, A and F stars. We searched for EB candidates using public releases of full-frame image (FFI) light curves from the first four observing cycles of TESS (sectors 1-55). The methodology for finding EBs presented in IJ21 has been improved upon for use in this work, as described in Section \ref{sec:eb_search}. Once identified, the EB light curves were analysed using the methodology described in IJ24 to obtain the properties of the systems. The catalogues of O/B/A/F stars and that of the EBs were made available to the public through CDS\footnote{Via anonymous ftp to \url{cdsarc.u-strasbg.fr} (130.79.128.5) or via \url{http://cdsweb.u-strasbg.fr/}}.

\subsection{Selection criteria}
\label{sec:sample}

The selection process is described in full detail in IJ21 (Section 2.1), and we followed all the same steps except for the different colour cuts. After initial selection steps, but before the colour cuts were made to select our desired (larger) temperature domain, the number of targets considered was reduced to $6.37\cdot 10^7$. We followed the colour cuts from \citet{garcia2022} that are based on a sample of A/F-type stars by \citet{Tkachenko2013, li2020}: $J - H < 0.24$ and $J - K < 0.30$. A total of $3.45\cdot 10^6$ targets survived the cut (a 94.6\% reduction). Finally, after we removed white dwarfs and giants, we retained a total of 3\,337\,497 targets, including all the 189\,981 O/B/early-A stars found by IJ21. 

\begin{figure}
\centering
\includegraphics[width=\hsize]{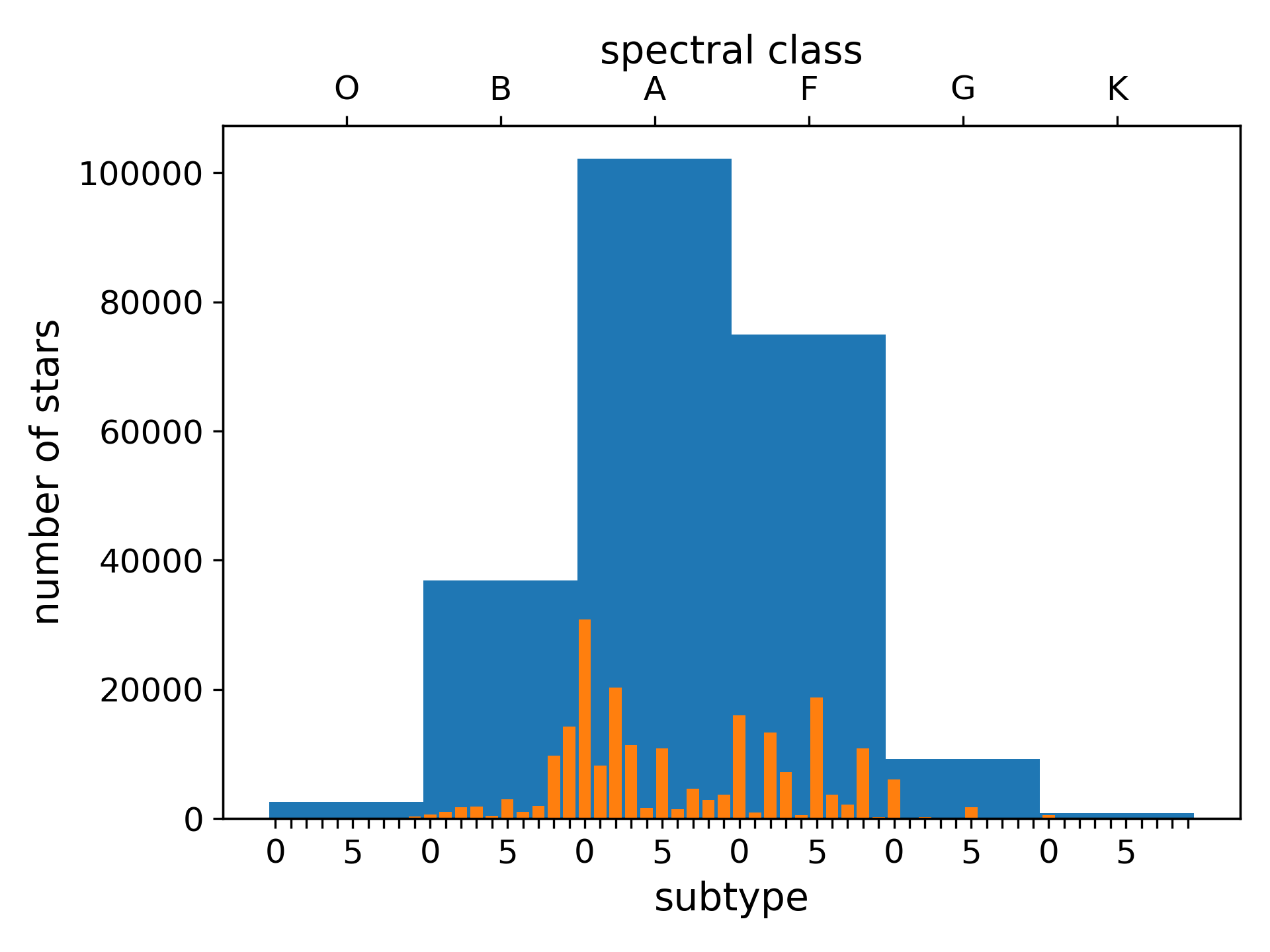}
    \caption{\texttt{SIMBAD} spectral types for 227556 targets in the O/B/A/F sample. Only the first occurrence of the spectral class letter and temperature subclass number was counted. This resulted in multiple artificially inflated peaks in the distribution of subclasses (narrow orange bars). The distribution is expected to follow the broad overall distribution of the wide blue bars.}
    \label{fig:spec_types}
\end{figure}

\begin{figure*}
\sidecaption
\includegraphics[width=12cm,clip]{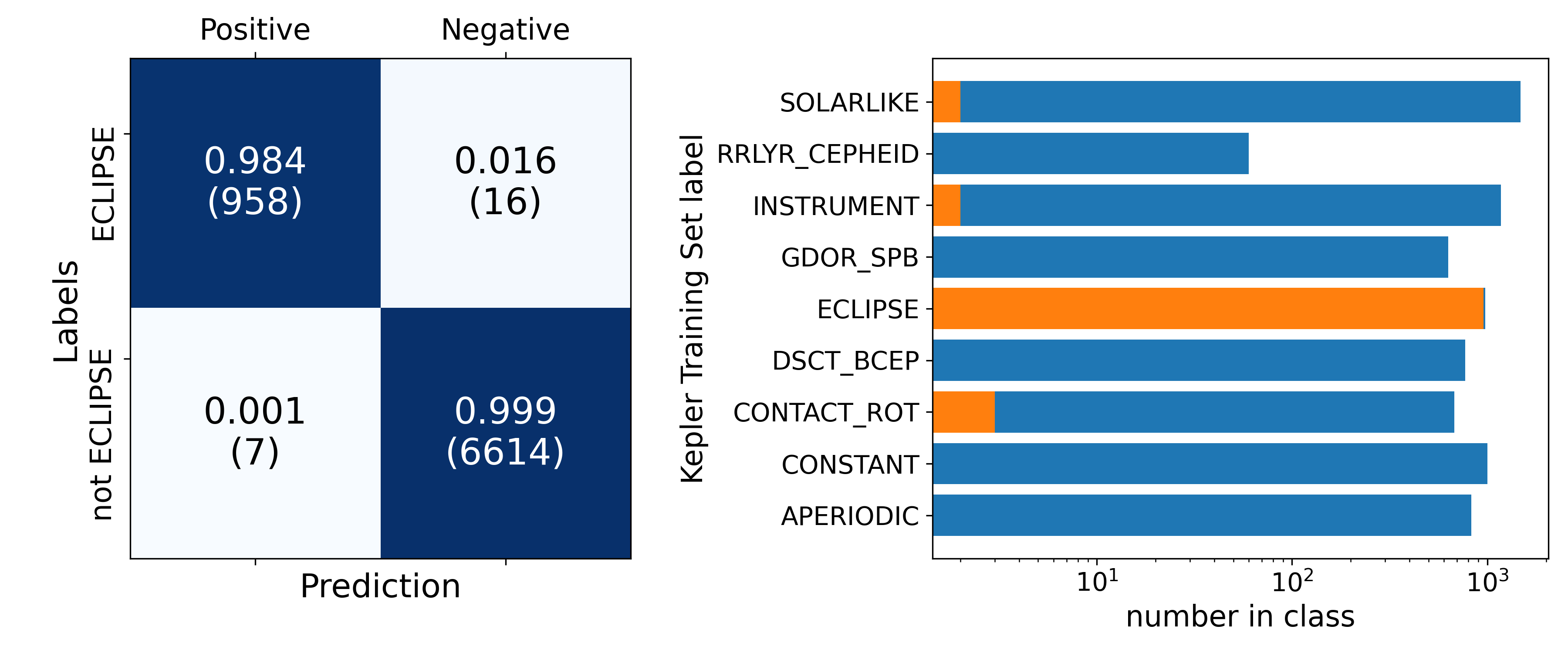}
    \caption{Confusion matrix (left) and per class break-down (right) of the \textit{Kepler} test set classifications using the newly implemented RF classifier of \texttt{ECLIPSR}. The confusion matrix shows the sensitivity and specificity on the diagonal and includes the absolute numbers in each category between brackets. The right panel shows to which classes the true and false positives belong, as well as the total number in each class, in log-scale.}
    \label{fig:kepler}
\end{figure*}

We queried the \texttt{SIMBAD}\footnote{\url{http://simbad.u-strasbg.fr/simbad/}} database for available spectral types of our selection and found that close to a quarter of targets (24\%) returned entries in that database, and roughly a quarter (28\%) of those had a spectral type specified. Figure \ref{fig:spec_types} shows the distribution of spectral types, for the main class and temperature subclass. Many spectral type entries in \texttt{SIMBAD} provide ranges for the subclass number; we only picked the first digit. This causes some subclasses to have an artificially large number of targets, and the subclass distribution differs substantially from the distribution of the main class. We can see a slight drop from A to F stars, and a steep drop going from F to G stars, indicating that our colour cuts were effective.

\subsection{TESS light curves}
\label{sec:tess_lc}

We obtained light curves from the Mikulski Archive for Space Telescopes (MAST)\footnote{\url{http://archive.stsci.edu/tess/all\_products.html}}, downloaded in December 2023. There are several public bulk FFI light curve data releases, of which we chose two for their complete coverage of observing cycles 1 through 4: the TESS Science Processing Operations Center (SPOC)\footnote{\href{https://dx.doi.org/10.17909/t9-wpz1-8s54}{DOI: 10.17909/t9-wpz1-8s54}} and the MIT Quick-Look Pipeline (QLP)\footnote{\href{https://dx.doi.org/10.17909/t9-r086-e880}{DOI: 10.17909/t9-r086-e880}}. Cross-matching the provided target lists with our selection resulted in 650\,000 targets with light curves from SPOC and 1.88 million from QLP. There is nearly complete overlap between these two sources: the total number of targets with a light curve is close to the larger number.

Depending on their position in the sky, each target may appear in any number of observing sectors, from 1 to a few dozen. In total, we downloaded $\thicksim$2 million SPOC and 5.4 million QLP sectors worth of light curves for analysis, approximately an average of 3 sectors per target.

\subsection{EB search}
\label{sec:eb_search}

The methodology described in IJ21 aims to rapidly find EB light curves that may exhibit intrinsic variability and is shown to be capable of this. Our goal in this work, to analyse a large ensemble of pulsating and non-pulsating EBs, aligned with this aim and we thus used the same methodology after giving it an upgrade and contributing to the publicly available implementation \texttt{ECLIPSR}\footnote{\href{https://github.com/LucIJspeert/eclipsr}{github.com/LucIJspeert/eclipsr}}. The method classifies light curves based on 6 derived quantities, which we call features here. The features are combined into a final score that informs the separation into positive and negative classifications. This final score is calculated based on Equation 2 in IJ21. Here we replaced this formula with a random forest (RF) classifier in an attempt to decrease the false positive rate (this addition was released with version 1.1.0 of the code).

IJ21 reported 5502 positive classifications and 85691 negatives, and further manually divided the positives into 3425 true positives (TPs) and 2077 false positives (FPs). Negatives were not further divided into true negatives (TNs) and false negatives (FNs). Using \texttt{ECLIPSR}, we computed the features for all light curves of the OBA-catalogue provided by IJ21. Merging the SPOC and QLP results was done by picking the features with the highest score value. We trained an RF classifier from scikit-learn \citep{sklearn} to distinguish between the reported TPs and FPs leading to their final classification of 3425 EBs and 2077 rejected candidates. Additional positive classifications were identified with this classifier within the group of negative classifications in the OBA-sample. We vetted these new positives into the TP and FP categories and used them to retrain the RF. Iterating this procedure until no further new positives were found yielded a list of 108 additional TPs and 136 FPs from the originally reported 85691 negatives. This procedure ensured that the divide between `EB' and `not EB' in the ultimately used training set was as clean as possible. Finally, the RF was trained on all 3533 targets now marked TP (and labelled `EB') and on the combined 2213 targets marked FP and remaining negatives (both now labelled `not EB'). The resulting RF classifier was published for use with \texttt{ECLIPSR}, along with additional code contributions. 

We applied the trained classifier to the same labelled \textit{Kepler} test set that is also used in IJ21, and was described in more detail by \citet{Audenaert2021}. This data was not seen by the classifier during training, so we used it to validate that our RF classifier was not overfitting on the training set. The output classifications are illustrated in Figure \ref{fig:kepler} and compared to the labels of the test set. We found a sensitivity of 98.4\% and a specificity of 99.9\%, within 0.1 percent of the values found by IJ21. The precision that we found shows a larger difference, 99.3\% with the RF classifier and 98.3\% without, halving the number of false positives for this dataset. Sensitivity is the true positive rate of the classification, defined as (TP/(TP+FN)). Specificity is the true negative rate, defined as (TN/(TN+FP)). Precision is the positive predictive value, defined as (TP/(TP+FP)).

The improved methodology was applied to the light curves in parts, divided per observing cycle. This reduces the chance of finding false positives due to misidentified data artefacts, and was found to perform better than stitching all light curves together with large (> 1 yr) gaps. We added the classification (0 for negative, 1 for positive) of the separate observing cycles, resulting in numbers from 0 to 3. All targets with a classification score higher than 0 are considered positives. We further added the classifications of the SPOC dataset and the QLP dataset to obtain our final sample of EB candidates. 

\subsection{Light curve analysis}

The main aim of our light curve analysis is to measure orbital eccentricity for the study of orbital circularisation. Additionally, the orbital period, scaled sum of radii and properties of further variability in the light curve in terms of frequencies are important parameters. We used the EB light curve analysis methodology described in IJ24, which has a publicly available implementation called \texttt{STAR SHADOW}\footnote{\href{https://github.com/LucIJspeert/star_shadow}{github.com/LucIJspeert/star\_shadow}}. No changes were made to their methodology (version 1.1.7a), as it is shown to be capable of finding the correct orbital periods and of extracting accurate eccentricities. It additionally provides a characterisation of the light curve in terms of sinusoids, which we used to summarise the intrinsic variability of our EBs. While not meant for classification, the method may stop analysis for systems where it cannot identify eclipses or are found to be unphysical in other ways. 

\texttt{STAR SHADOW} works without manual intervention to go from raw light curves of EBs to the output parameters. It reports eccentricity, argument of periastron, inclination, sum of scaled radii, radius ratio and surface brightness ratio, in addition to frequencies, amplitudes and phases of sinusoidal signals found to be present in superposition with the eclipses. The first step in the algorithm is to perform a full iterative prewhitening procedure, building a sinusoidal model of the light curve. The orbital period is determined from combined statistics of a periodogram and phase dispersion measure, and the known frequencies of sinusoids in the light curve model. Orbital harmonic sinusoids are coupled to the orbital period, forming a harmonic model of the eclipses. Subtracting this yields accurate parameters of the remaining sinusoids that appear at frequencies other than the harmonics (regardless of the level of convergence of the subsequent physical eclipse model fit). The binary parameters are found from a geometric analysis of the harmonic model of the eclipses, and these are used as the starting point of the following fit. Finally, a simple physical model of the eclipses is fit to the light curve simultaneously with a set of sinusoids describing the full variability content of the light curve. 

The procedure adopted here is not meant to replace physical modelling with codes like Wilson-Devinney \citep{WD1971, Wilson1979, Wilson1990, Wilson2008} or Phoebe 2 \citep{Prsa2016}. Unlike the above-mentioned codes that prioritise the accuracy of the input physics and have the purpose of high-precision analysis of individual binary systems, \texttt{STAR SHADOW} was developed to deliver fundamental characteristics of binary systems (e.g., orbital periods, eccentricities) with or without intrinsic variability. As is demonstrated in IJ24 and further in this work, the precision with which binary characteristics are inferred is largely sufficient for statistical studies of large (thousands to tens of thousands) populations of binaries.

We chose to run the analysis pipeline on each separate observing cycle, instead of on the stitched light curves. This was done to avoid issues arising from alias frequencies or other adverse effects due to year-long gaps in the light curve. The final parameter results, for example, the orbital periods, were combined by unweighted average over all separate measurements (including both datasets), after checking for consistency within the errors. Error-values proportionally reduce in size. In case of inconsistent measurement values, we instead picked the values associated with the model that lead to the lowest standard deviation of the residuals. There are a number of parameters, like the total number of frequencies found and noise levels, where we have simply taken the maximum or minimum values over all cycles. For the five most dominant frequencies we have taken the values of the model with the lowest standard deviation of the residuals; these are selected after removal of all harmonic sinusoids constituting the harmonic model of the eclipses. 

The methodology in IJ24 is meant to be used with detrended light curves of EBs, while the data from our sources may still have instrumental effects remaining, and our selection of EB candidates may not all be EBs. To ensure a clean sample of characterised EBs, we therefore manually checked the output results for correctness. In the case that an eccentricity measurement was obtained, we looked at the identification of primary and secondary eclipse: correct positioning of eclipse edges and midpoints ensures an accurate determination of the parameters. This means the check was not a judgement of the correct classification as an EB. On the contrary, while many analysis results were deemed inaccurate, the vast majority of checked light curves showed eclipses. Artefacts in the data were often the cause of bad analysis results.

\subsection{\textit{Gaia} catalogue}

In addition to the parameters we retained from the TIC, we used temperatures and various other parameters from the \textit{Gaia} catalogue \citep{Gaia2023}. The \textit{Gaia} catalogue provides a cross-match with the Two Micron All-Sky Survey \citep[2MASS, ][]{2mass2006}, and all of our targets have a 2MASS identifier since we have used band-passes from this survey in our colour selection. Relevant parameters were found in the \textit{Gaia} documentation pages (\href{https://gea.esac.esa.int/archive/documentation/GDR3//Gaia_archive/chap_datamodel/}{gea.esac.esa.int}, mainly chapter 20.2.1). We used the \textit{Gaia} advanced search option to query the database with the Astronomical Data Query Language (ADQL); the query is added in Appendix \ref{apx:gaia}. 

When merged with our list of targets, we found that the \textit{Gaia} data contained a number of duplicates compared to our TIC numbers. We removed these by only keeping the brightest target based on \texttt{phot\_g\_mean\_mag}. We combined the `ESP-HS' parameters, that are more suitable for hot stars (and among which are notably the effective temperatures), with the `GSP-Phot' parameters by replacing the latter by the former where available. We divided the sample according to \texttt{SIMBAD} spectral class and made a kernel density estimate (KDE) of the temperatures in each bin (see Figure \ref{fig:kde_temps}). If the spectral classifications are to be taken at face value, this tells us that the temperature measurements below about 7000\,K cannot distinguish the G-type stars in our sample from the F-types. However, this can be ascribed to our colour selection cutting off the high-temperature tail of the G-type distribution (the total number of G-type stars is 8 times lower than F-type). This is corroborated by the histograms of the same data, which show the absolute numbers of targets instead of densities (Figure \ref{fig:hist_temps}). The A-type stars are well separated, although we see overlap between temperature regimes from both sides. The B stars overlap largely with the A-type temperatures, but they do make it well past the 10kK marker where the A-type stars taper off. The few (probably late) O-type stars that both had a temperature measurement and spectral class are seen at the highest temperatures, although these are still in the temperature regime for B-types. The figure is cut off at 35kK due to the very low number of stars present beyond that point. 

\section{The EB catalogue}

In the SPOC data, a total of 12\,344 EBs were identified, of which 5480 with a classification score of 2 and 119 with a score of 3. This amounts to 1.91\% of the total number of targets in this dataset. For the QLP data, 67\,392 EBs were found, 26\,381 with a score of 2 and 457 with a score of 3. This is equal to 3.59\% of the number of targets analysed. Combining the two datasets, we get the final total of 69\,057 EBs (3.67\% of the total number of targets). There is a large overlap in positive classifications, which is reassuring, but the part that does not overlap shows that the methodology might give rise to a large number of FNs. For our purpose, we prefer a less complete but more pure sample over a more complete but more contaminated one. Moreover, our approach of separately analysing the different observing cycles and two nearly completely overlapping datasets mitigates this to some extent, as sources have multiple chances of being detected. 

IJ21 found 457 EB candidates in their 14\,970 SPOC targets and 5418 among their 91\,142 QLP targets, using \texttt{ECLIPSR} (version 1.0.2). This constitutes 3.05\% and 5.94\%, respectively. After manually vetting each candidate, they end up with samples of 378 (2.52\%) and 3387 (3.72\%) EBs for each dataset. Comparing these percentages to our results, obtained with version 1.1.0, we see that they are both lower than the initial amounts and slightly lower than the vetted amounts. Therefore the level of contamination in our new samples is assumed to be low.

\begin{table}
	\centering
	\caption{Classification results per subdivision of the data.}
	\label{tab:classification}
	\begin{tabular}{c r r r}
	\hline
	  dataset & N$^\circ$ targets & N$^\circ$ positives & N$^\circ$ total EBs \\
	\hline
	SPOC c1 & 293\,788 & 4980 (1.70\%) & 8006 \\
    SPOC c2 & 230\,523 & 4002 (1.74\%) & 6632 \\
    SPOC c3 & 294\,004 & 5057 (1.72\%) & 8053 \\
    SPOC c4 & 231\,300 & 4023 (1.74\%) & 6617 \\
    Total SPOC & 646\,380 & 12\,344 (1.91\%) & 18\,173 \\
    QLP c1 & 1\,087\,385 & 33\,426 (3.07\%) & 40\,138 \\
    QLP c2 & 557\,055 & 16\,881 (3.03\%) & 22\,239 \\
    QLP c3 & 1\,022\,794 & 28\,808 (2.82\%) & 37\,980 \\
    QLP c4 & 508\,924 & 15\,572 (3.06\%) & 18\,920 \\
    Total QLP & 1\,878\,846 & 67\,392 (3.59\%) & 69\,046 \\
    Total & 1\,879\,195 & 69057 (3.67\%) & - \\
	\hline
	\end{tabular}
	\tablefoot{The middle columns have the number of targets and positive classifications per dataset and per observing cycle (c1-c4). The last column contains the number of EBs of the final total, that are in a given observing cycle.}
\end{table}

In Table \ref{tab:classification} we summarise the number of EBs found in each part of the dataset. The second column has the number of targets in that respective part of the dataset and the third column contains the number of EBs found in that part of the dataset. The right most column instead contains the number of EBs of the final total (69\,057), that have a light curve in each subdivision of the dataset, so including targets that were missed. If we `correct' the percentages calculated above with the numbers in the last column, we obtain 2.81\% for SPOC and 3.67\% for QLP; this is how many EBs we could in principle have found in that portion of the data. Both percentages are still close to the ones of the vetted sample by IJ21. However, these numbers are before checking for duplicate ephemerides in nearby sources that are likely caused by contamination. 

\subsection{Filtering steps}

The manual check of light curve analysis output was applied to 37\,741 systems that have orbital parameters specified for one or more of the data subsets. Of those, 19\,198 were found to be correct by measure of the determined positions of primary and secondary eclipse start, minimum, and end points. A further 4519 were judged to have an accurate period and a tangential part of the eccentricity that could be used for analysis within its estimated error, as the eclipse minima (but not the eclipse edges) are placed close to the correct positions. We do not use the latter category in our later analysis. The relatively low fraction of correct results does not reflect the number of misclassified EB candidates; on the contrary, it was estimated that on the order of a few percent of cases involved light curves not of EBs. The latter includes cases of pulsations mistaken as eclipses, but these were a small minority. 

Whereas IJ21 reported many cases of pulsations in light curves confused for eclipses, our upgrade to the classification algorithm seems to have effectively diminished this source of contamination. We compare the orbital periods within a proximity of 1 degree angular distance on the sky: if the period agrees within 0.1\% we give all corresponding EB signals a number designating the duplicate group they are assigned to. If signals were correctly analysed, the signal with the largest eclipse depth is designated as the target where the signal originates. The same is done if signals fall in the category of having a usable period and tangential eccentricity, but the former is preferred. Both the designation of the source where the EB signal is deemed to originate and the membership designation of the duplicate groups are stored in the catalogue. Among the 62\,928 targets with a measured orbital period, we identify 5605 duplicate groups containing a total of 12\,836 targets.

In dividing our characterised EB sample into high and low eccentricity, we apply one further filter. The high eccentricity systems are required to have tangential eccentricity values inconsistent with zero within 3 $\sigma$. The selections are made as follows:
\begin{itemize}
  \item `low e': $|e \cos(\omega)| < 0.02$, $e < 0.1$,
  \item `high e': $|e \cos(\omega)| >= 0.02$, $|e \cos(\omega)| > 3 \sigma_{e \cos(\omega)}$.
\end{itemize}
This excludes only 50 targets after all other filters have been applied. The remaining count after removing duplicates, incorrect analysis results, and high eccentricities consistent with zero, is 14\,573 characterised EBs (0.78\% of the original number of targets). By the selection for high eccentricity, 33\% of systems are eccentric. In the further analysis, we focused on this filtered sample of characterised EBs.

\subsection{General properties of the sample}

\begin{figure}
\centering
\includegraphics[width=\hsize]{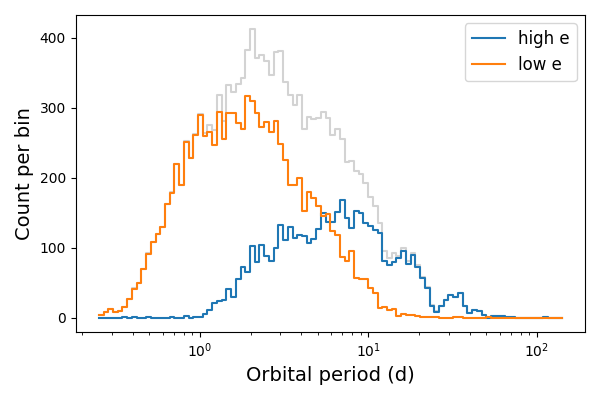}
    \caption{Logarithmic period distribution of the selected 14\,573 EBs. The overall distribution is grey, the distribution of systems of low eccentricity is orange and that of systems with high eccentricity is blue. The period bins have equal width in log-space.}
    \label{fig:hist_period}
\end{figure}

Figure \ref{fig:hist_period} shows the logarithmic period distribution of EBs over the complete period range up to a year: it strongly peaks around 2-3 days and falls off sharply after 25 days. We see the signature of the distribution of time series lengths, showing up as a trough at 27 days and a secondary peak between 27 and 54 days. The vast majority of light curves are about 27 days in length with a subset of $\thicksim$54 days in length and a minority of longer ones. For periods well below a day, we also lose sensitivity: the detection methods of \texttt{ECLIPSR} are less well suited for very short period eclipses than for example a box least squares algorithm, and further limited by the cadence of 30 minutes in the first two cycles and 10 minutes in the third and fourth cycles. Nevertheless, we do expect the low end of the true period distribution to be truncated at a certain point. The eventual coalescence of the binary dictates a limit on the size of both stars relative to the orbital separation, which strongly correlates with the orbital period. On the main sequence (MS), which our sample is selected for, this also correlates with effective temperature.

\begin{figure}
\centering
\includegraphics[width=\hsize]{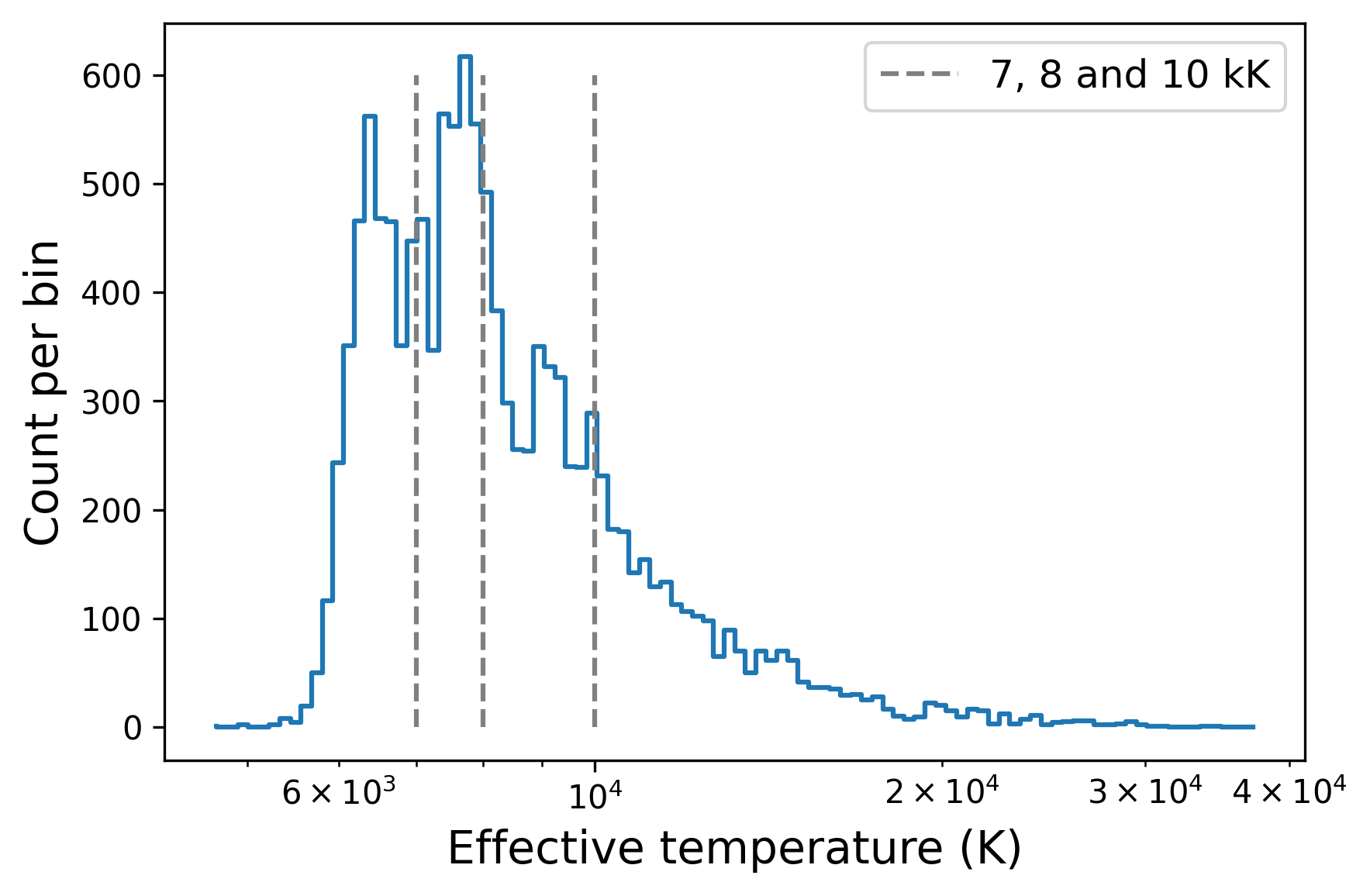}
    \caption{Logarithmic effective temperature distribution of the selected 12\,743 EBs with \textit{Gaia} $T_{\rm eff}$ value available. The three dividing lines for the two low and the two high-temperature bins at 7000, 8000 and 10000\,K are indicated. We note that temperature bins have equal width in log-space.}
    \label{fig:hist_teff}
\end{figure}

The majority of characterised EBs have an effective temperature value in our cross match with the \textit{Gaia} catalogue; 1880 targets lack this measurement, so when discussing temperatures the reduced sample size is 12\,743. The distribution of \textit{Gaia} temperatures shown in Figure \ref{fig:hist_teff} is a rather jagged line peaking at 6500K and with minima near 7000 and 8500\,K, and two further accumulations at 7500 and 9000\,K. It is not expected that these are real features, but rather that they are artefacts of the determination of the $T_{\rm eff}$ values and of mixing \texttt{gspphot} with \texttt{esp-hs} temperatures. We divide our sample based on effective temperature, following \citet{Torres2010} and \citet{Justesen2021}, with dividing lines at 7000, 8000 and 10\,000\,K. This leaves each temperature bin with a similar number of about 3000. 

\begin{figure}
\centering
\includegraphics[width=\hsize]{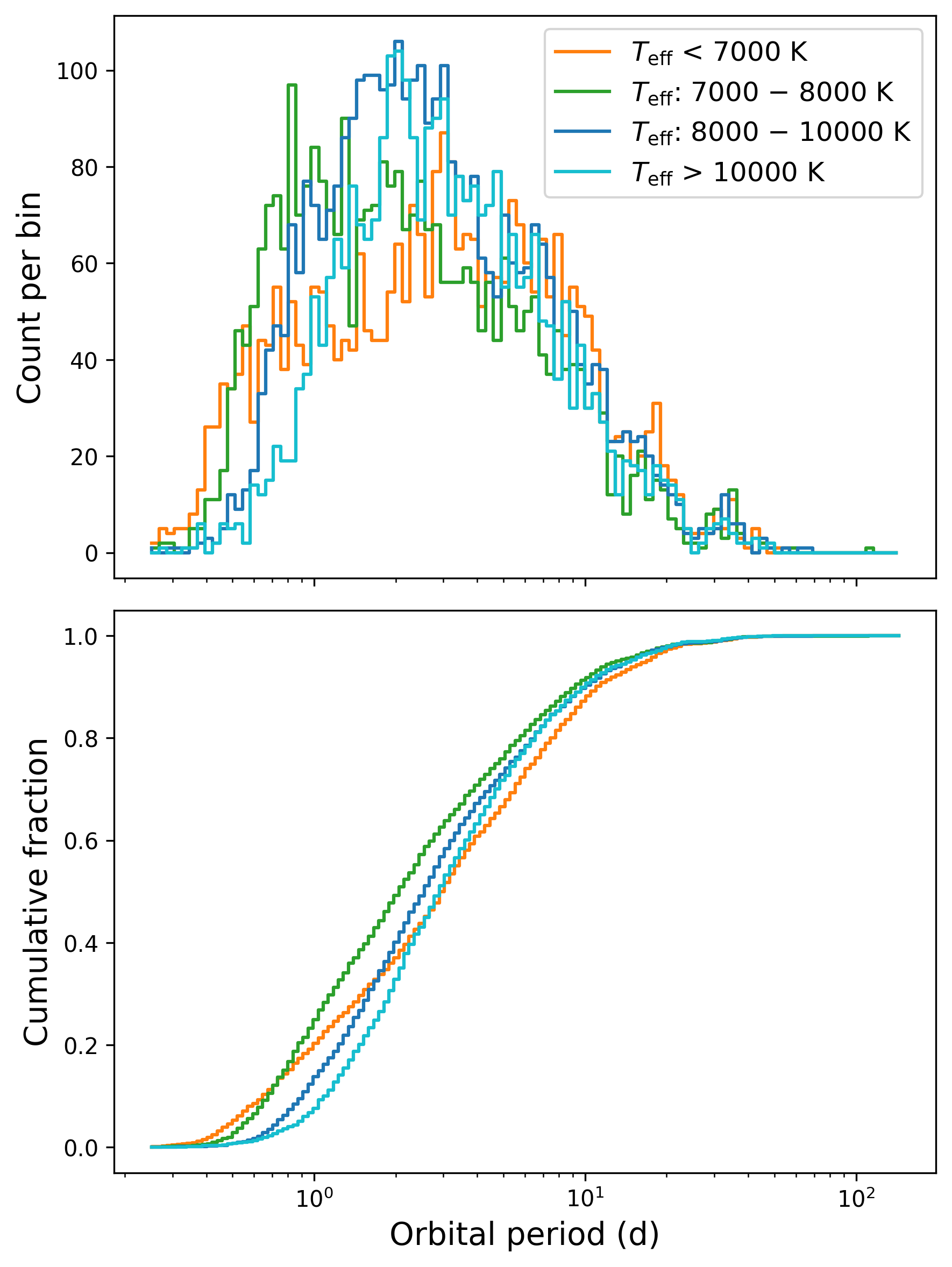}
    \caption{Logarithmic orbital period distribution of the EBs for four $T_{\rm eff}$ bins. The top panel shows histograms of the distribution, while the bottom panel shows the cumulative fraction of EBs with period for each temperature range. The period bins have equal width in log-space.}
    \label{fig:hist_period_teff}
\end{figure}

Dividing the targets according to temperature results in different overall period distributions. Figure \ref{fig:hist_period_teff} shows the period distributions in each temperature range. Generally, the lower temperature systems reach shorter orbital periods than the higher temperatures, although the lowest temperature bin shows a flatter distribution than the others. The cumulative fraction of the lowest $T_{\rm eff}$ range crosses over from highest to lowest between about 0.7 to 2.6 days orbital period. Kolmogorov–Smirnov tests (KS-tests) of the four period distributions show that they are significantly different, with p-values staying below $10^{-9}$. Taking the fifth percentile of each distribution for an objective measure of their short period limit gives 0.53, 0.60, 0.79, and 0.95 days ($\pm\sim0.06$) from low to high $T_{\rm eff}$.

\begin{figure*}
\sidecaption
\includegraphics[width=12cm,clip]{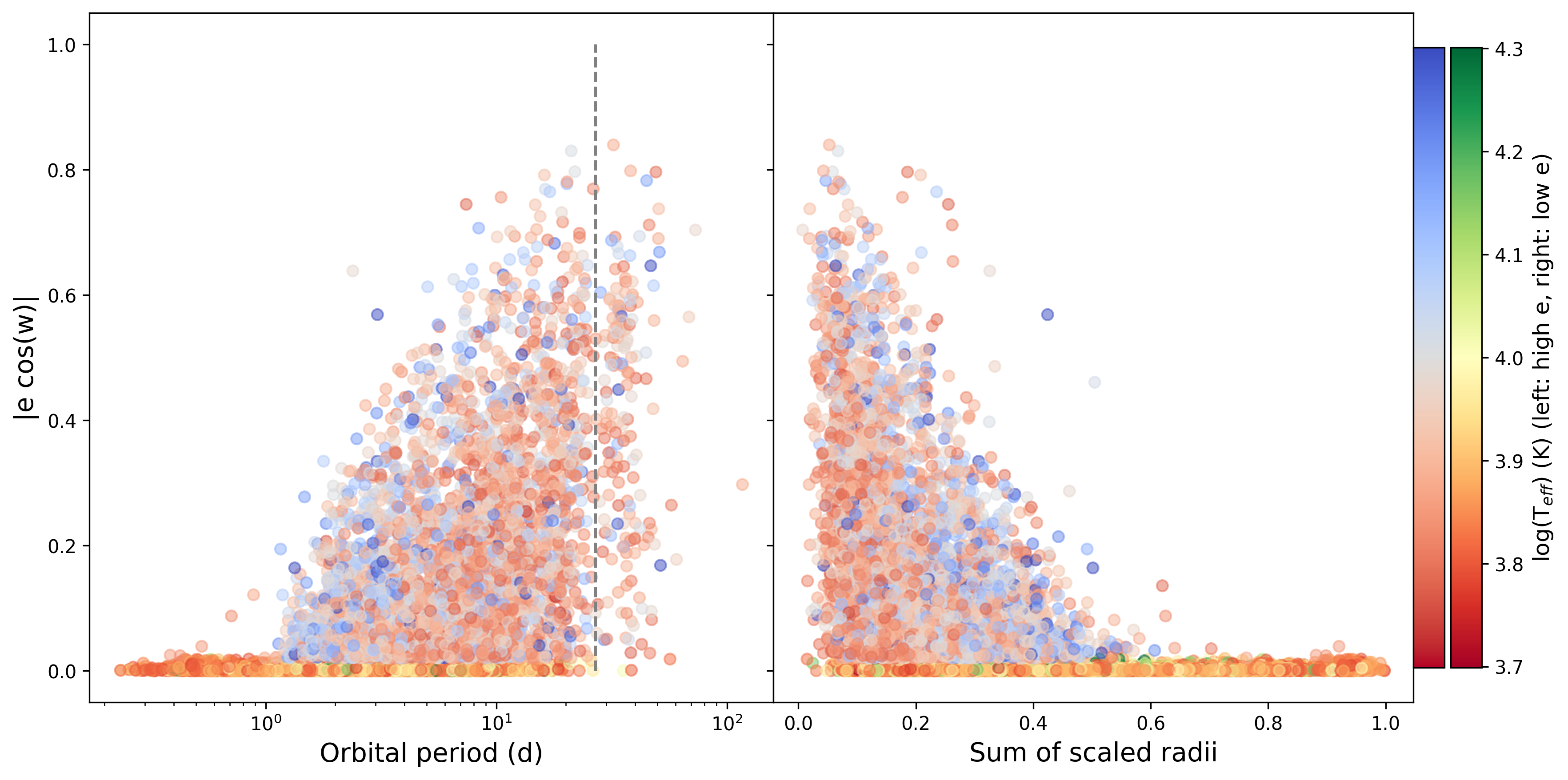}
    \caption{Tangential component of eccentricity as a function of orbital period and sum of scaled radii. Effective temperatures are indicated by the colour gradient. The grey dashed line indicates the dip in the orbital period distribution at 27 days. Systems with high and low eccentricity are shaded with different colour maps (left and right colour bars, respectively).}
    \label{fig:e-p}
\end{figure*}

For the reason of accuracy, and analogous to many other observational studies \citep[e.g.,][]{vaneylen2016, Justesen2021, Zanazzi2022}, we take the absolute value of the tangential component of eccentricity $|e \cos(\omega)|$ as a proxy for the eccentricity of a system. Doing this might label eccentric systems as circular, but this can only happen in a small part of the parameter space where the major axis of the orbit is near alignment with our line of sight. In Figure \ref{fig:e-p} we plot $|e \cos(\omega)|$ as a function of orbital period and as a function of the sum of radii. The colour scale indicates effective temperature, with different colour maps for high and for low eccentricity. The overall trend is a clear strong dependence of the eccentricity on the abscissa in both cases. Measurements of the sum of radii are likely to be overestimated above about 0.7; many of these systems were seen to have wide eclipses with strong ellipsoidal variability that shapes the eclipses more like a sine wave. The analysis methodology regularly puts the eclipse edges too far away from the eclipse centre. For our current analysis, this is inconsequential, as nearly all of the systems at higher values have circularised. Figure \ref{fig:e-p_log} provides a version of the same plot with logarithmically scaled $|e \cos(\omega)|$.

\section{The eccentricity distribution}

\begin{figure}
\centering
\includegraphics[width=\hsize]{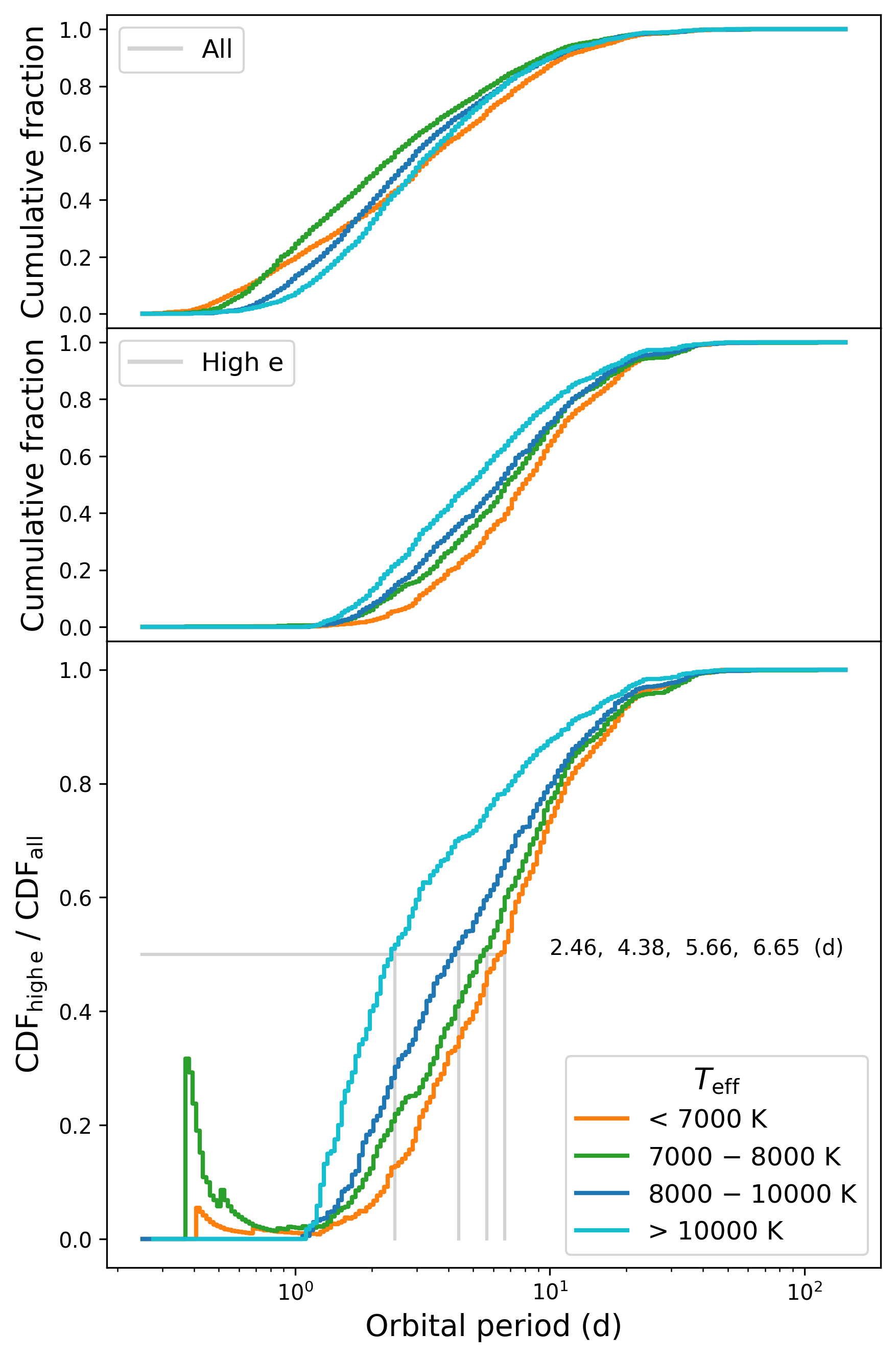}
    \caption{Cumulative distributions of period and their ratio for each temperature bin. The top panel is the CDF over all systems, in the middle is the CDF for eccentric systems and the bottom shows the ratio of the CDF for high eccentricity and the CDF for all systems. Measurements of the period at a fraction of 0.5 are indicated in the plot.}
    \label{fig:cdf_period}
\end{figure}

We see a similar large overlap between systems that are eccentric and systems that are circular as noted by \citet{Zanazzi2022}, and following them we measure an envelope period $P_{\rm env}$ as well as an overall circularisation period $P_{\rm circ}$ where most EBs circularise (see also, for example, \citet{Mazeh2008}). The envelope period traces the point where the outer envelope of the most eccentric systems reaches a circularised state. We further add a measurement of the so called `cold core' $P_{\rm core}$ which traces the longest periods and separations where circular systems are found. However, our method of obtaining these critical period measurements differs from that by \citet{Zanazzi2022}. We compute the cumulative period distribution of all systems in each temperature bin separately, by taking the cumulative sum of the logarithmic histogram of periods with 200 bins. This is repeated for the eccentric systems, and the two cumulative distribution functions (CDFs) are divided by each other. This gives the fraction of the probability of finding an eccentric system below a given period over the probability of finding any system below that period. $P_{\rm circ}$ is taken as the period at which this fraction equals 0.5, as illustrated in Figure \ref{fig:cdf_period}. The probability of finding an eccentric system at or below this period is half that of finding any system at or below this period. $P_{\rm env}$ and $P_{\rm core}$ are both measured similarly, but different from $P_{\rm circ}$. The envelope period is determined by taking the 5th percentile of eccentric systems, corresponding to the point below which 5\% of the eccentric systems are found. Conversely, the core period is determined from the 95th percentile of circular systems, corresponding to the period below which we find 95\% of all circular systems. These measurements and the accompanying period distributions are shown in Figures \ref{fig:period_perc_1} and \ref{fig:period_perc_2}. For an estimate of the measurement error on the percentile values, we consider the standard error on the mean of a sample for inspiration: $\sigma/\sqrt{n}$, where $\sigma$ is the standard deviation and $n$ the sample size. Since we are looking at a distribution that is approximately normal in log-space, instead we recall that the geometric mean is equal to the arithmetic mean in log scale. The geometric mean has standard error $G\cdot\sigma_{\rm log}/\sqrt{n}$ \citep{Norris1940}, with $G$ the geometric mean and $\sigma_{\rm log}$ now computed from the sample in log-space. We are computing percentiles far away from the median, in the tails of the distribution, so we are more sensitive to outliers. Hence, as a rough estimate, we reduce the effective sample size to the amount left in the tail that the percentile cuts from the distribution (5\% of the sample size for the 5th and 95th percentiles). This results in the following approximation for the statistical error in the measured percentile values:

\begin{equation}
    \sigma_{P_k} = \frac{\exp(P_k(log(x_i)))\, \sigma_{\rm log}}{\sqrt{\left(0.5 - abs\left(0.5 - \frac{k}{100}\right)\right)n}}, 
    \label{eq:perc_err}
\end{equation}

\noindent with $x_i$ the parameter values and $k$ the percentage at which the percentile $P_k$ is calculated. This equation ignores the asymmetry of the errors in linear space. To calculate errors in the values for the circularisation period determined from the division of two CDFs, we follow the same logic of percentiles (with the same caveats) since the measurements are closely related. The effective sample size is now much larger, as we are near the middle of the distributions. We take the sample size $n$ to be the number of eccentric systems in consideration and compute Equation \ref{eq:perc_err} at $k=50$ (the median), for which the effective sample size is 50\% of the total. The resulting errors are larger than the bin widths used in the determination of the CDFs and changing the number of bins does not significantly impact our results.

In the low temperature systems, we find $P_{\rm env} = 2.51\pm0.26$ days, $P_{\rm circ} = 6.65\pm0.27$ days, and $P_{\rm core} = 8.3\pm0.7$ days. Between $T_{\rm eff}$ of 7000 and 8000\,K, we measure $P_{\rm env} = 1.98\pm0.24$ days, $P_{\rm circ} = 5.66\pm0.27$ days, and $P_{\rm core} = 6.8\pm0.5$. For systems within 8000-10\,000\,K, the values are $P_{\rm env} = 1.90\pm0.20$ days, $P_{\rm circ} = 4.38\pm0.22$ days, and $P_{\rm core} = 7.7\pm0.5$. Finally, for the high-temperature EBs, we have $P_{\rm env} = 1.62\pm0.17$ days, $P_{\rm circ} = 2.46\pm0.17$ days, and $P_{\rm core} = 8.2\pm0.6$. To verify that our measurements of $P_{\rm circ}$ are statistically distinct, we perform KS-tests of the period distributions at high eccentricity in each temperature regime. All pairs of distributions have p-values well below 0.05, indicating that they differ significantly. The largest p-value found is 0.012, for the distributions in the two middle temperature bins. This reinforces the error estimates obtained for the $P_{\rm circ}$ values, which also show a significant difference between the temperature regimes. The relatively high p-value for the period distributions in the middle two temperature ranges also substantiates the fact that their tails ($P_{\rm env}$) are statistically indistinguishable according to their errors. 

We observe two trends in the critical periods: $P_{\rm env}$ and $P_{\rm circ}$ decrease with temperature, while $P_{\rm core}$ first decreases and then comes back up nearly to where it started. Moreover, the decrease in $P_{\rm env}$ is less strong than that of $P_{\rm circ}$, but still significant over the whole range of temperatures. The differing trends also mean that the EBs above 10\,000\,K circularise over the largest range of periods of 6.5 days, and systems within 7000-8000\,K circularise within the smallest range of 4.8 days width.

The rate of circularisation is predicted by the tidal theory to strongly depend on the radius of a star scaled by the orbital separation $R_1/a$, and therefore \citet{Justesen2021} focus mainly on the scaled orbital separation $a/R_1$ that can be measured for EBs. The analysis methodology we employ is shown to produce rather inaccurate values of the radius ratio (see IJ24), so we opt to instead use the much more reliable measurements of the sum of scaled radii, and invert it to get another measure of the scaled orbital separation $a/(R_1 + R_2)$. This parameter is less directly related to the theoretically derived equations of circularisation time scales. However, as \citet{Claret1995} note, these are often expressed in terms of the primary component only. Taking into account the contributions of both components means inversely adding the time scales \citep[][Equation 10]{Claret1995}, reducing the total circularisation time. If one component has a much longer time scale, its contribution to the final time scale is small. Since larger stars (in terms of $R/a$) result in much shorter time scales, adding the contribution of a small companion has little effect on that time scale. Therefore, the inverse of the sum of scaled radii should be a reasonable proxy for $a/R_1$ to evaluate tidal circularisation. 

\begin{figure}
\centering
\includegraphics[width=\hsize]{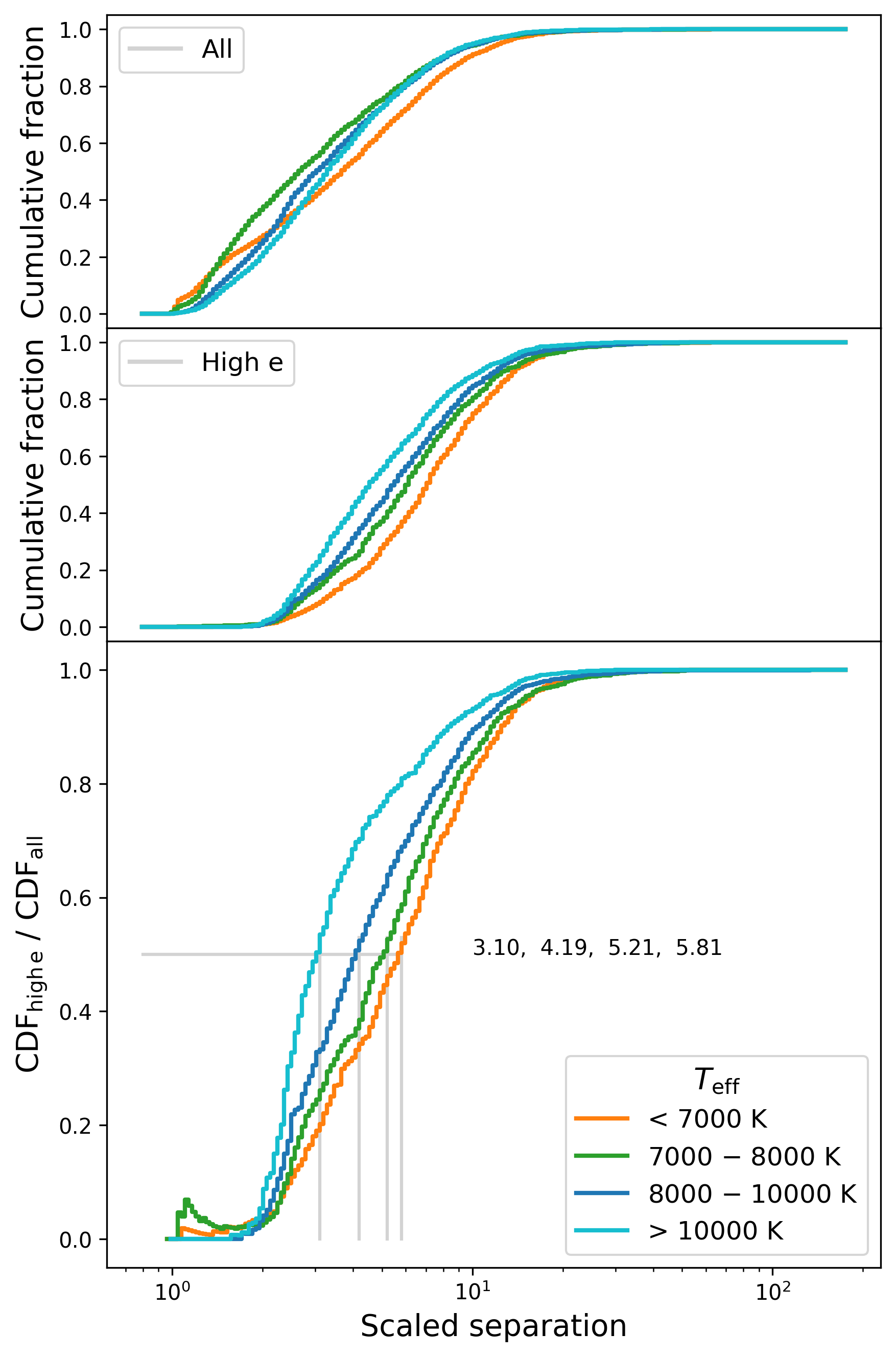}
    \caption{Same as in Figure \ref{fig:cdf_period}, but for the scaled orbital separation. The top panel is the CDF over all systems, in the middle is the CDF for eccentric systems and the bottom shows the ratio of the CDF for high eccentricity and the CDF for all systems. Measurements of the separation at a fraction of 0.5 are indicated in the plot.}
    \label{fig:cdf_rsum}
\end{figure}

We abbreviate $(a/(R_1 + R_2))_{\rm crit}$ as $(a/R_s)_{\rm crit}$ in the following, for brevity. Measurements of $(a/R_s)_{\rm crit}$ are performed analogously to those for the critical periods, and shown in the corresponding Figures \ref{fig:cdf_rsum}, \ref{fig:rsum_perc_1}, and \ref{fig:rsum_perc_2}. We find critical scaled separations in the lowest temperatures of $(a/R_s)_{\rm env} = 2.84\pm0.23$, $(a/R_s)_{\rm circ} = 5.81\pm0.19$, and $(a/R_s)_{\rm core} = 8.2\pm0.5$. Systems in the lower middle effective temperature bin give values of $(a/R_s)_{\rm env} = 2.54\pm0.24$, $(a/R_s)_{\rm circ} = 5.21\pm0.18$, and $(a/R_s)_{\rm core} = 6.8\pm0.3$. The upper middle temperatures have measurements of $(a/R_s)_{\rm env} = 2.47\pm0.19$, $(a/R_s)_{\rm circ} = 4.19\pm0.14$, and $(a/R_s)_{\rm core} = 7.2\pm0.4$. Lastly, for the highest temperature EBs we observe $(a/R_s)_{\rm env} = 2.37\pm0.17$, $(a/R_s)_{\rm circ} = 3.10\pm0.11$, and $(a/R_s)_{\rm core} = 7.3\pm0.4$. Again, KS-tests of the distributions of $(a/R_s)$ at high eccentricity show that they are significantly different, with the largest p-value of 0.002 between the middle two temperature regimes. We see similar trends in these separations as for the periods, although $(a/R_s)_{\rm env}$ does not decrease as much as $P_{\rm env}$ over the four temperature regimes, and $(a/R_s)_{\rm core}$ does not increase as far back to its starting point in the lowest temperatures. The decrease of $(a/R_s)_{\rm env}$ with temperature is only marginally significant, with a difference of $2.03\sigma$ between the two furthest values, while for $P_{\rm env}$, this difference is $3.4\sigma$.

The obtained values of critical periods and scaled separations are summarised in Table \ref{tab:crit}. We note a few peculiarities about the shapes of the various distributions. The period distribution of circular systems in the 7000-8000\,K bin has a markedly steep low-period end; in other words, it is skewed towards short periods. The next higher $T_{\rm eff}$ category shows this to a lesser degree, while the others are more symmetrical. The skewness of the distribution of circular systems within 7000-8000\,K is even more pronounced in scaled separation, looking almost triangular. We also see an excess of low scaled-separation systems among the lowest temperatures. The distribution of scaled separations for $T_{\rm eff}$ above 10\,000\,K is seemingly more skewed towards the low end than any of the other temperature ranges.

Table \ref{tab:crit_sub} contains additional statistics analogous to Table \ref{tab:crit}, for subdivisions of the four temperature ranges. The main three effective temperature boundaries of our four ranges are kept, and each range is further split up into three according to a roughly equal sample size of 1000 systems. The new temperature boundaries are given in the first column of Table \ref{tab:crit_sub}, and the sample size is in the second column. The contents of both tables are visualised in Figures \ref{fig:crit_period_subbins} and \ref{fig:crit_separation_subbins} for the critical orbital periods and orbital scaled separations, respectively. Error bars on the measurements are larger in proportion to the smaller sample sizes, and the majority of the measurements is indistinguishable from the `parent' temperature range. Most of the visible trends are already captured by the rougher partitioning, with the exception of a seemingly significant downward trend of $(a/R_s)_{\rm core}$ with increasing temperature above 10\,000\,K.

\section{Comparison to other studies}

We compare our results to two recent studies of large samples of EBs. \citet{Zanazzi2022} has a sample of 524 EBs with temperatures between 4500 and 7000\,K, corresponding to our lowest $T_{\rm eff}$ range. Using their technique of fitting a parametric function to $|e \cos(\omega)|$ as a function of orbital period, they find $P_{\rm circ} = 6.2_{-0.8}^{+1.4}$. This matches what we found for our lowest temperature regime ($P_{\rm circ} = 6.65\pm0.27$). Fitting the parametric function to the binned maximum $|e \cos(\omega)|$ values, the envelope period they obtain is $P_{\rm env} = 3.2_{-0.3}^{+0.6}$, not in agreement with our, more precise value ($P_{\rm env} = 2.51\pm0.26$). The same author argues that the envelope period was missed by \citet{Meibom2005} because of the difference in sample size. \citet{Zanazzi2022} simulates a smaller dataset and finds that they could indeed not distinguish the envelope and circularisation periods if the sample size is below about 50. The difference becomes mildly significant for a sample size of 100. The difference we find for the envelope period may also arise due to the difference in sample size: our characterised EB sample within this temperature range has approximately 6 times more systems. 

\begin{table}
	\centering
	\caption{Summary of critical values of period and scaled separation for the characterised EBs.}
	\label{tab:crit}
	\begin{tabular}{c l l l l}
	\hline
    $T_{\rm eff}$ & < 7 kK & 7 - 8 kK & 8 - 10 kK & > 10 kK \\
	\hline
	$N_{\rm EB}$ & 3067 & 3175 & 3521 & 2934 \\
    $P_{\rm env}$ (d) & $2.51{\scriptstyle\pm0.26}$ & $1.98{\scriptstyle\pm0.24}$ & $1.90{\scriptstyle\pm0.20}$ & $1.62{\scriptstyle\pm0.17}$ \\
    $P_{\rm circ}$ (d) & $6.65{\scriptstyle\pm0.27}$ & $5.66{\scriptstyle\pm0.27}$  & $4.38{\scriptstyle\pm0.22}$ & $2.46{\scriptstyle\pm0.17}$ \\
    $P_{\rm core}$ (d) & $8.3{\scriptstyle\pm0.7}$ & $6.8{\scriptstyle\pm0.5}$ & $7.7{\scriptstyle\pm0.5}$ & $8.2{\scriptstyle\pm0.6}$ \\
    $\frac{a}{R_s}_{\rm env}$ & $2.84{\scriptstyle\pm0.23}$ & $2.54{\scriptstyle\pm0.24}$ & $2.47{\scriptstyle\pm0.19}$ & $2.37{\scriptstyle\pm0.17}$ \\
    $\frac{a}{R_s}_{\rm circ}$ & $5.81{\scriptstyle\pm0.19}$ & $5.21{\scriptstyle\pm0.18}$ & $4.19{\scriptstyle\pm0.14}$ & $3.10{\scriptstyle\pm0.11}$ \\
    $\frac{a}{R_s}_{\rm core}$ & $8.2{\scriptstyle\pm0.5}$ & $6.8{\scriptstyle\pm0.3}$ & $7.2{\scriptstyle\pm0.4}$ & $7.3{\scriptstyle\pm0.4}$ \\
	\hline
	\end{tabular}
\end{table}

\citet{Zanazzi2022} notes that convective damping of the equilibrium tide is only able to circularise systems up to $\thicksim$2 days on the pre-main-sequence (PMS) and little beyond that \citep{Goodman1997, Goodman1998, Barker2020, Zanazzi2021}. This is due to an overestimation of the effective viscosity by \citet{Zahn1977} \citep[see, e.g.,][]{Penev2009, Duguid2020}. Our $P_{\rm env}$, lower than the one found by \citet{Zanazzi2022}, likewise supports this level of effectiveness of convective damping. We similarly cannot rule out the slightly higher critical periods ($\thicksim$3-4 days) calculated for the mechanism of resonance locking of the dynamical tide on the PMS for solar-type stars \citep{Zanazzi2021}. \citet{Zanazzi2022} further expresses the presence of a `cold core' of circular binaries out to long orbital periods, but does not measure a critical period for them. It is mentioned that circular systems are found for orbital periods as long as $\thicksim$10-20 days. We confirm this from our sample, with a few cases even beyond that. Our measurement of the core period of $P_{\rm core} = 8.3\pm0.7$ suggests that the circular systems at longer periods are in the tail end of the distribution, as 95\% of circular EBs are found below $P_{\rm core}$. 

It is pointed out by \citet{Goodman1998}, \citet{Ogilvie2007}, \citet{Barker2010}, \citet{Barker2011}, and \citet{Barker2020} that nonlinear wave-breaking increases the effect of dynamical tides in MS stars with radiative cores and convective envelopes. \citet{Goodman1998} calculate a period of circularisation of 4 to 6 days on the MS with the inclusion of nonlinear wave-breaking and evolving resonant frequencies. This comes close to our measured circularisation period of the bulk of systems, $P_{\rm circ}$, with $T_{\rm eff}$ below 7000\,K. Since this mechanism is effective over the lifetime of the MS, different stellar ages will result in a spread of circularised and eccentric systems across a range of periods. This makes it more compatible with observations than a mechanism only effective during the PMS phase. However, the highest value of 6 days falls short of $P_{\rm core}$, and the lower end of the period range is too long to explain the eccentric systems close to $P_{\rm env}$.

Tidal theory has to be able to explain the presence of circularised systems over a wide range of orbital periods in coexistence with eccentric systems. The measurements of $P_{\rm env}$ and $P_{\rm core}$ for low temperatures indicate that this range is around 5.8 days wide. If the difference is due to age alone, \citet{Zanazzi2022} introduced a parameter $\eta$ denoting the strength of dependence on the orbital period of the circularisation time scale (their Equation 4). It is further shown that an estimate for the upper bound of $\eta$ (2.4) is lower than that of any tidal dissipation mechanism. Giving a wider age difference between the youngest and oldest stars of the sample increases the $\eta$ found, as does decreasing the period range. We take the same $t_{\rm young} = 1\,Gyr$, which is lower than deemed likely for Galactic disk stars, but in line with some of the clusters of \citet{Meibom2005}. We increase the upper age to $t_{\rm old} = 10\,Gyr$, a relatively high estimate according to \citet{Kilic2017}. If we now take a period interval reduced by one $\sigma$ from both sides and compute $\eta$, we find a value of 2.3 as an estimate for the upper bound. The value for $\eta$ strongly depends on the $t_{\rm young}$ taken for the sample. This indicates that a spread in age alone is unable to reproduce the observed population with current theories.

Two other pathways are offered as an explanation by \citet{Zanazzi2022}: circularisation at birth of the binary as a result of interactions with a disk, and selective circularisation of binaries with specific properties (like fast rotation). While circularisation at birth helps explain some aspects of the observations, it fails to account for the observed distribution of eccentricities at a given orbital period. Furthermore, while selective circularisation for binaries with certain properties is promising, there is no clear consensus as to the main mechanism. Tidally excited inertial waves \citep{Ogilvie2007} are suggested to be more efficient in synchronising and subsequently circularising fast rotators. The calculations by \citet{Barker2022} support the idea of differential circularisation over different rotation rates, but the critical periods are too high to explain $P_{\rm env}$ and $P_{\rm circ}$. Even for the slower rotators it is found that the critical period, which is nearly constant on the MS, is about 7 days. \citet{Barker2022} further shows that the viscous dissipation of turbulent flows by Zahn had initially made use of an unphysically high effective viscosity resulting in much higher critical periods. This illustrates the effect that simplifying assumptions made for the input physics or initial conditions can have on the predictions of a tidal theory mechanism. We surmise that such assumptions in other mechanisms, for example, near-synchronicity or equal mass ratio, could cause spread in the predicted critical periods for stars with varying initial properties. This spread would be multiplicative with that expected from the spread in binary ages.

The sample by \citet{Justesen2021} comprises 809 EBs ranging in temperature from 3000 to 50\,000\,K, and the overall distribution of $T_{\rm eff}$ is concentrated between 5000-9000\,K. They select slightly different temperature ranges: the lowest one (<4500\,K) falls outside our sample, the second lowest corresponds roughly to the lower half of our lowest range, while their third bin ranges from 6250 to 8000\,K and overlaps with our two lowest domains. Their two highest temperature bins are the same as ours at 8000-10\,000\,K and the last bin includes everything above that. Furthermore, they work mostly with the quantity of scaled orbital separation, but calculate it as $a/R_1$ instead of our $a/(R_1 + R_2)$. However, having individual temperatures for their binaries they also select systems with $T_{\rm eff}$ within 1000\,K for approximately equal mass ratios. This means that it is not too inaccurate to halve their $a/R_1$ values for comparison, as the companion would be of similar size as the primary. Nevertheless, this means that our values for the scaled separation come from a larger mix of radius ratios. \citet{Justesen2021} do provide measurements of circularisation period alongside the circularisation separation in their Table 4. These are obtained analogously to \citet{Zanazzi2022}, by fitting a (similar) parametric function to the values of $|e \cos(\omega)|$.

The systems between 4500 and 6250\,K are measured by \citet{Justesen2021} to have $P_{\rm circ} = 9.0_{-2.0}^{+5.5}$ and $(a/R_1)_{\rm circ} = 18.0_{-0.86}^{+3.1}$ ($9.0_{-0.4}^{+1.6}$ when halved). To better compare our measurements, we compute values for the same temperature regime: $P_{\rm circ} = 7.32$ and $(a/R_s)_{\rm circ} = 6.84$. Even though the circularisation period agrees within $1\sigma$, the scaled separation measurement is substantially smaller, which we ascribe to the larger relative number of cooler stars their sample has within this $T_{\rm eff}$ bin. The better agreement between the measurements in the next temperature bin corroborates this idea, although it is not ruled out that the fitting method is simply more sensitive to the `cold core' of circular systems. 
From 6250-8000\,K the critical values are close to ours: $P_{\rm circ} = 5.5_{-0.29}^{+3.3}$ and $(a/R_1)_{\rm circ} = 11.4_{-0.48}^{+3.5}$ ($5.7_{-0.24}^{+1.8}$), where we found $P_{\rm circ} = 5.85$ and $(a/R_s)_{\rm circ} = 5.35$. 
In the systems within 8000-10\,000\,K, they find: $P_{\rm circ} = 5.7_{-1.9}^{+2.4}$ and $(a/R_1)_{\rm circ} = 8.4_{-1.9}^{+5.0}$ ($4.2_{-1.0}^{+2.5}$), while our values in this range are $P_{\rm circ} = 4.38$ and $(a/R_s)_{\rm circ} = 4.19$, well within the error estimates.
In the highest temperature range, the authors measure $P_{\rm circ} = 2.8_{-0.13}^{+0.84}$ and $(a/R_1)_{\rm circ} = 3.8_{-0.3}^{+4.7}$ ($1.9_{-0.15}^{+2.4}$). This is in line with what we find for the scaled separation at $(a/R_s)_{\rm circ} = 3.10\pm0.11$, but at the edge of our reported $2\sigma$ error margin for the critical period $P_{\rm circ} = 2.46\pm0.17$. 

\citet{Justesen2021} find good agreement with the theory below 6250\,K, but we note that the theory presented in \citet{ZahnBouchet1989} was used for these temperatures. As stated above, this theory was later updated with more realistic values for the effective viscosity and found to be unable to circularise binaries in orbits much longer than 2 days. The high $P_{\rm circ}$ measurements below 6250\,K could be well explained by the selective inertial wave mechanism proposed by \citet{Barker2022}. However, the envelope period that we compute for temperatures below 6250\,K of $P_{\rm env} = 3.00$ is in tension with the efficiency of this mechanism. It is likely that this measurement would increase if our sample contained more targets below 6000\,K, but it seems implausible that it would increase to the $\sim7$ day circularisation period that the slowest rotators achieve in this formalism. A further extension of the sample to lower temperatures would be necessary to confirm this.

In the hotter stars between 7000\,K and 10\,000\,K, \citet{Justesen2021} report an increased amount of circularisation compared to the theory of radiative dissipation by Zahn \citep[using][Equation 5.9]{Zahn1977}. Their theoretical calculations only produce circularised systems to 1-2 days and scaled separations of 3-4 (1.5-2 $(a/R_s)$), which they calculated using half the MS lifetime as stellar age. Our results, therefore, support the notion that non-resonant radiative dissipation of the dynamical tide is an insufficient mechanism to circularise systems with radiative envelopes. Resonance locking of the dynamical tide is proposed as a mechanism for increased tidal dissipation in the two problematic temperature domains. Indeed, if we look at the calculations by \citet{Zanazzi2021} for resonance locking in stars of 1, 1.5, and 2 solar masses, we see circularisation up to 3 days for the lower masses and 4 days for the high mass end ($(a/R_s) \thicksim4$). These masses align nicely with our middle two temperature bins. In the 8000-10\,000\,K range this matches the circularisation period and $(a/R_s)_{\rm circ}$. However, the critical periods of the eccentricity envelope agree well with the non-resonant theory for temperatures above 7000\,K. This would support the idea that the non-resonant mechanism could be effective on the PMS, while another mechanism (or mechanisms), like resonance locking, is needed for selective additional circularisation on the MS. That would explain the high values for $P_{\rm core}$ in terms of a significant difference in tidal dissipation experienced by systems with different initial conditions.

For the highest temperatures (> 10\,000\,K), our bulk circularisation period value is more compatible with the theoretical values of 1-2 days than the one found by \citet{Justesen2021}. Our scaled separation is disproportionally less compatible with the theoretical $(a/R_1)$ of 2-4 ($(a/R_s)$ of 1-2), although still close in absolute value. Our sample sizes in the two highest temperature bins are two orders of magnitude larger, so our measurements should be statistically more robust. Moreover, looking at the measurements of envelope period and scaled separation, these seem quite consistent with the non-resonant theory in this case. However, in line with the lower temperature regimes, and in contrast with the conclusions by \citet{Justesen2021}, the period and scaled separation of the circular core need an increased amount of circularisation to be explained.

\citet{Justesen2021} consider the existence of a hot (18\,000\,K) circular binary at large separation of $(a/R_1) = 8.6$. Taking half of that separation for comparison to our scaled separations, we find this binary well within our $(a/R_s)_{\rm core}$ value of $7.3\pm0.4$, indicating that it is less special than the sample it originates from might suggest.\footnote{The binary in question is actually present in our catalogue, but was flagged as incorrectly analysed due to the presence of data artefacts.} 

For the hottest stars in our sample (early B and O), a portion is expected to have undergone binary interaction. \citet[][Figure 3]{deMink2014} show that around 11\% of systems detectable as binaries could have a rejuvenated (i.e. looking like a younger star) companion resulting from mass transfer. Our EBs are not selected through radial velocity monitoring, but we expect that the selection of EBs would result in similar enough ratios compared to those found from an idealised radial velocity criterion. These post-interaction systems would have circularised during the mass-transfer phase, so this could potentially explain circular systems at longer periods than the tidal mechanisms. We find that 19\% of systems with $T_{\rm eff}$ above 10\,000\,K having an orbital period above $P_{\rm core}$ are circular. This fraction decreases quickly for longer periods, reaching 11\% at $P > 11$ days. These longest-period circular binaries could therefore be explained by interaction, although around and below the cold core period of 8.2 days it is likely not the sole explanation.

\section{Pulsations}

\begin{figure}
\centering
\includegraphics[width=\hsize]{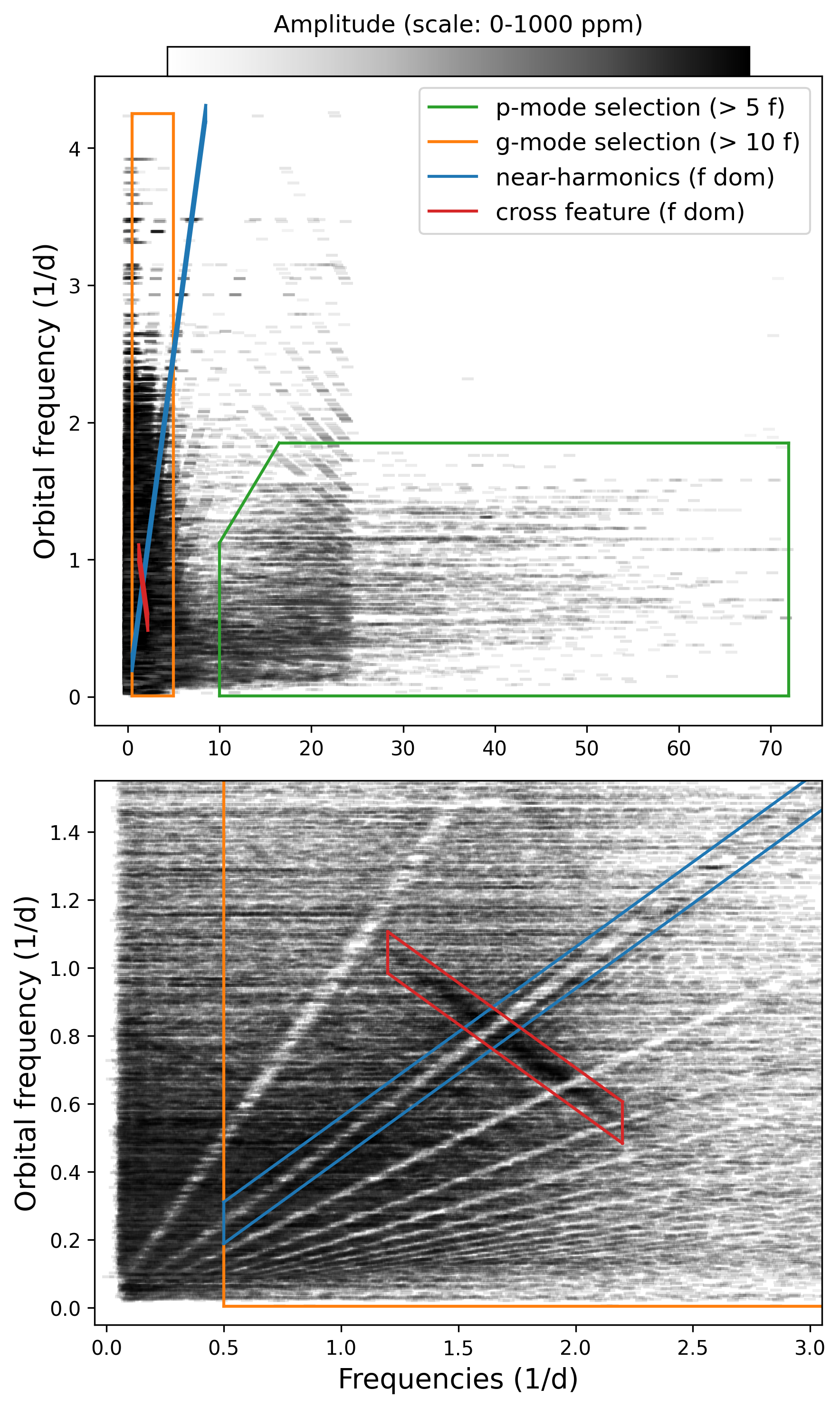}
    \caption{Overview of the extracted significant frequencies with the orbital frequency on the vertical axis. Each horizontal line contains the frequencies (dashes) of one EB, coloured by amplitude. The boxed-in regions demarcate some of the selection criteria going into the four selected groups.}
    \label{fig:freqs}
\end{figure}

A first in the study of close binary circularisation is the bulk analysis of light curves in terms of constituent sinusoids. 
We explore some of our findings in the wealth of data provided by our light curve analysis. First, we explain the selection processes going into the four categories of systems we will discuss. One of the most informative representations of this data is the plot of orbital frequency versus the frequency of the extracted significant sinusoids, scaled in colour by their amplitudes. This plot is shown in Figure \ref{fig:freqs}, and includes marked regions representing the selections. All sinusoidal signals discussed here were obtained from a full iterative prewhitening procedure as implemented by IJ24, before a physical light curve model of the eclipses was fitted. This prevents us from interpreting signals that may be introduced into the light curve by the residuals of such a fit. The frequencies plotted exclude the exact orbital harmonics that form an empirical model of the eclipses. This means we have no direct grasp on features like the dynamical tide, or the equilibrium tide for that matter. Also removed are the frequencies we call the `near-harmonics', shown as lighter lines in the bottom panel. They were selected within a boundary of $1.5/t_{\rm tot}$ around the harmonics, where $t_{\rm tot}$ is the time base of observations. This was done because they were a mix of real and artificial signals. Even so, we see multiple additional less prominent dark lines parallel to the lines of exact harmonics, mainly around the second harmonic.

The real signals close to harmonic frequencies are thought to be comprised of pulsations that may have a tidal origin. We capture part of it by selecting the systems that have their dominant frequency near the second harmonic ($2 f_{\rm orb}$), which had the darkest line in the $f_{\rm orb}$-$f_{\rm puls}$ plot. These targets are too contaminated by artificial signals to do much analysis without further filtering. The artificial signals found at these frequencies could have several origins, connected to the method of sinusoid extraction. One possibility is that the orbital period is not obtained accurately enough and, therefore, the eclipses are not completely removed, leading to leftover power near the harmonics. A second option is that the finite harmonic series used to describe the eclipses is unable to reach a near-constant flux level out-of-eclipse, therefore injecting variability that is then necessarily described by frequencies close to the harmonics. \citet{Hey2024} find similar signals in thousands of new gravity-mode pulsators observed by TESS, discovered from \textit{Gaia} DR3 light curves. We leave this group for closer study at a later time.

The second feature that catches the eye in the bottom panel of Figure \ref{fig:freqs} is the dark `cross' of frequencies going down from left to right at the same (but negative) slope as the second harmonic. This is not an isolated feature, as we see a weaker version of it at the first harmonic and at the same frequency range. Yet another slope (corresponding to the fourth harmonic) is found to the left of the main cross at the same orbital frequency range, vaguely visible in the same plot. Lastly, multiple comparable structures can be identified below the main cross feature. All of these structures have in common that they are mirrored around a frequency of $\thicksim$1.645\,d$^{-1}$ compared to the harmonics. We only select systems with their dominant frequency within the red parallelogram. A fifth of the 607 selected targets have a light curve in both the QLP and SPOC datasets. Comparing the two reveals that the likely cause for the cross feature is an artefact from the QLP data reduction pipeline. Picking the frequencies of these targets only from the SPOC data and plotting it in the $f_{\rm orb}$-$f_{\rm puls}$ plane reveals the absence of any cross feature. We find numerous cases where the ellipsoidal variability, present in the SPOC light curve, is suppressed in the QLP light curves. The ellipsoidal variability can in some cases even be inverted with respect to the median flux level in the QLP data. Two examples are given in Figure \ref{fig:ev_gone}.

We attempt to select g-mode pulsations from among the many frequencies found in their variability regime. For this selection, we remove additional frequencies close to the harmonics by increasing the boundary to $2.5/t_{\rm tot}$. The selection then only retains sources with more than 10 significant frequencies between 0.5 and 5 d$^{-1}$, a requirement met by 2887 sources ($\thicksim$20\% of the characterised sample). The p-mode pulsators are relatively easy to pick out in Figure \ref{fig:freqs}, although we take a conservative lower limit to their frequency selection of 10\,d$^{-1}$. More than 5 significant frequencies are required to be present within the green box to qualify as a p-mode oscillator, resulting in 677 sources ($\thicksim$4.6\% of the characterised sample). Included are 270 cases that are also in the g-mode category, thus labelled hybrids.

The described selections lead to the identification of g-mode pulsators, p-mode pulsators, and hybrids. We now further characterise these groups. Plots of the $T_{\rm eff}$ and $P_{\rm orb}$ distributions of two of the selected groups, believed to predominantly consist of pulsators, are included in Appendix \ref{apx:puls}. The g-mode pulsators are present at all temperatures above $\thicksim$5500-6000\,K, but have an overdensity at 7000\,K compared to the whole sample. The peak corresponds to the expected position of $\gamma$ Dor pulsators, and the temperature distribution continues up to the SPB domain, as previously also found from \textit{Gaia} DR3 sparsely sampled light curves \citep{DeRidder2023, Aerts2023}. A selection of examples of g-mode pulsators is shown in Figure \ref{fig:puls_g_mode}. They have an apparent bias to shorter orbital periods below about 8.4 days, as compared to the 16.2 days 95th percentile of all characterised EBs, although this could be due to the smaller sample size. A large portion of g-mode pulsators do have a dominant frequency in the close vicinity of an orbital harmonic, which might indicate many are plagued by the data analysis issues mentioned above, or otherwise point to a coupling with the tides. Since the frequencies resulting in the selection of these systems are in a more problematic area of the parameter space, it is difficult to make strong statements.

\begin{figure}
\centering
\includegraphics[width=\hsize]{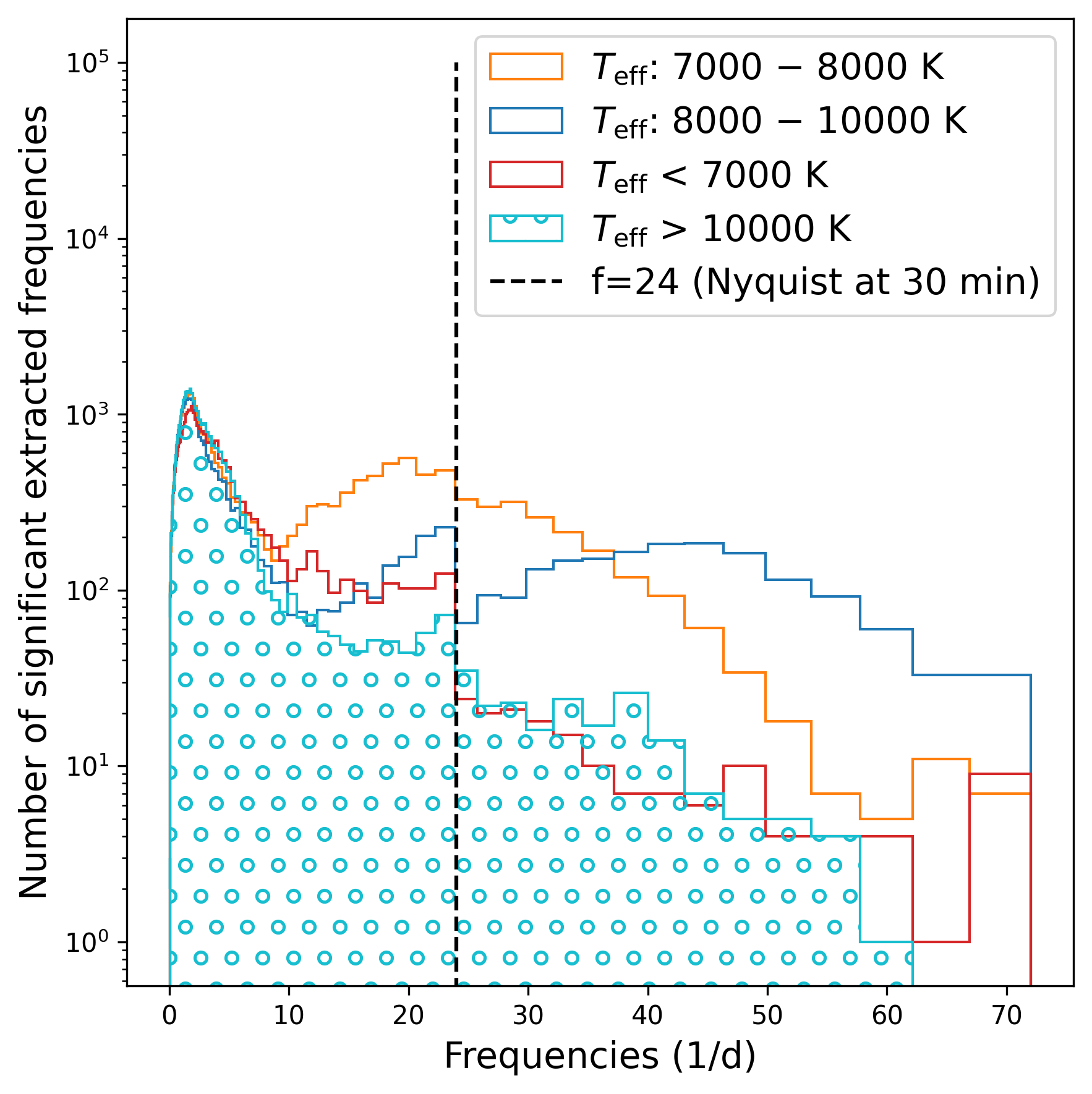}
    \caption{Histogram of the extracted significant frequencies in each temperature range. Bins are scaled logarithmically in size to better view the lower densities at high frequency. Included are all targets in the groups of p-mode pulsators and g-mode pulsators.}
    \label{fig:hist_freqs}
\end{figure}

The temperatures of the detected p-mode pulsators fall squarely between 7000 and 8000\,K, with few below and a modest amount up to 9000\,K. This is in good agreement with the instability strip observationally determined by \citet{Murphy2019} for $\delta$ Scuti pulsators. A selection of examples of p-mode pulsators is shown in Figure \ref{fig:puls_p_mode}. The p-mode pulsators also show a lower overall period distribution, which again could be sample size. Figure \ref{fig:hist_freqs} shows the distribution of all significant frequencies for pulsators in the p-mode and g-mode groups, separated by temperature domain. This shows a clear bulge of p-modes, that is emphasised in the same plot for just the p-mode pulsators (Figure \ref{fig:hist_freqs_p}). The frequencies in the higher temperature bin, 8000-10\,000\,K, are noticeably higher than in the range 7000-8000\,K. Additional plots are also provided for the hybrids (Figure \ref{fig:hist_freqs_h}) and g-mode pulsators (Figure \ref{fig:hist_freqs_g}).

\begin{table}
	\centering
	\caption{Fractions of eccentric to all binary systems in each temperature range, within orbital periods of 0.5 to 4 days.}
	\label{tab:frac}
	\begin{tabular}{c l l l l}
	\hline
    $T_{\rm eff}$ & < 7 kK & 7 - 8 kK & 8 - 10 kK & > 10 kK \\
	\hline
	All EBs & $9.8{\scriptstyle\pm0.8}\%$ & $9.6{\scriptstyle\pm0.7}\%$ & $15.1{\scriptstyle\pm0.9}\%$ & $27.4{\scriptstyle\pm1.4}\%$ \\
    g-modes & $5.4{\scriptstyle\pm1.2}\%$ & $6.0{\scriptstyle\pm1.1}\%$  & $12.9{\scriptstyle\pm1.7}\%$ & $23.1{\scriptstyle\pm2.3}\%$ \\
	\hline
	\end{tabular}
\end{table}

We determine eccentricity fractions between 0.5 and 4 days orbital period, a range selected to incorporate the majority of pulsators, while excluding the distribution tails with low numbers. The fractions of eccentric systems to the total for pulsators are $12.3\pm0.8\%$ for g-modes only, $9.5\pm2.1\%$ for hybrids and $7.6\pm1.5\%$ for p-modes only. This is compared to the whole sample between the same periods which gives $15.5\pm0.4\%$, and restricted to the middle two temperature bins it is $12.4\pm0.6\%$. Errors are determined using the square root of N, the number of selected cases, and the appropriate error propagation. The p-mode only and hybrid pulsators are mainly present in the middle two temperature bins and are compared to the overall sample in the range 7000-10\,000\,K. They show a reduction of 4.8\% at $3.2\sigma$ significance for p-modes and 2.9\% at $1.4\sigma$ significance for hybrids. The fraction varies with temperature, so for the larger group of g-mode (only) pulsators we include the fractions of eccentric systems as a function of temperature in Table \ref{tab:frac}. All show a significant reduction between 2.1 and 4.4\%. If the tides excite oscillations, these can in turn be dissipated within the star. The energy coming from the orbit that goes into driving the oscillation then has an additional path to being dissipated and the circularisation process could be enhanced. With gravity as the restoring force of the oscillation, one likely needs resonance locking with stellar eigenmodes for this to be effective in the long term \citep{Witte2002}. In the inertial regime, where the Coriolis force is dominant, oscillations could be excited and dissipated within the convective envelope over a broad range of frequencies \citep{Ogilvie2007}. Finding lower fractions of eccentric systems among pulsators might indicate the involvement of such processes.

\section{Conclusions}

We have presented a large new all-sky EB catalogue of 69\,000 targets observed by TESS in its first four years of observations. To search for EBs in millions of TESS light curves we implemented an RF classifier to augment the \texttt{ECLIPSR} algorithm. We analysed the light curves with automated methodology \texttt{STAR SHADOW} to characterise the EBs, and obtained orbital parameters for over 37\,000 of them. These results were manually vetted for correctness, to get a clean sample of 14\,500 for statistical analysis. We supplemented the catalogue with \textit{Gaia} effective temperatures and found that our initial photometric colour cuts have been effective at selecting sources in the regime of F-, A-, B-, and O-type stars, having temperatures above $\thicksim$6000\,K. 

The verified EB sample was statistically characterised in terms of the critical periods and scaled orbital separations to obtain empirical constraints on the theory of close binary tides. For this purpose, we followed \citet{Zanazzi2022} who introduced a measurement of the eccentricity envelope in the eccentricity-period diagram that traces the most eccentric systems, complementing the measurement of where the bulk of eccentric systems circularises. Taking this even further, we introduced a third measure that traces only the circular systems out to their furthest regions. These critical quantities are referred to with the subscripts `env' for the eccentric envelope, `circ' for the circularisation of the bulk, and `core' for the circularised `cold core'. We proposed a different way to obtain these measurements, based on statistical methods, rather than fitting a parametric function to our data. The envelope and core quantities are the 5th and 95th percentiles of the eccentric and the circular population, respectively. The circularisation of the bulk is taken to be the period or separation below which the probability of finding an eccentric binary is equal to half the probability of finding any binary. This measurement is sensitive to the fraction of eccentric systems, which is about 33\% in our sample. We do not expect to be strongly biased either way by our methodology, although it could be that proportionally more circular systems were removed in vetting.

We find agreement between theory and observations in the case of $P_{\rm env}$, with values between 2.5 and 1.6 days across the temperature range of our EBs. The turbulent dissipation of the equilibrium tide is shown to circularise systems up to $\thicksim$2 days on the PMS for stars with convective envelope \citep{Barker2022}. This leaves some room for other mechanisms in our lowest temperature regime, but if the age of the most eccentric systems is an appreciable fraction of the MS lifetime, this could explain the observed envelope period. Our lowest temperature range coincides with the one of the sample by \citet{Zanazzi2022}, and our critical period measurements are close to those derived by him. The envelope period of our larger sample is slightly below his and supports some of his conclusions. Predicted PMS circularisation of convective envelope stars adequately explains the circularised systems up to the envelope period. Circularisation on the PMS is, however, not sufficient to explain all circular binaries. 

At the higher temperatures, we find close agreement between $P_{\rm env}$ and the predicted orbital periods for circularisation due to radiative dissipation of 1-2 days. This seemingly contradicts the findings by \citet{Justesen2021}, who only measured circularisation of the bulk of their EBs and conclude on a deficit of predicted circularisation in the range 6250-10\,000\,K. Theoretical predictions for the hot stars in their work include half of the MS lifetime: too much to compare with the envelope of the most eccentric systems. Instead, we interpret our own $P_{\rm circ}$ measurements, which are much higher than 1-2 days in all but the hottest stars. We, therefore, support \citet{Justesen2021} in their conclusion that non-resonant radiative dissipation of the dynamical tide is insufficient to explain the circularisation of hot stars (except for those above 10\,000\,K). Going back to the lowest temperatures, $P_{\rm circ}$ is likewise much larger than $P_{\rm env}$. A formation scenario where stars are preferentially born circular is deemed unlikely to explain this, so it must be concluded that here, too, additional circularisation is needed to explain observations. The calculations by \citet{Barker2022} show that an inertial wave mechanism could successfully circularise low-mass binaries out to long periods of $\thicksim$7-11 days. This has the potential to explain the large spread in circularisation of EBs as well, as it is a strong function of stellar rotation rate. The issue is that the mechanism may be too efficient to explain the most eccentric systems. 

Circularised systems are found to exist between 10-20 days, as also noted by several other authors \citep[e.g., ][]{Justesen2021, Zanazzi2022}. $P_{\rm core}$ reveals that the bulk of circular systems occurs below $\thicksim$7-8 days; only 5\% reach a circularised state at longer periods. Even though this objective statistic is less than half of the period of some circular systems, it is still well above the circularisation periods in all categories. This poses a challenge for tidal theories, as they need to explain the large difference between the closest binaries that are eccentric and the widest binaries that have circularised. While the extent of the cold core is shown to depend on temperature, its change is not a monotonic function as was found for the eccentricity envelope and measure of bulk circularisation. 

Our results may have some bearing on the role of pulsations. We find a clear population of sources with p-modes, that correspond to the temperature regime of $\delta$ Scuti pulsators. The population with g-modes is found over the whole temperature range of F- to B-type stars, with an overdensity at 7000\,K. This is just below the temperatures of the p-mode excitation, where $\gamma$ Dor pulsators are expected. A similar result was obtained for single stars by \citet{Hey2024}. We find lower fractions of eccentric stars among pulsators, within the same orbital period ranges. The reduction in percentage is around 2.1-4.4\% for g-mode and 4.8\% for p-mode pulsators, and may hint at the involvement of pulsations in circularisation.

It has been speculated that differences in initial parameters can cause a spread in the predictions of circularisation in certain mechanisms, accumulating with the age spread of the sample. Inertial waves can produce long-period circular systems with fast rotation and much shorter periods in slow rotators \citep{Barker2022}. Resonance locking of gravity waves has been suggested \citep[by, e.g.,][]{Justesen2021} as a selective process for enhanced dissipation that could circularise certain stars before others. A single process that is so efficient as to circularise all systems up to the core period does not suffice, as in doing so it fails to explain all the eccentric systems at shorter periods. Likewise, a process that only efficiently works on the PMS would not succeed at explaining the large spread of periods where both eccentric and circular systems exist. In a sense, this shifts the debate of circularisation from which process can most efficiently circularise binaries to which process can produce the largest spread in circularisation periods.

The upcoming PLATO mission \citep{PLATO2022} brings a balance of a larger field of view than \textit{Kepler} while observing for longer than TESS has done in most areas of the sky. This means it will enable in-depth asteroseismic studies of many EBs. Accumulating large numbers of binary systems that have been characterised to the fullest extent could prove pivotal in completing the picture of tidal circularisation and the involvement of the various types of pulsations in its mechanisms.

\section*{Data availability}

The O/B/A/F-type catalogue and the summarising catalogue of our TESS analysis results are available at the CDS via anonymous ftp to \url{cdsarc.u-strasbg.fr} (130.79.128.5) or via \url{http://cdsweb.u-strasbg.fr/}.

\begin{acknowledgements}

The research leading to these results has received funding from the KU\,Leuven Research Council (grant C16/18/005: PARADISE), from the Research Foundation Flanders (FWO) under grant agreements 1124321N (Aspirant Fellowship to LIJ), G089422N (AT), as well as from the BELgian federal Science Policy Office (BELSPO) through PRODEX grant PLATO. 
CA acknowledges funding by the European Research Council under grant ERC SyG 101071505. Funded by the European Union. Views and opinions expressed are however those of the author(s) only and do not necessarily reflect those of the European Union or the European Research Council. Neither the European Union nor the granting authority can be held responsible for them. 
We thank T. Van Reeth and Z. Guo for useful discussions on pulsating binaries.
This paper includes data collected by the TESS mission, which are publicly available from the Mikulski Archive for Space Telescopes (MAST). Funding for the TESS mission is provided by NASA’s Science Mission directorate. We acknowledge the use of TESS High-Level Science Products (HLSP) produced by the TESS Science Processing Operations Center (SPOC) (\href{https://dx.doi.org/10.17909/t9-wpz1-8s54}{DOI: 10.17909/t9-wpz1-8s54}) and the Quick-Look Pipeline (QLP) at the TESS Science Office at MIT (\href{https://dx.doi.org/10.17909/t9-r086-e880}{DOI: 10.17909/t9-r086-e880}).

This work makes use of Python (Python Software Foundation. Python Language Reference, version 3.7. Available at \href{http://www.python.org}{www.python.org}) and the Python packages Numpy \citep{numpy}, Scipy  \citep{scipy}, Matplotlib \citep{matplotlib}, and Pandas \citep{pandas, pandas2010}.
 
\end{acknowledgements}

%
%

\bibliographystyle{aa}
\bibliography{references}

\begin{appendix}
\section{\textit{Gaia} data query}
\label{apx:gaia}

\begin{verbatim}
SELECT 
    table.tic_id, table.tmass_id,
    gaia.designation, gaia.source_id, gaia.bp_rp,
    tmass.original_ext_source_id,
    gaia_p.teff_gspphot, gaia_p.teff_gspphot_lower, gaia_p.teff_gspphot_upper,
    gaia_p.logg_gspphot, gaia_p.logg_gspphot_lower, gaia_p.logg_gspphot_upper,
    gaia_p.distance_gspphot, gaia_p.distance_gspphot_lower, gaia_p.distance_gspphot_upper,
    gaia_p.ebpminrp_gspphot, gaia_p.ebpminrp_gspphot_lower, gaia_p.ebpminrp_gspphot_upper,
    gaia_p.teff_gspspec, gaia_p.teff_gspspec_lower, gaia_p.teff_gspspec_upper,
    gaia_p.logg_gspspec, gaia_p.logg_gspspec_lower, gaia_p.logg_gspspec_upper,
    gaia_p.teff_esphs, gaia_p.teff_esphs_uncertainty,
    gaia_p.ebpminrp_esphs, gaia_p.ebpminrp_esphs_uncertainty,
    gaia_p.logg_esphs, gaia_p.logg_esphs_uncertainty,
    gaia_p.vsini_esphs, gaia_p.vsini_esphs_uncertainty,
    gaia_p.ebpminrp_esphs, gaia_p.ebpminrp_esphs_uncertainty,
    gaia_p.teff_msc1, gaia_p.teff_msc1_upper, gaia_p.teff_msc1_lower,
    gaia_p.teff_msc2, gaia_p.teff_msc2_upper, gaia_p.teff_msc2_lower,
    gaia_p.logg_msc1, gaia_p.logg_msc1_upper, gaia_p.logg_msc1_lower,
    gaia_p.logg_msc2, gaia_p.logg_msc2_upper, gaia_p.logg_msc2_lower,
    gaia_p.distance_msc, gaia_p.distance_msc_upper, gaia_p.distance_msc_lower 
FROM 
    gaiadr3.gaia_source AS gaia
JOIN 
    gaiadr3.tmass_psc_xsc_best_neighbour AS tmass ON gaia.source_id = tmass.source_id
JOIN
    gaiadr3.astrophysical_parameters AS gaia_p ON gaia.source_id = gaia_p.source_id
JOIN 
    user_name.obaf_ids AS table ON tmass.original_ext_source_id = table.tmass_id
\end{verbatim}


\section{Statistics of temperature subdomains}
\label{apx:extra_stats}

\begin{table*}
\centering
\caption{Summary of critical values of period and scaled separation for the characterised EBs. Here, subdomains of the original temperature bins are used that result in samples of approximately 1000 systems per bin. The contents of the table are visualised in Figures \ref{fig:crit_period_subbins} and \ref{fig:crit_separation_subbins}.}
\label{tab:crit_sub}
\begin{tabular}{c l l l l l l l}
    \hline
    $T_{\rm eff}$ (K) & $N_{\rm EB}$ & $P_{\rm env}$ (d) & $P_{\rm circ}$ (d) & $P_{\rm core}$ (d) & $\frac{a}{R_s}_{\rm env}$ & $\frac{a}{R_s}_{\rm circ}$ & $\frac{a}{R_s}_{\rm core}$ \\
    \hline 
    < 6380 & 1035 & $3.14\pm0.48$ & $7.32\pm0.43$ & $8.2\pm1.2$ & $3.25\pm0.40$ & $6.84\pm0.31$ & $8.37\pm0.92$ \\
    6380 - 6650 & 1002 & $2.68\pm0.50$ & $6.24\pm0.49$ & $8.9\pm1.3$ & $2.99\pm0.45$ & $5.50\pm0.33$ & $8.34\pm0.93$ \\
    6650 - 7000 & 1030 & $2.19\pm0.44$ & $5.31\pm0.47$ & $7.4\pm1.1$ & $2.52\pm0.38$ & $5.21\pm0.33$ & $7.89\pm0.84$ \\
    7000 - 7350 & 1023 & $2.05\pm0.49$ & $6.04\pm0.57$ & $6.70\pm0.86$ & $2.54\pm0.47$ & $5.35\pm0.37$ & $7.15\pm0.68$ \\
    7350 - 7700 & 1076 & $1.99\pm0.43$ & $5.14\pm0.47$ & $6.77\pm0.81$ & $2.59\pm0.44$ & $4.93\pm0.34$ & $6.60\pm0.56$ \\
    7700 - 8000 & 1076 & $1.98\pm0.37$ & $5.66\pm0.43$ & $6.62\pm0.80$ & $2.54\pm0.34$ & $5.35\pm0.26$ & $6.45\pm0.55$ \\
    8000 - 8440 & 1207 & $1.96\pm0.34$ & $5.31\pm0.38$ & $7.10\pm0.79$ & $2.57\pm0.33$ & $4.54\pm0.24$ & $6.87\pm0.56$ \\
    8440 - 9200 & 1190 & $1.91\pm0.35$ & $4.11\pm0.38$ & $7.70\pm0.87$ & $2.41\pm0.32$ & $4.30\pm0.23$ & $7.14\pm0.61$ \\
    9200 - 10000 & 1124 & $1.88\pm0.36$ & $3.39\pm0.36$ & $8.36\pm0.98$ & $2.45\pm0.34$ & $3.76\pm0.24$ & $7.55\pm0.66$ \\
    10000 - 11000 & 976 & $1.88\pm0.34$ & $3.18\pm0.34$ & $8.8\pm1.2$ & $2.58\pm0.33$ & $3.46\pm0.22$ & $8.21\pm0.85$ \\
    11000 - 13000 & 958 & $1.60\pm0.28$ & $2.30\pm0.28$ & $7.8\pm1.1$ & $2.40\pm0.28$ & $3.10\pm0.18$ & $7.08\pm0.73$ \\
    > 13000 & 1000 & $1.61\pm0.29$ & $2.23\pm0.25$ & $7.54\pm0.93$ & $2.25\pm0.28$ & $2.86\pm0.17$ & $6.36\pm0.59$ \\
    \hline
\end{tabular}
\end{table*}

\begin{figure*}
\resizebox{\hsize}{!}
    {\includegraphics{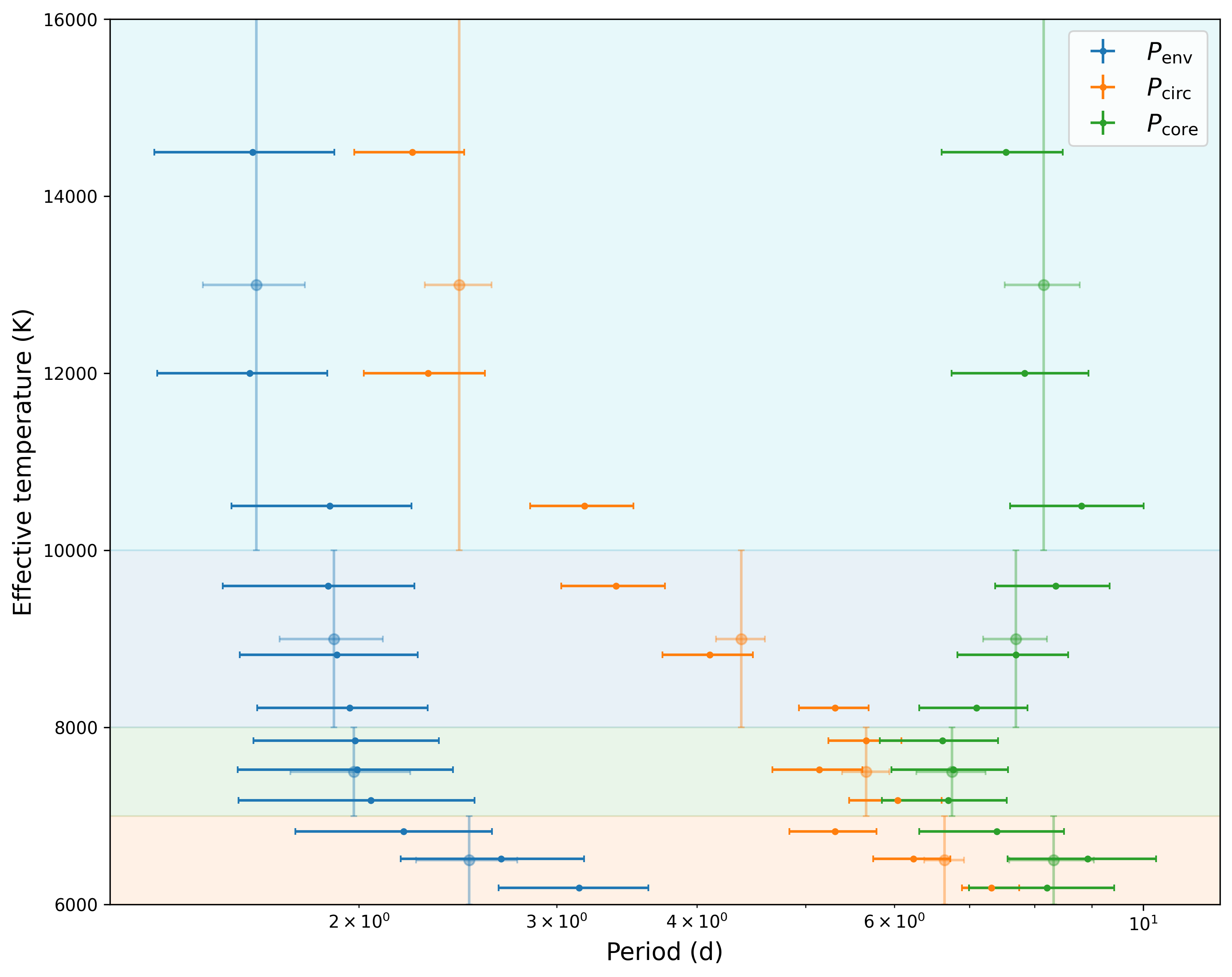}}
    \caption{Visualisation of critical orbital period measurements in each temperature range. The colour of the shaded areas corresponds to the four main temperature domains, as do the vertical bars on the associated, lighter-coloured measurement points. The points with only horizontal error bars correspond to the effective temperature subdivisions and are plotted in the centre of the range.}
    \label{fig:crit_period_subbins}
\end{figure*}

\begin{figure*}
\resizebox{\hsize}{!}
    {\includegraphics{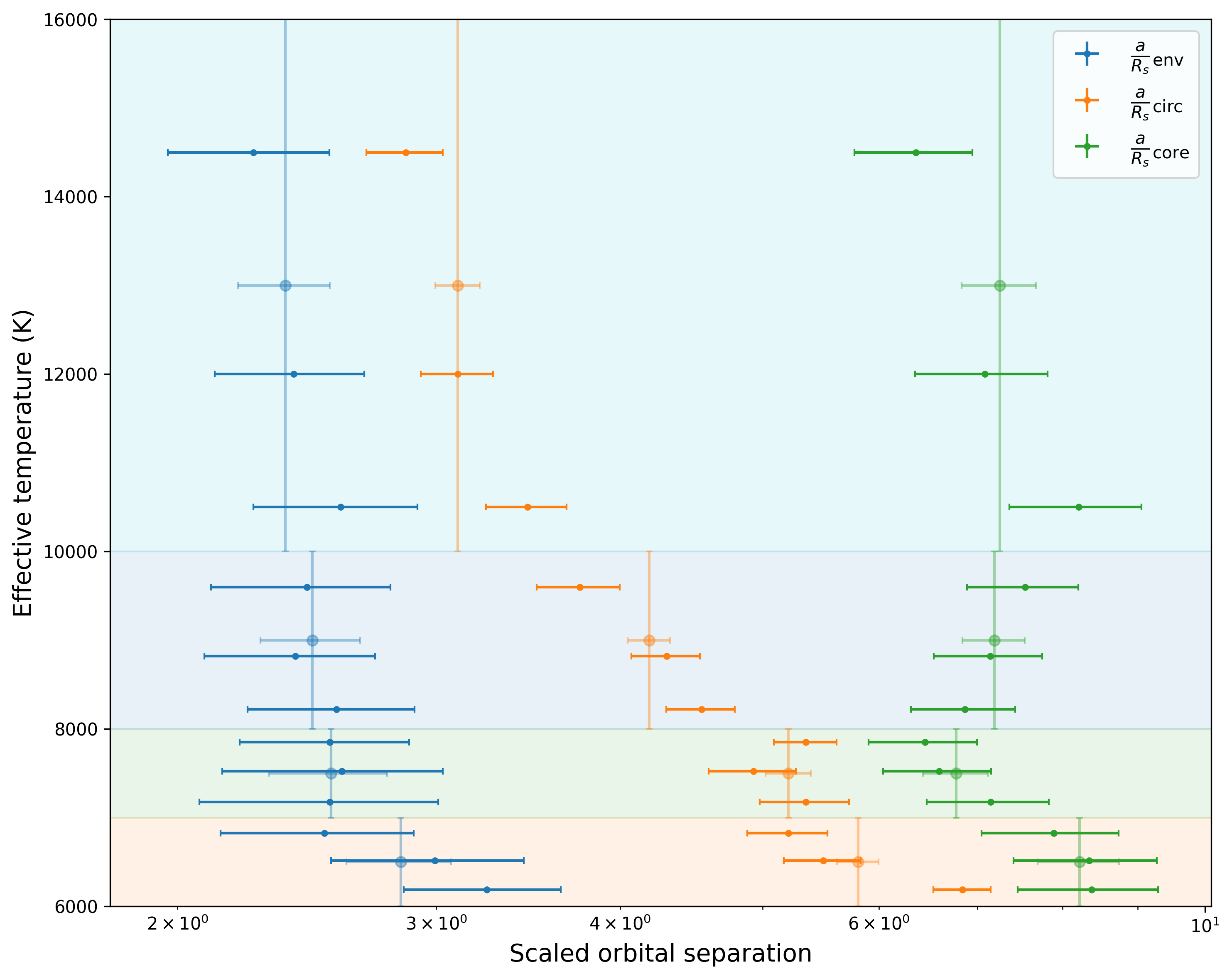}}
    \caption{Visualisation of critical scaled orbital separation measurements in each temperature range. The colour of the shaded areas corresponds to the four main temperature domains, as do the vertical bars on the associated, lighter-coloured measurement points. The points with only horizontal error bars correspond to the effective temperature subdivisions and are plotted in the centre of the range.}
    \label{fig:crit_separation_subbins}
\end{figure*}

\clearpage

\section{Additional distribution figures}
\label{apx:dists}

\begin{figure*}
\resizebox{\hsize}{!}
    {\includegraphics[width=\hsize,clip]{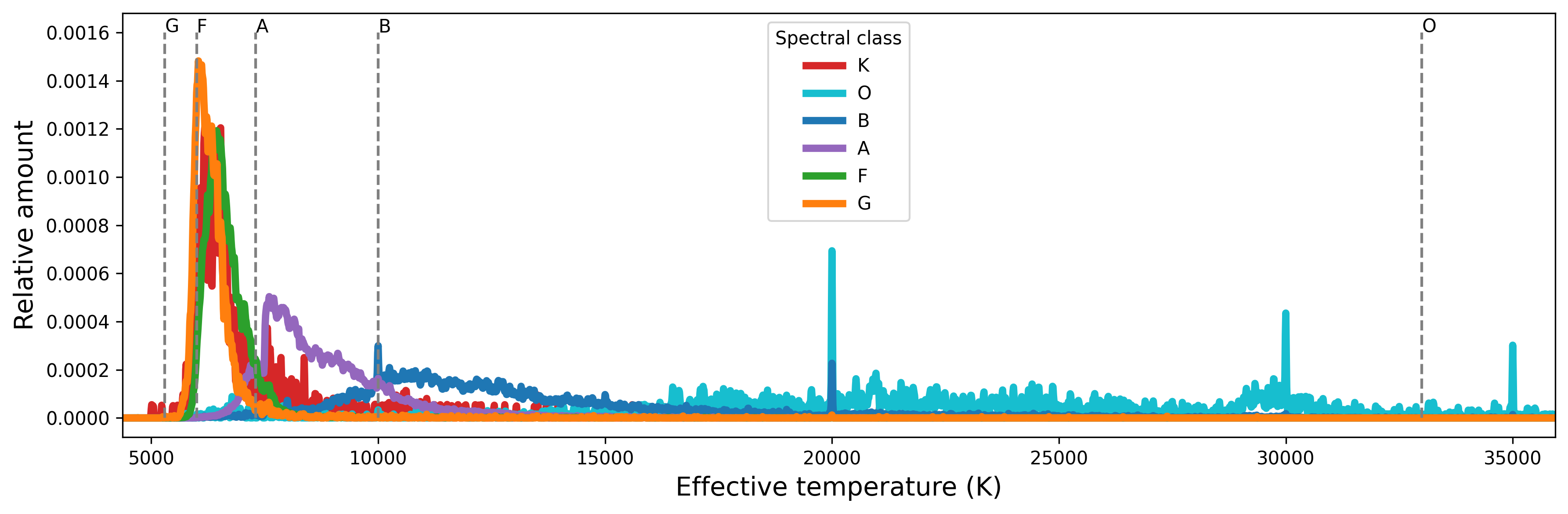}}
    \caption{KDE of the \textit{Gaia} effective temperatures divided into the \texttt{SIMBAD} spectral classes. This includes all sources in the O/B/A/F catalogue for which the temperatures and spectral classes are available. The temperature axis is linear and cuts off above 35\,000\,K as there are very few targets above that.}
    \label{fig:kde_temps}
\end{figure*}

\begin{figure*}
\resizebox{\hsize}{!}
    {\includegraphics[width=\hsize,clip]{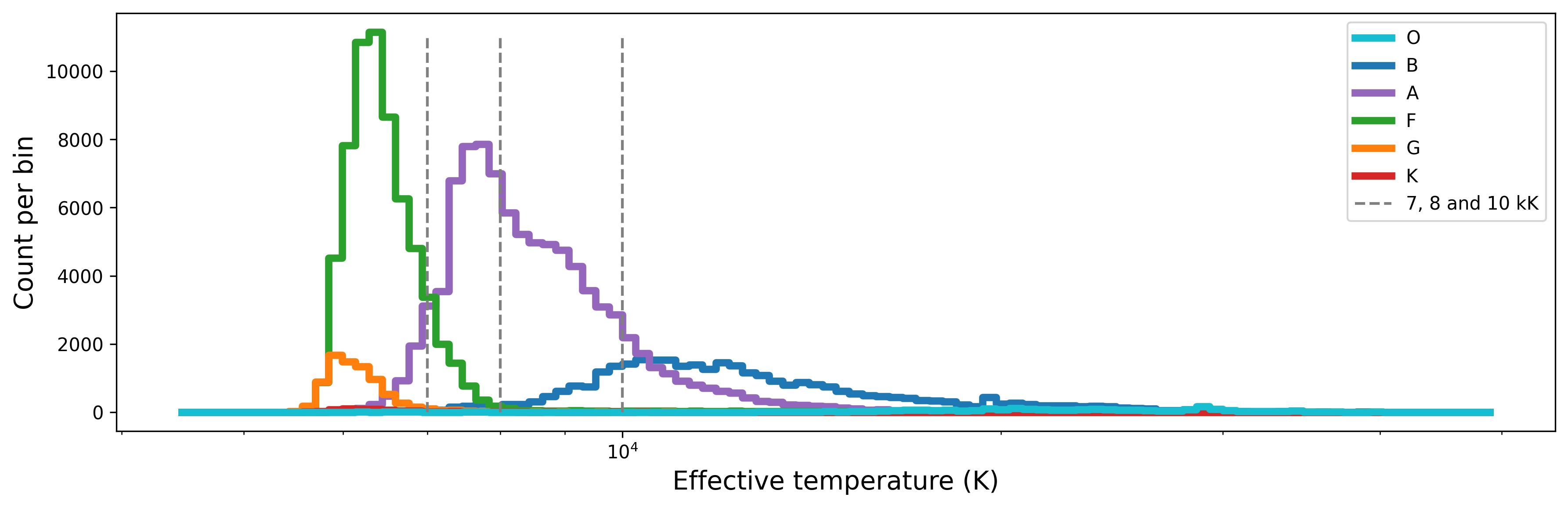}}
    \caption{Histogram of the \textit{Gaia} effective temperatures divided into the \texttt{SIMBAD} spectral classes. This includes all sources in the O/B/A/F catalogue for which the temperatures and spectral classes are available. The temperature axis is logarithmic and temperature bins have equal width in log-space.}
    \label{fig:hist_temps}
\end{figure*}

\begin{figure*}
\resizebox{\hsize}{!}
    {\includegraphics[width=\hsize,clip]{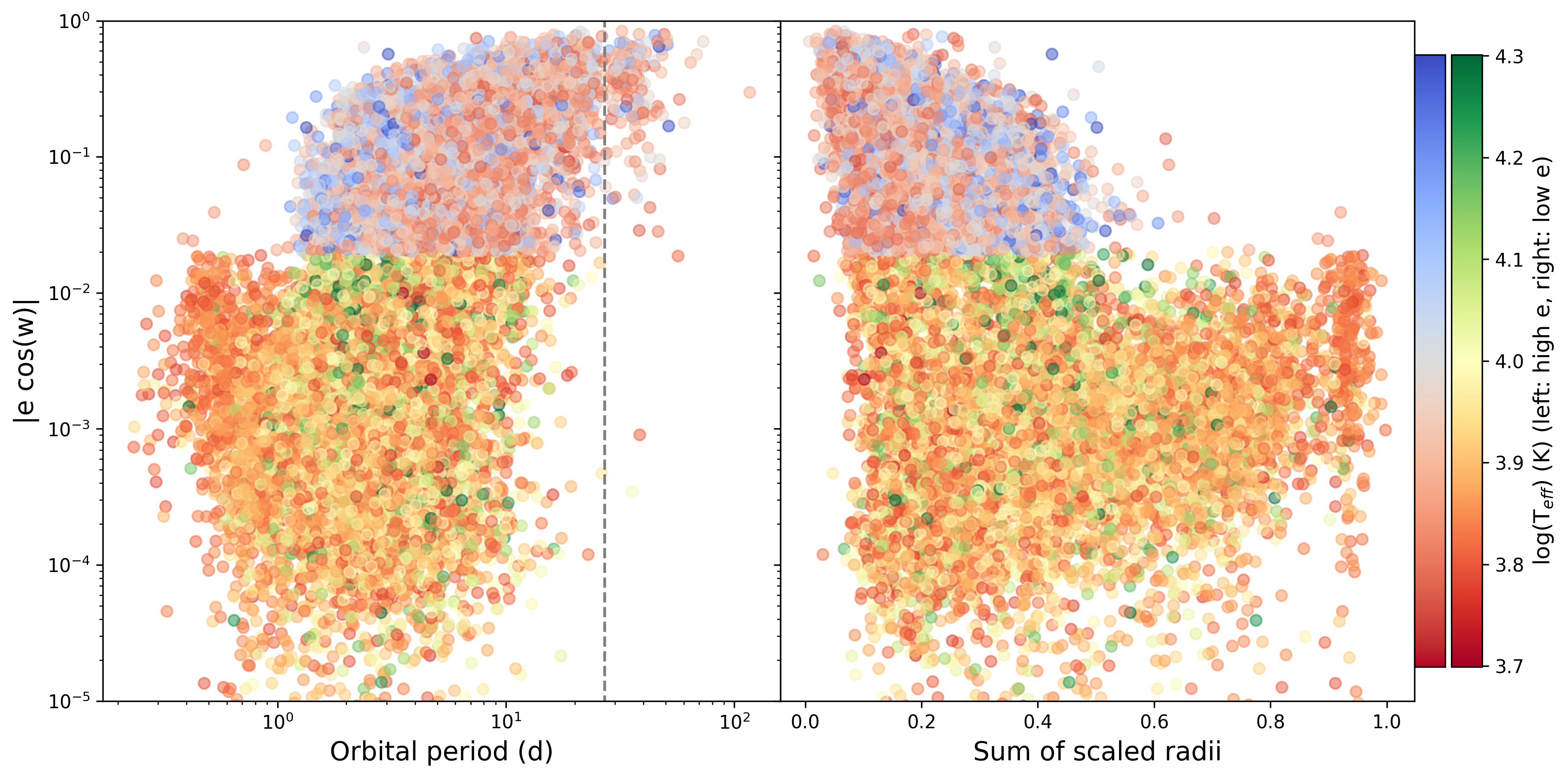}}
    \caption{Logarithmically scaled tangential component of eccentricity as a function of orbital period and sum of scaled radii. Effective temperatures are indicated by the colour gradient. The grey dashed line indicates the dip in the orbital period distribution at 27 days. Systems with high and low eccentricity are shaded with different colour maps (left and right colour bars, respectively). The vertical scale is cut off below $10^{-5}$.}
    \label{fig:e-p_log}
\end{figure*}

\begin{figure}
\centering
\includegraphics[width=\hsize]{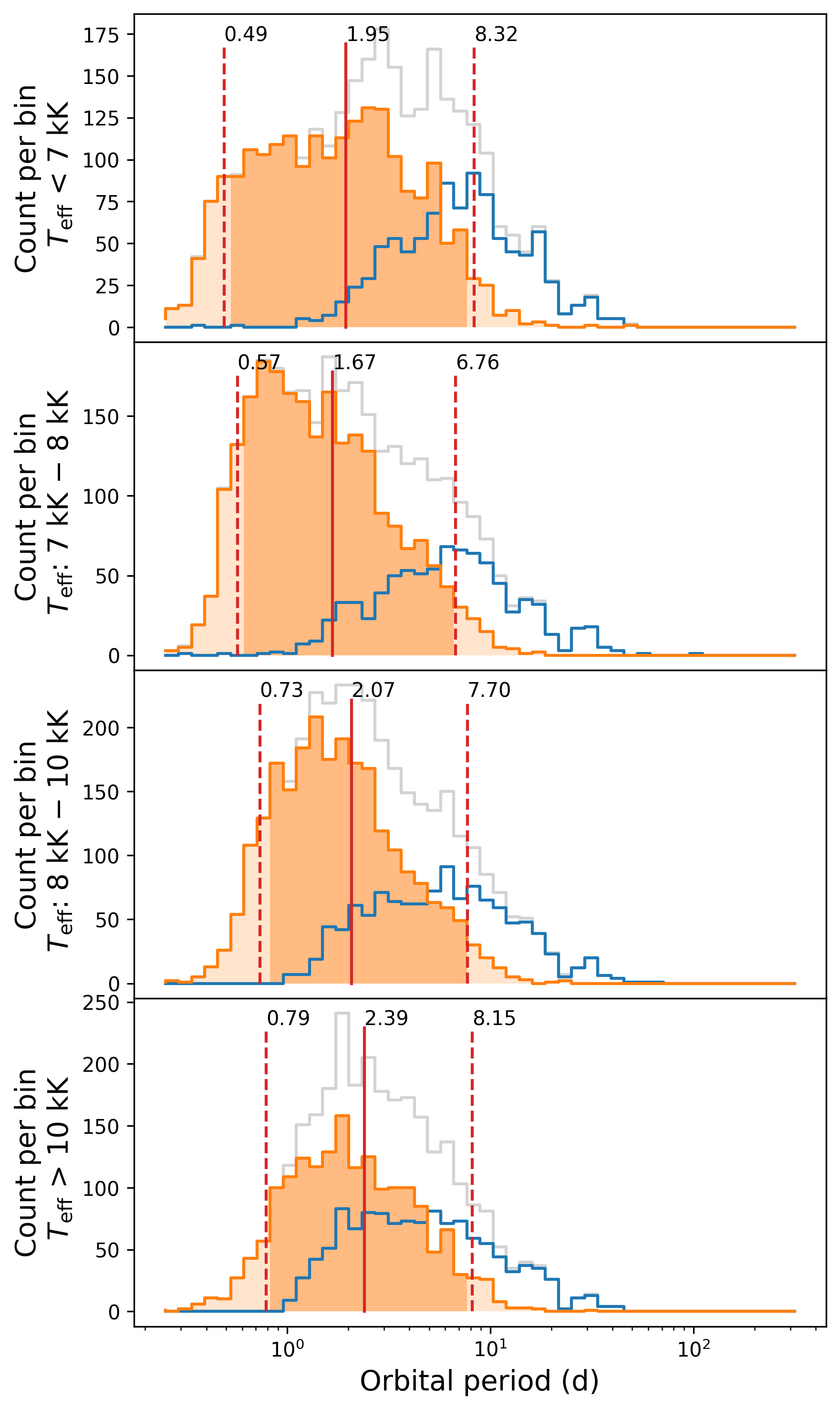}
    \caption{Logarithmic distribution functions of the orbital period of eccentric and circular EB systems. This includes all 14\,573 correctly characterised EBs. Each panel shows a different $T_{\rm eff}$ bin, for the dividing temperatures of 7000, 8000 and 10\,000\,K. The overall distribution is grey, the distribution of systems of low eccentricity is orange and that of systems with high eccentricity is blue. Indicated on the circular distributions are the 5th percentile, the median, and the 95th percentile.}
    \label{fig:period_perc_1}
\end{figure}

\begin{figure}
\centering
\includegraphics[width=\hsize]{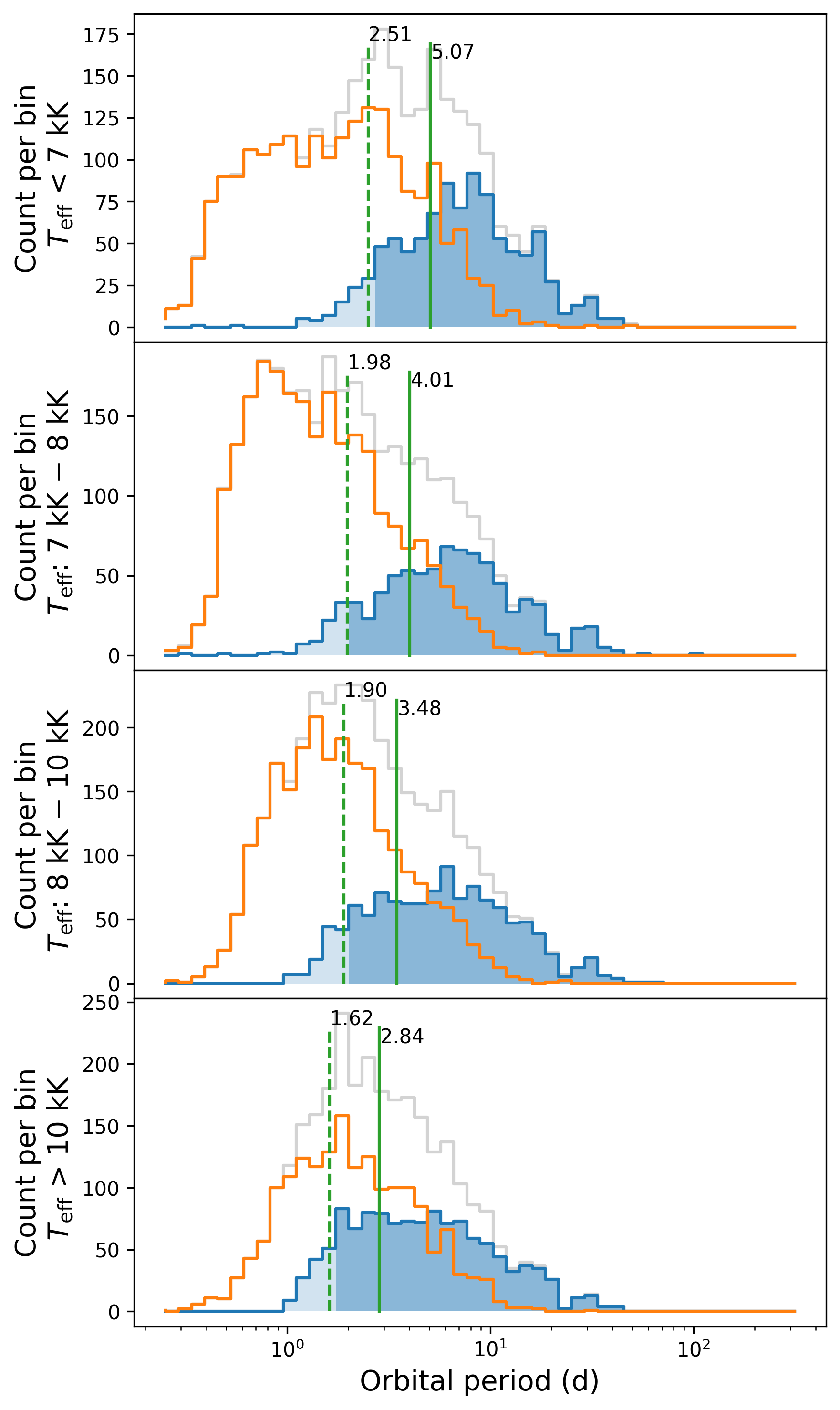}
    \caption{Logarithmic distribution functions of the orbital period of eccentric and circular EB systems. This includes all 14\,573 correctly characterised EBs. Each panel shows a different $T_{\rm eff}$ bin, for the dividing temperatures of 7000, 8000 and 10\,000\,K. The overall distribution is grey, the distribution of systems of low eccentricity is orange and that of systems with high eccentricity is blue. Indicated on the eccentric distributions are the 5th percentile, and the 25th percentile.}
    \label{fig:period_perc_2}
\end{figure}

\begin{figure}
\centering
\includegraphics[width=\hsize]{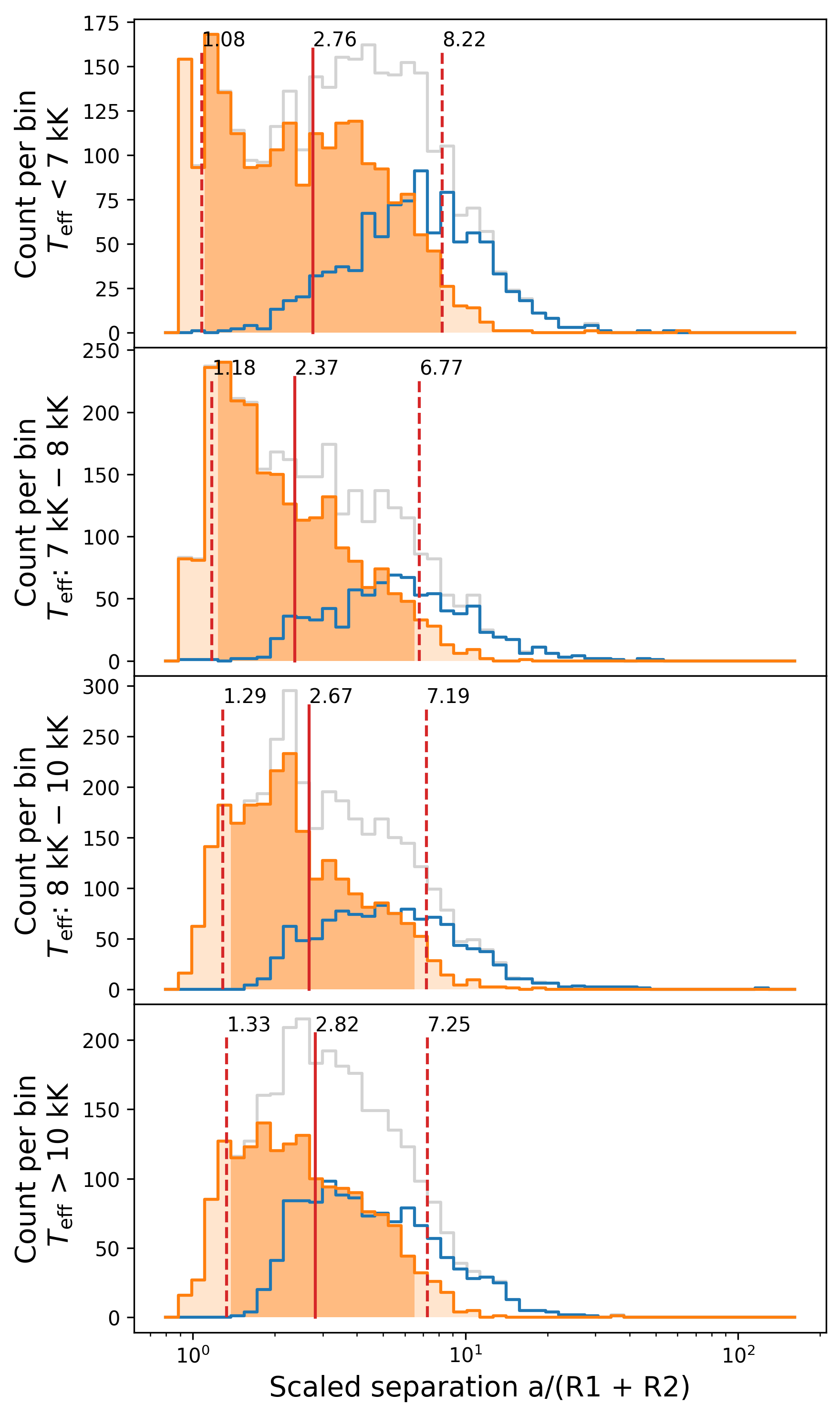}
    \caption{Logarithmic distribution functions of the scaled orbital separation of eccentric and of circular EB systems. This includes all 14\,573 correctly characterised EBs. Each panel shows a different $T_{\rm eff}$ bin, for the dividing temperatures of 7000, 8000 and 10\,000\,K. The overall distribution is grey, the distribution of systems of low eccentricity is orange and that of systems with high eccentricity is blue. Indicated on the circular distributions are the 5th percentile, the median, and the 95th percentile.}
    \label{fig:rsum_perc_1}
\end{figure}

\begin{figure}
\centering
\includegraphics[width=\hsize]{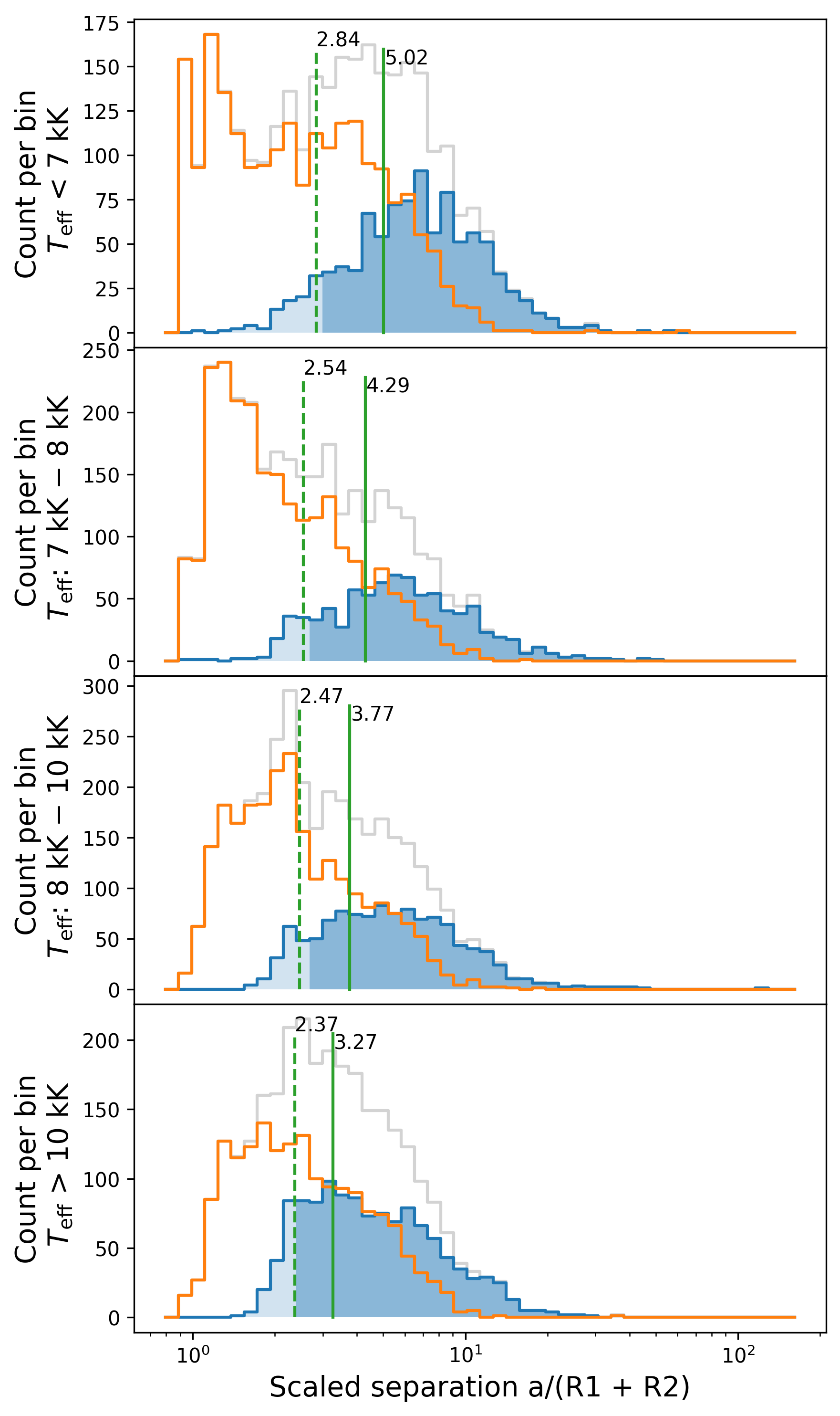}
    \caption{Logarithmic distribution functions of the scaled orbital separation of eccentric and of circular EB systems. This includes all 14\,573 correctly characterised EBs. Each panel shows a different $T_{\rm eff}$ bin, for the dividing temperatures of 7000, 8000 and 10\,000\,K. The overall distribution is grey, the distribution of systems of low eccentricity is orange and that of systems with high eccentricity is blue. Indicated on the eccentric distributions are the 5th percentile, and the 25th percentile.}
    \label{fig:rsum_perc_2}
\end{figure}

\clearpage

\section{Additional pulsation figures}
\label{apx:puls}

\begin{figure}
\centering
\includegraphics[width=\hsize]{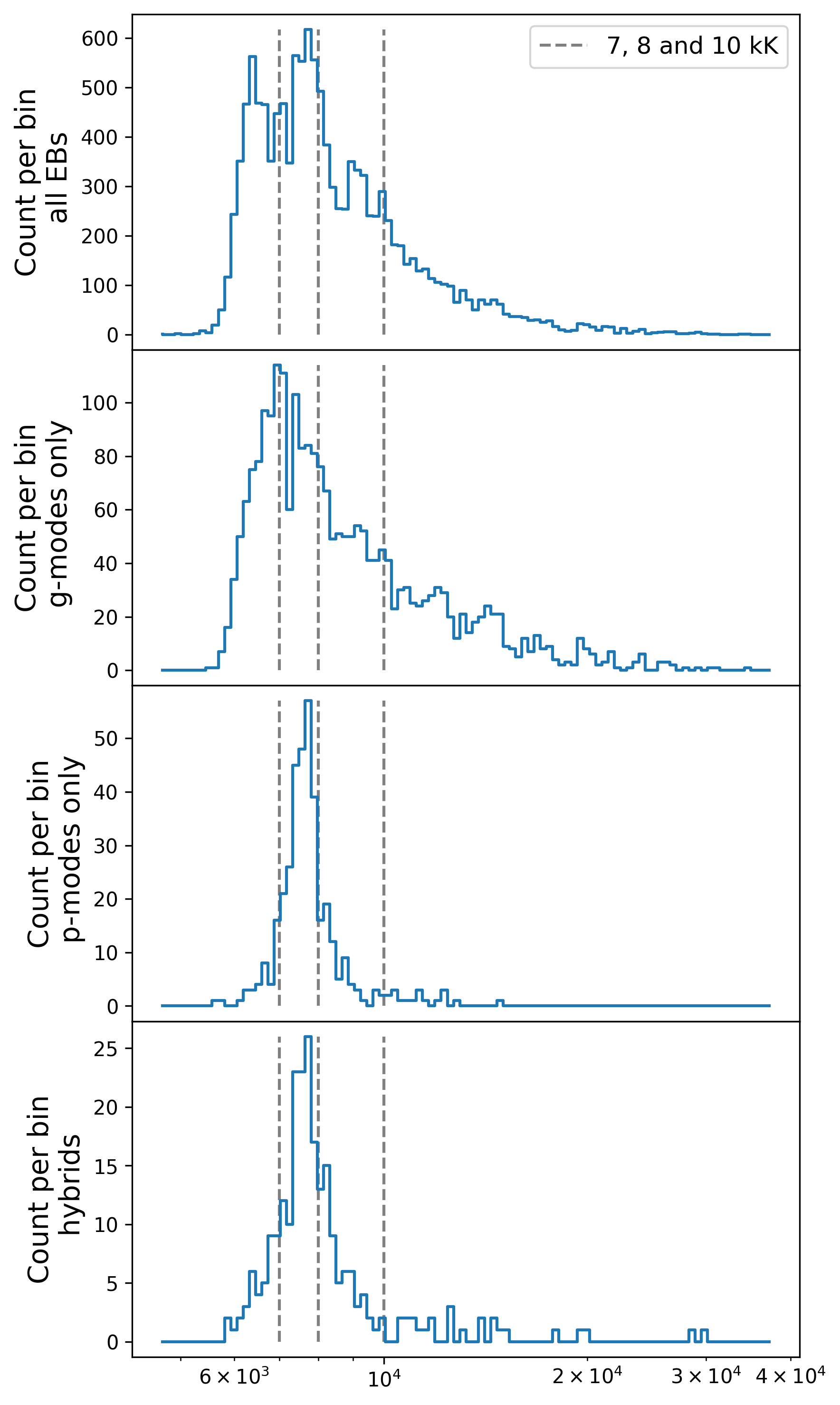}
    \caption{Histogram of the effective temperature for the pulsators. From top to bottom: all EBs, g-mode pulsators only, p-mode pulsators only, hybrid pulsators}
    \label{fig:hist_teff_puls}
\end{figure}

\begin{figure}
\centering
\includegraphics[width=\hsize]{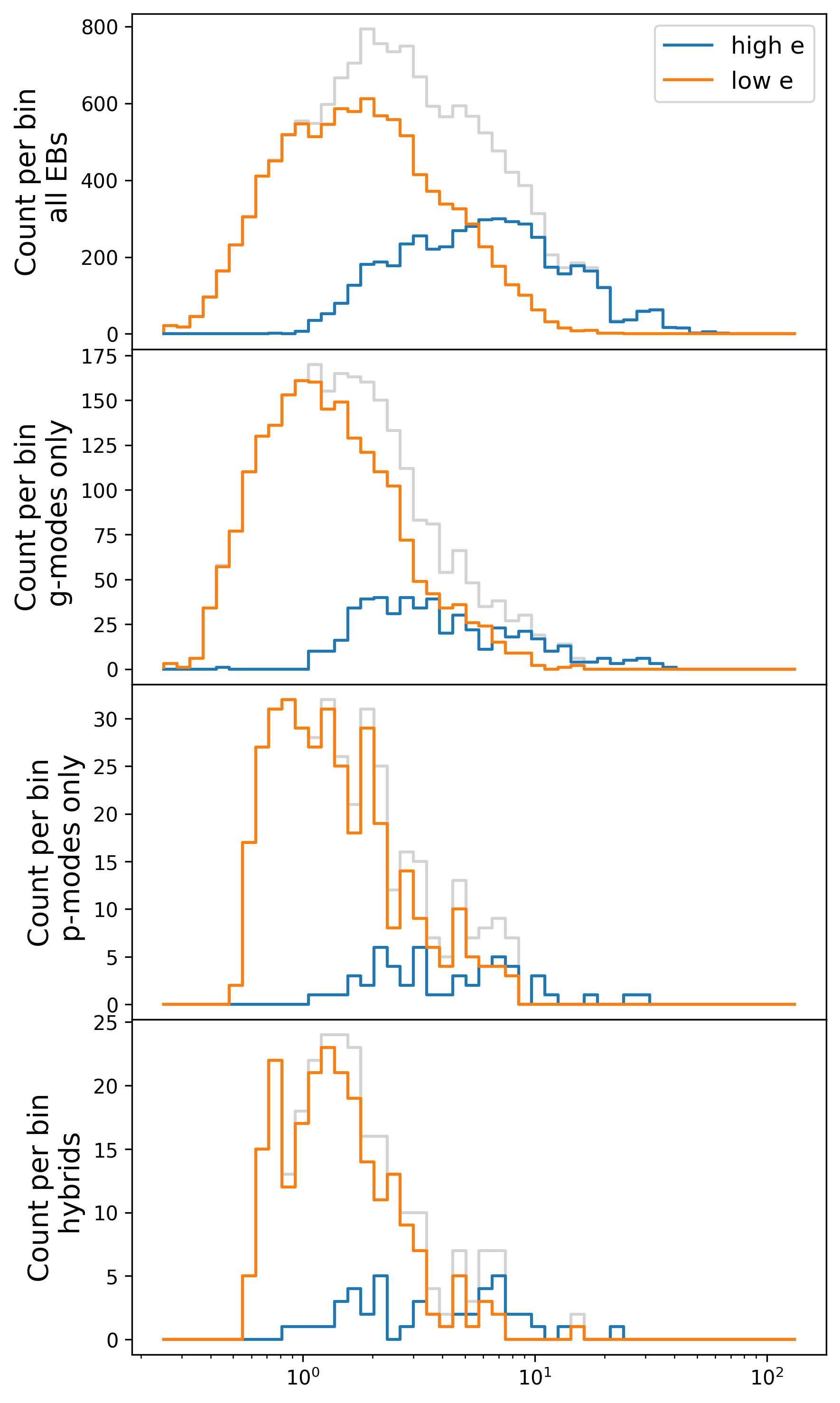}
    \caption{Histogram of the orbital period for the pulsators. The overall distribution is grey, the distribution of systems of low eccentricity is orange and that of systems with high eccentricity is blue. From top to bottom: all EBs, g-mode pulsators only, p-mode pulsators only, hybrid pulsators.}
    \label{fig:hist_period_puls}
\end{figure}

\begin{figure*}
\resizebox{\hsize}{!}
    {\includegraphics[width=\hsize,clip]{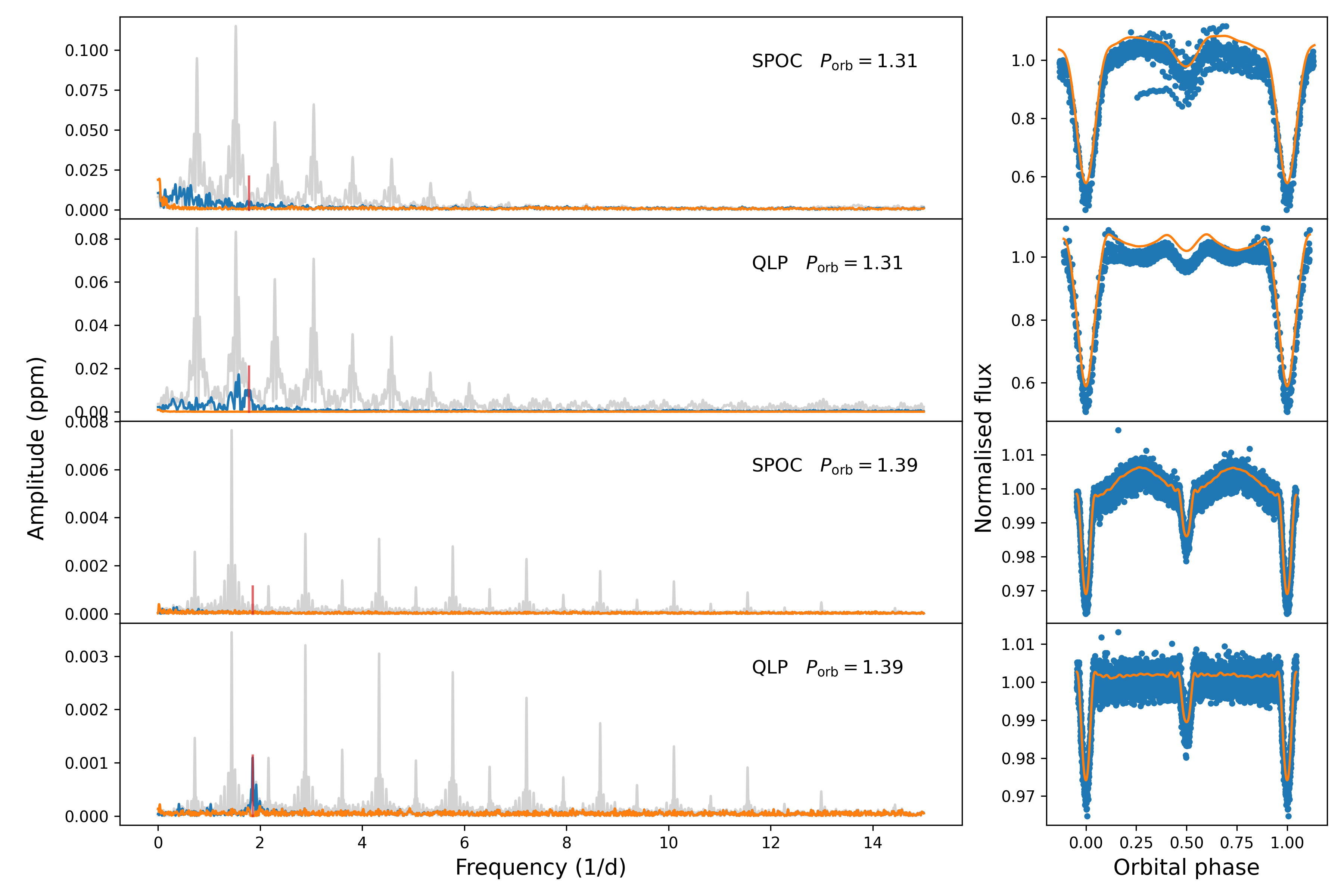}}
    \caption{Two examples of the data reduction artefact in QLP, compared to the SPOC. Top: TIC 393672467, bottom: TIC 180974778. On the left are periodograms of amplitudes with in grey the original data, in blue the data minus harmonics, and in orange the final residuals. Red lines indicate the independent frequency of the highest amplitude. On the right are the light curves folded by the orbital period. The orange curve is a harmonic model of sinusoids based on multiples of the orbital frequency.}
    \label{fig:ev_gone}
\end{figure*}

\begin{figure*}
\resizebox{\hsize}{!}
    {\includegraphics[width=\hsize,clip]{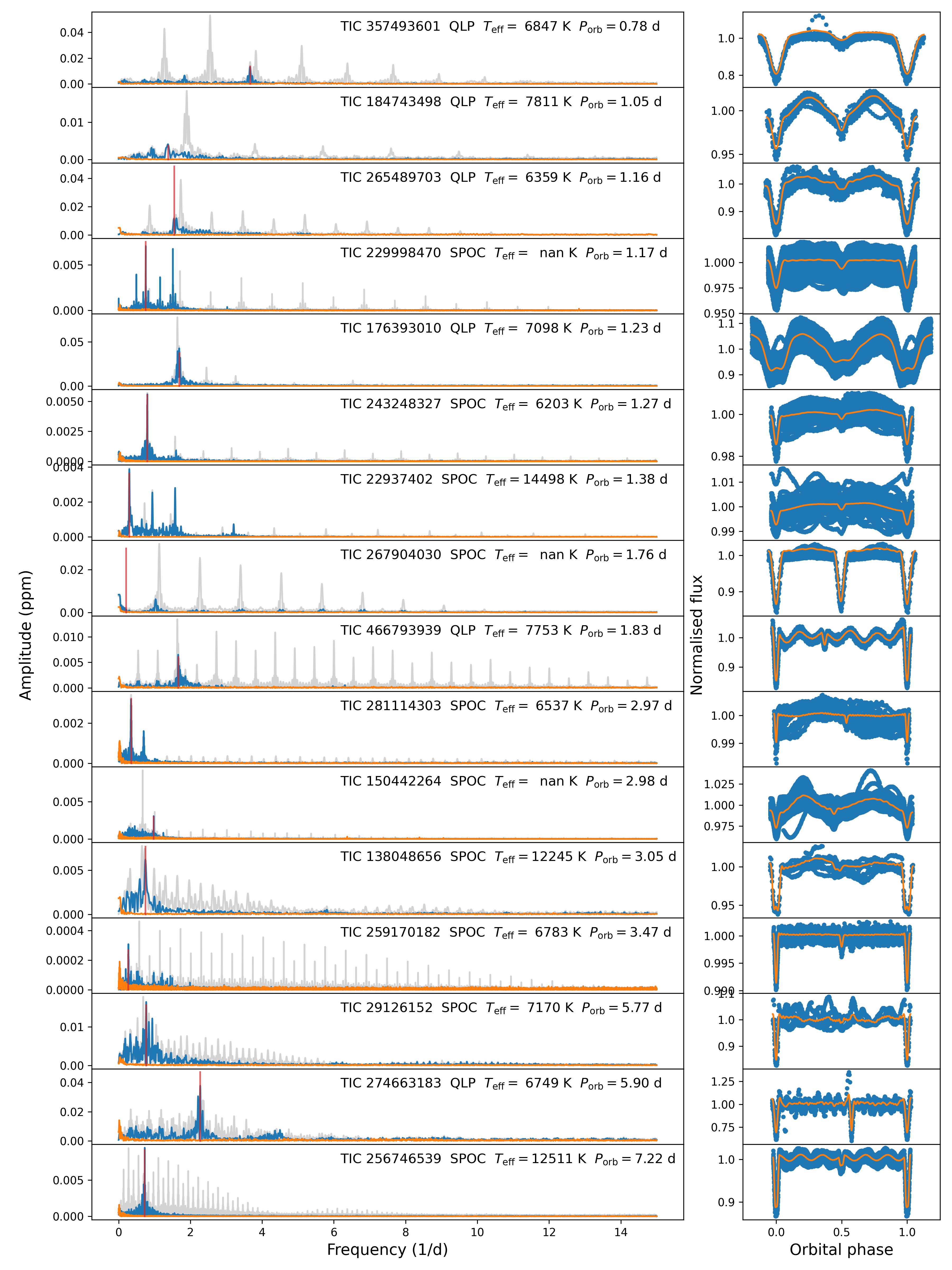}}
    \caption{A selection of EBs from the g-mode (only) pulsators of variability. On the left are periodograms of amplitudes with in grey the original data, in blue the data minus harmonics, and in orange the final residuals. Red lines indicate the independent frequency of the highest amplitude. On the right are the light curves folded by the orbital period. The orange curve is a harmonic model of sinusoids based on multiples of the orbital frequency. From top to bottom the orbital period increases.}
    \label{fig:puls_g_mode}
\end{figure*}

\begin{figure*}
\resizebox{\hsize}{!}
    {\includegraphics[width=\hsize,clip]{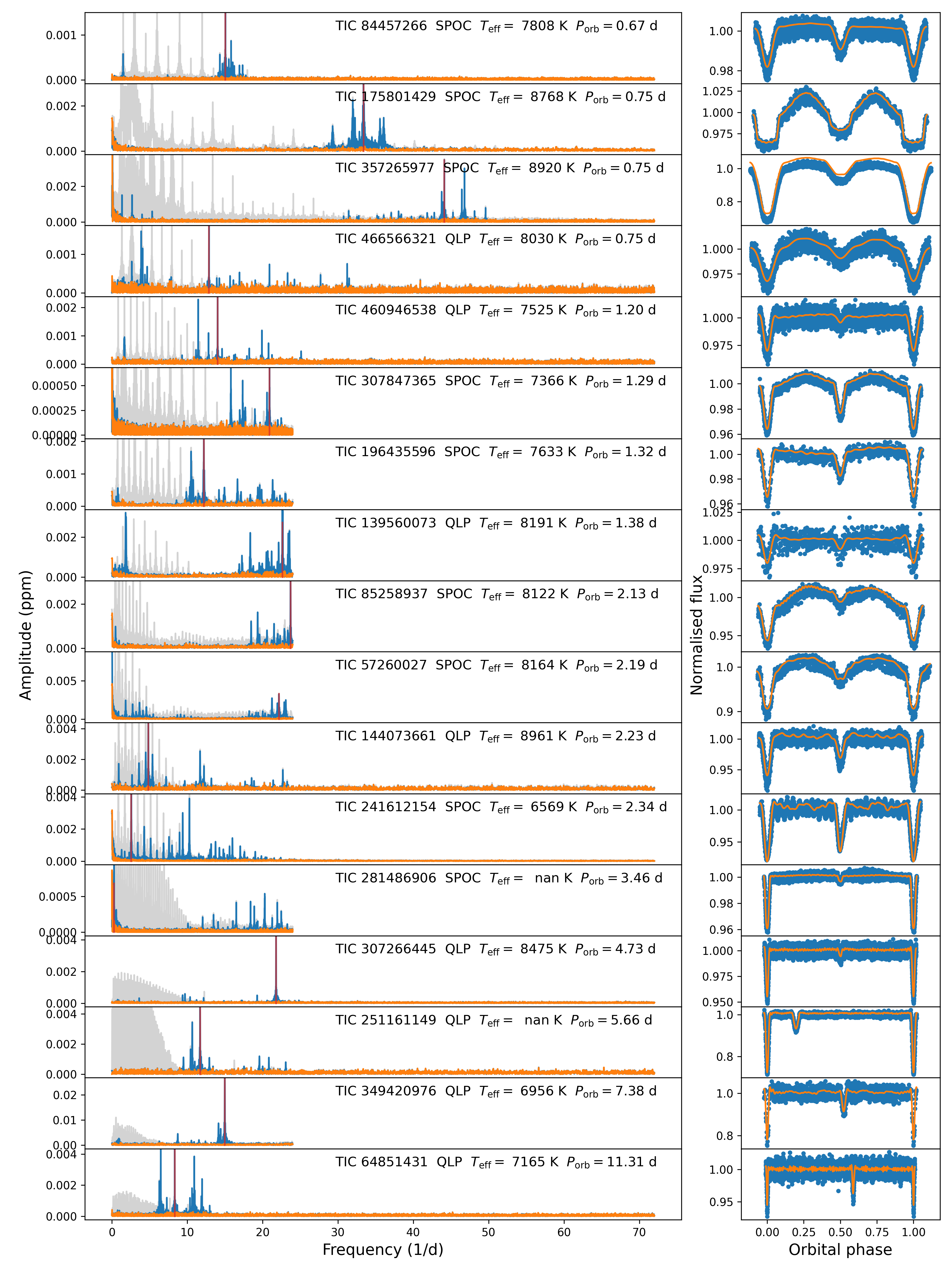}}
    \caption{A selection of EBs from the p-mode (only) pulsators of variability. On the left are periodograms of amplitudes with in grey the original data, in blue the data minus harmonics, and in orange the final residuals. Red lines indicate the independent frequency of the highest amplitude. On the right are the light curves folded by the orbital period. The orange curve is a harmonic model of sinusoids based on multiples of the orbital frequency. From top to bottom the orbital period increases.}
    \label{fig:puls_p_mode}
\end{figure*}


\begin{figure}
\centering
\includegraphics[width=\hsize]{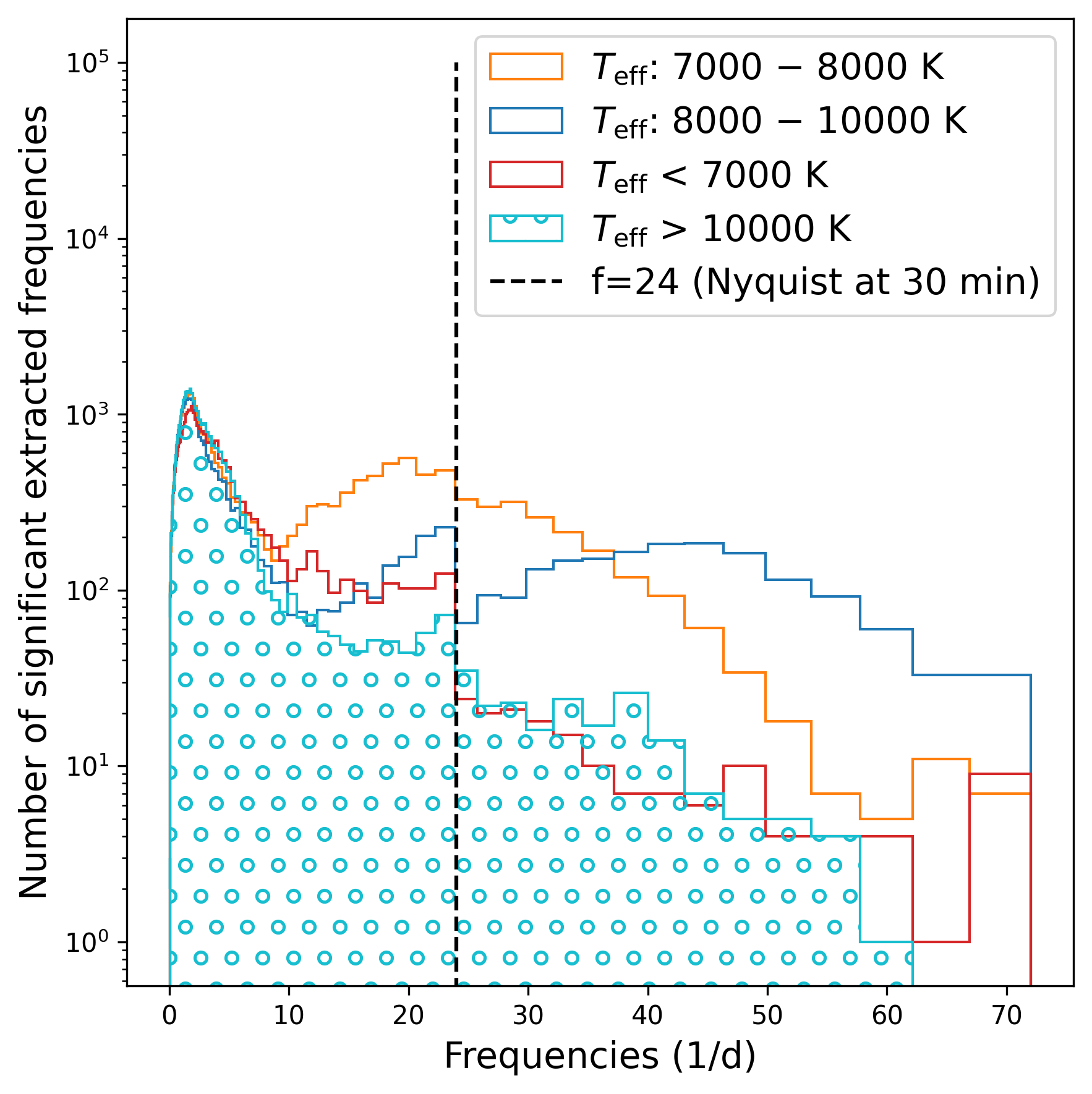}
    \caption{Histogram of the extracted significant frequencies in each temperature range for all pulsators. Bins are scaled logarithmically in size to better view the lower densities at high frequency. Included are just those targets in the group of p-mode pulsators that are not in the g-mode pulsators. Same as Figure \ref{fig:hist_freqs}.}
    \label{fig:hist_freqs_apx}
\end{figure}

\begin{figure}
\centering
\includegraphics[width=\hsize]{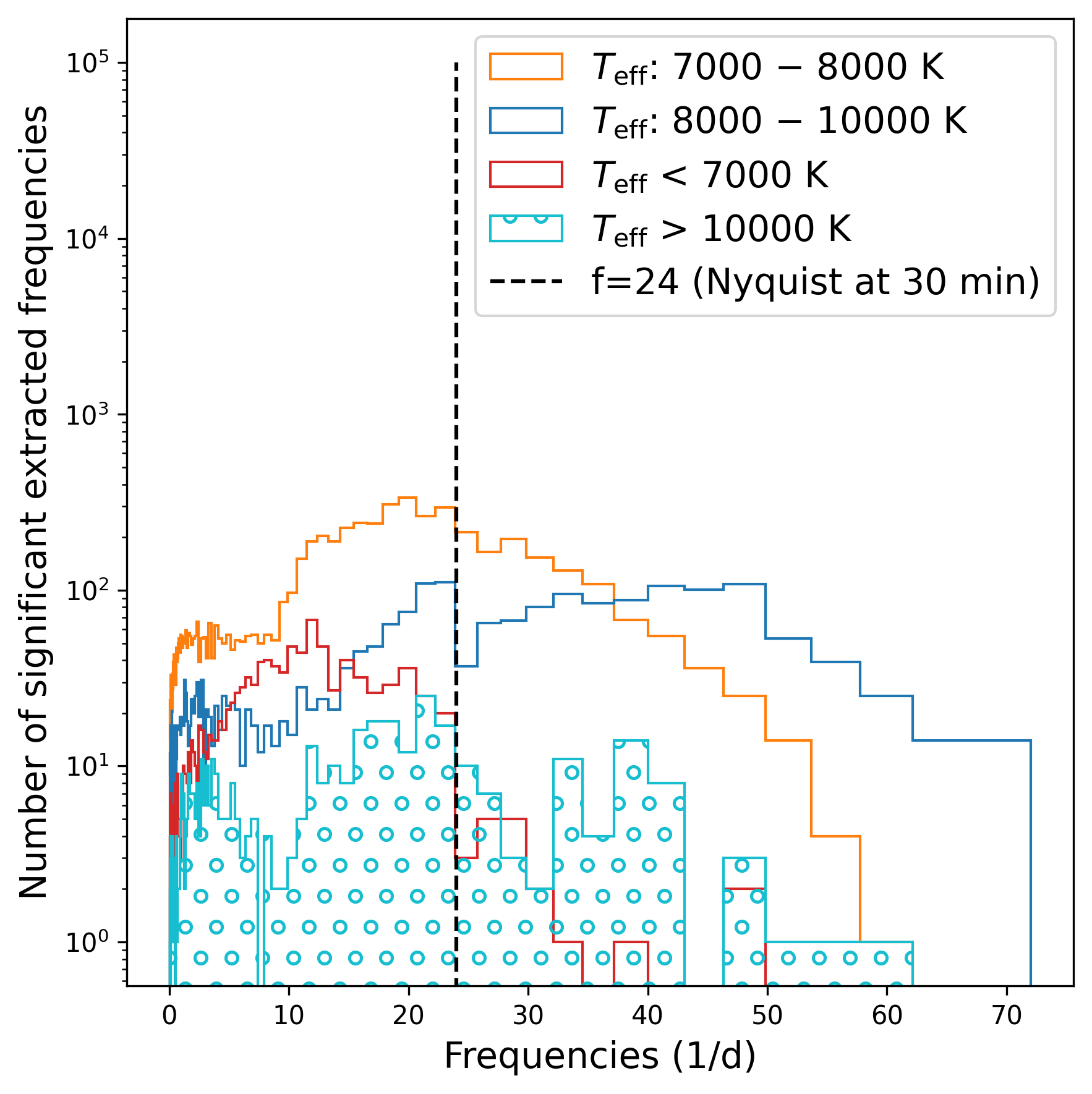}
    \caption{Histogram of the extracted significant frequencies in each temperature range for p-mode pulsators. Bins are scaled logarithmically in size to better view the lower densities at high frequency. Included are just those targets in the group of p-mode pulsators that are not in the g-mode pulsators.}
    \label{fig:hist_freqs_p}
\end{figure}

\begin{figure}
\centering
\includegraphics[width=\hsize]{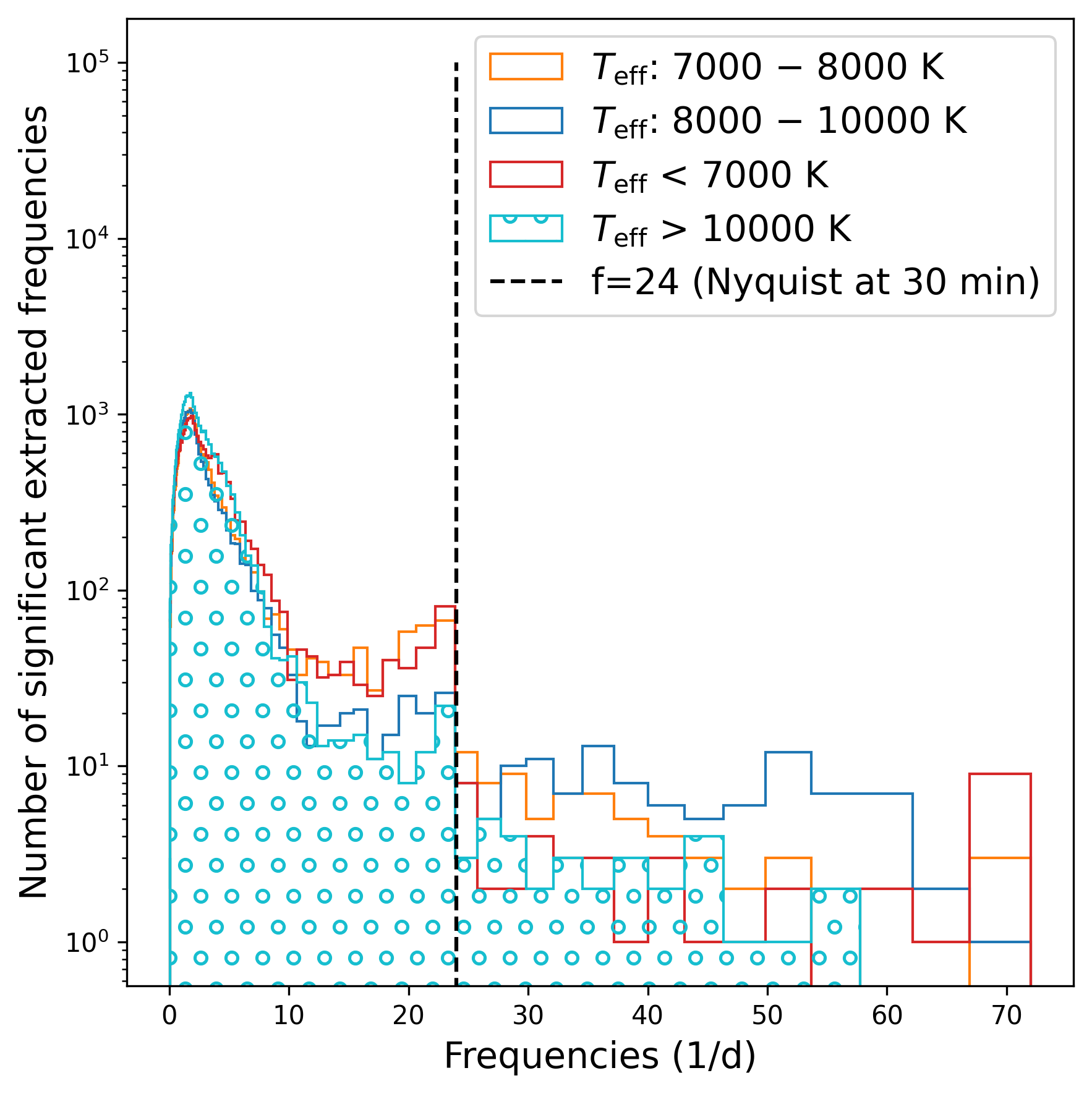}
    \caption{Histogram of the extracted significant frequencies in each temperature range for g-mode pulsators. Bins are scaled logarithmically in size to better view the lower densities at high frequency. Included are just those targets in the group of g-mode pulsators that are not in the p-mode pulsators.}
    \label{fig:hist_freqs_g}
\end{figure}

\begin{figure}
\centering
\includegraphics[width=\hsize]{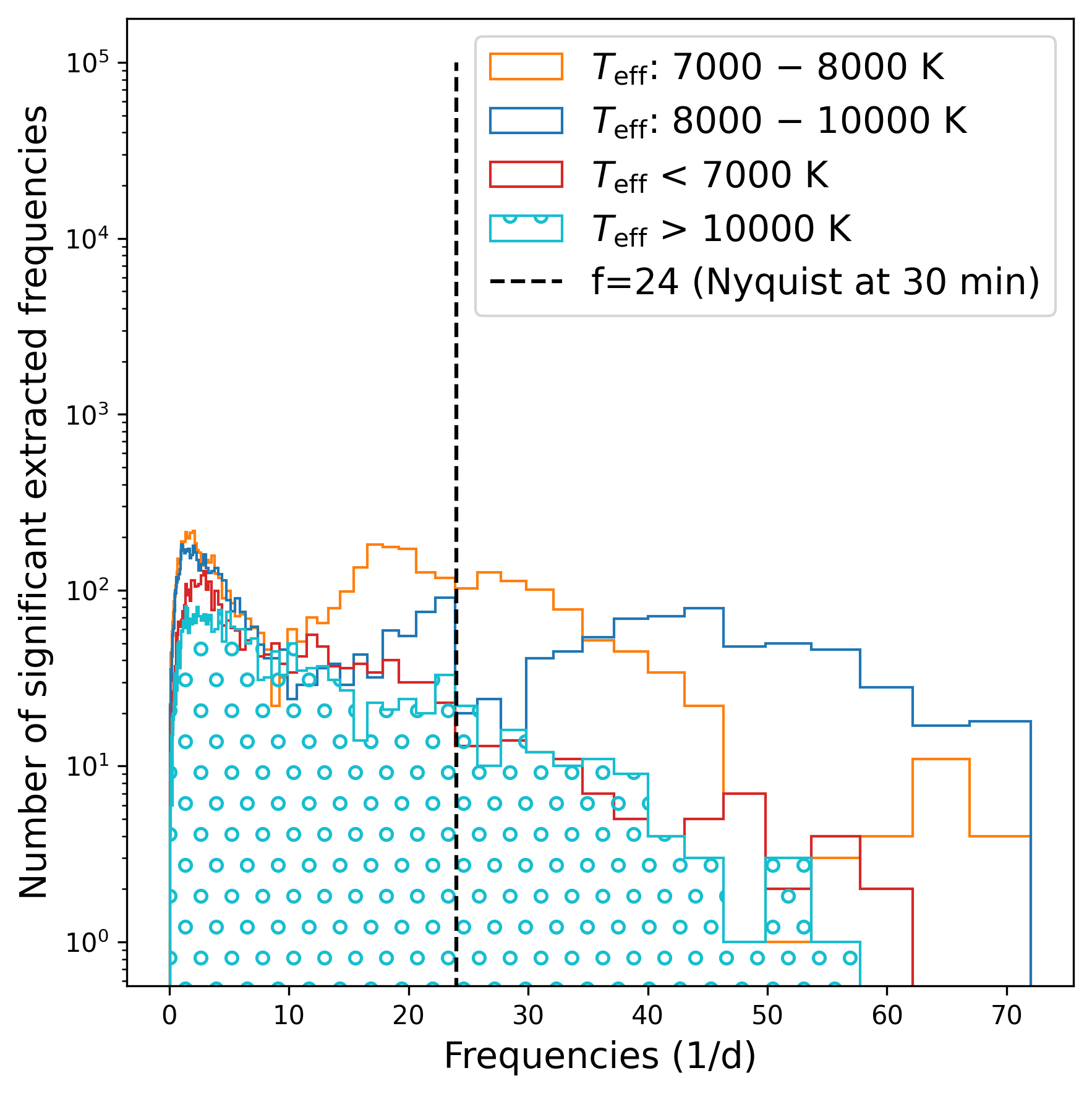}
    \caption{Histogram of the extracted significant frequencies in each temperature range for hybrids. Bins are scaled logarithmically in size to better view the lower densities at high frequency. Included are just those targets in the group of p-mode pulsators that are also in the g-mode pulsators.}
    \label{fig:hist_freqs_h}
\end{figure}

\begin{figure}
\centering
\includegraphics[width=\hsize]{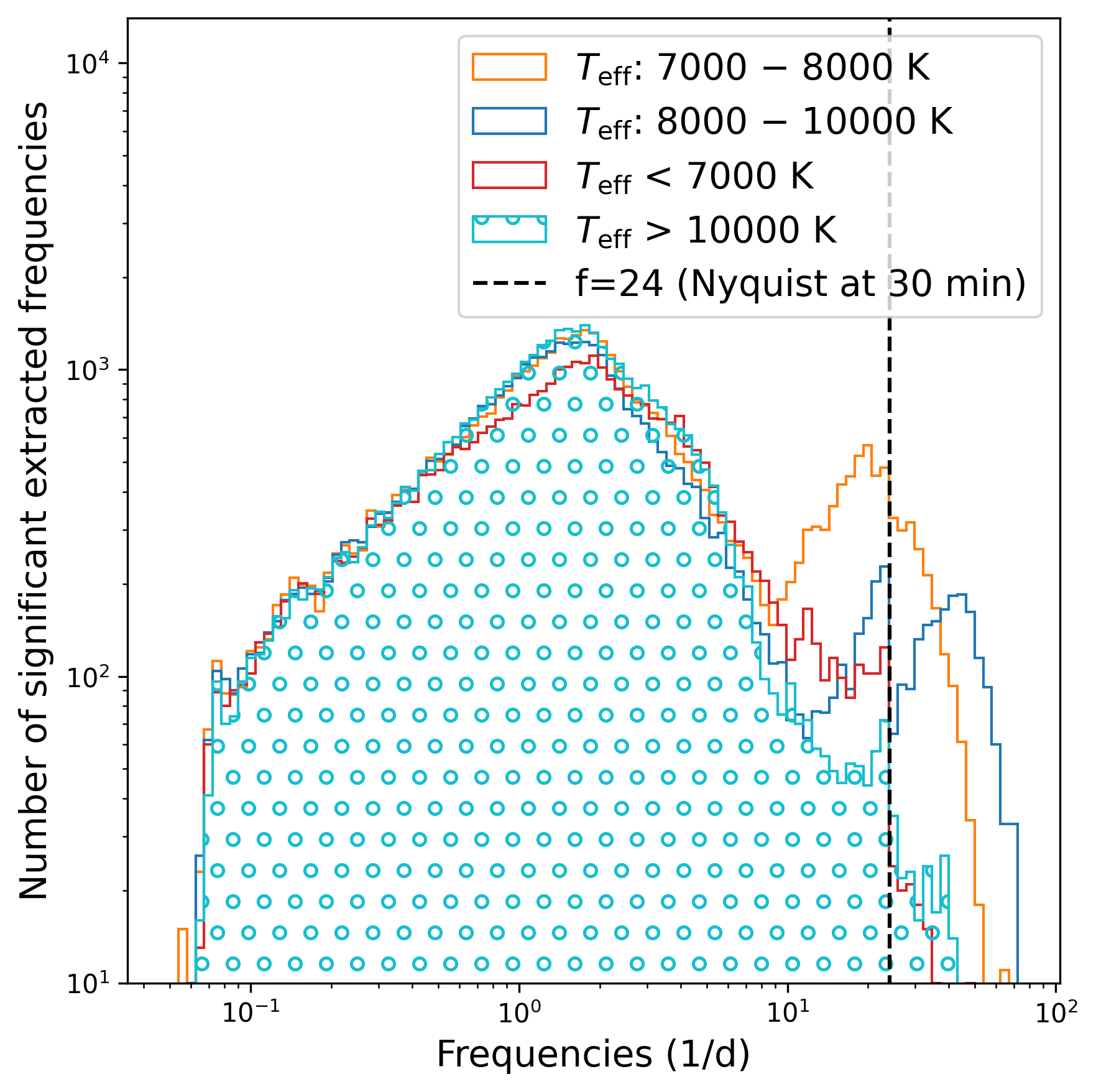}
    \caption{Histogram of the extracted significant frequencies in each temperature range for all pulsators. Bins are scaled logarithmically in size to better view the lower densities at high frequency. Included are just those targets in the group of p-mode pulsators that are not in the g-mode pulsators. Same as \ref{fig:hist_freqs_apx}, but with a logarithmic frequency axis.}
    \label{fig:hist_freqs_log}
\end{figure}

\begin{figure}
\centering
\includegraphics[width=\hsize]{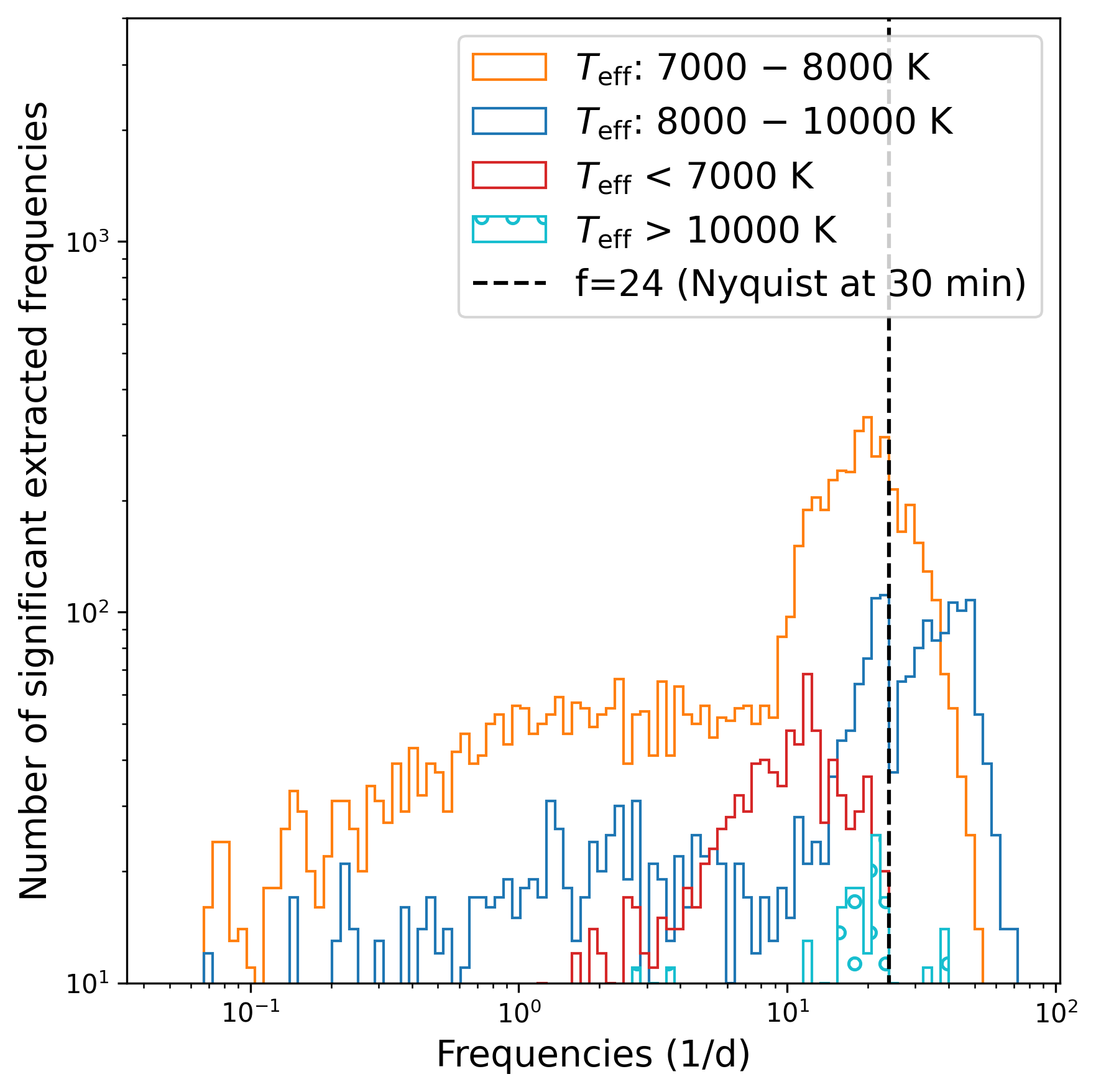}
    \caption{Histogram of the extracted significant frequencies in each temperature range for p-mode pulsators. Bins are scaled logarithmically in size to better view the lower densities at high frequency. Included are just those targets in the group of p-mode pulsators that are not in the g-mode pulsators. Same as \ref{fig:hist_freqs_p}, but with a logarithmic frequency axis.}
    \label{fig:hist_freqs_p_log}
\end{figure}

\begin{figure}
\centering
\includegraphics[width=\hsize]{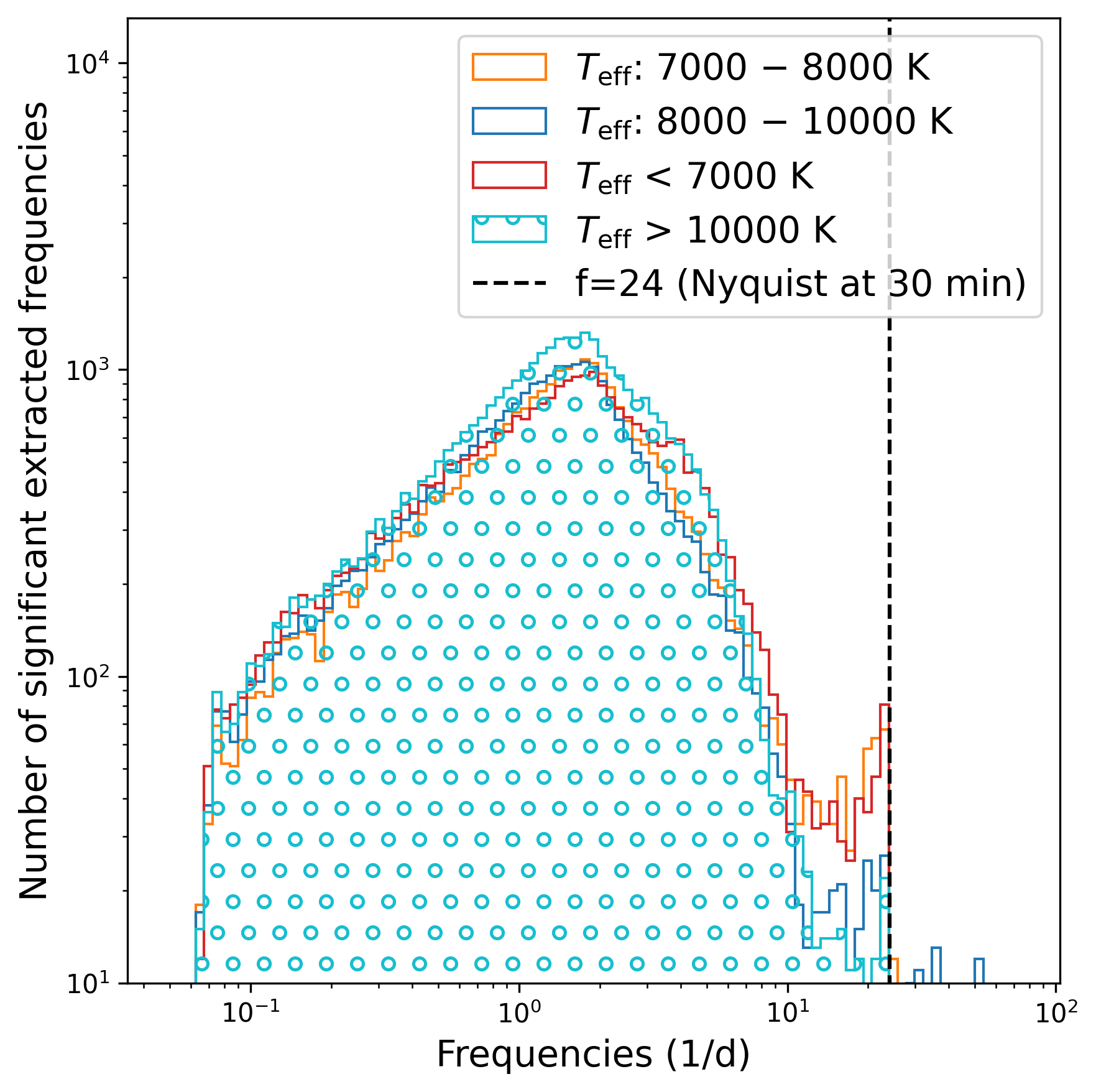}
    \caption{Histogram of the extracted significant frequencies in each temperature range for g-mode pulsators. Bins are scaled logarithmically in size to better view the lower densities at high frequency. Included are just those targets in the group of g-mode pulsators that are not in the p-mode pulsators. Same as \ref{fig:hist_freqs_g}, but with a logarithmic frequency axis.}
    \label{fig:hist_freqs_g_log}
\end{figure}

\begin{figure}
\centering
\includegraphics[width=\hsize]{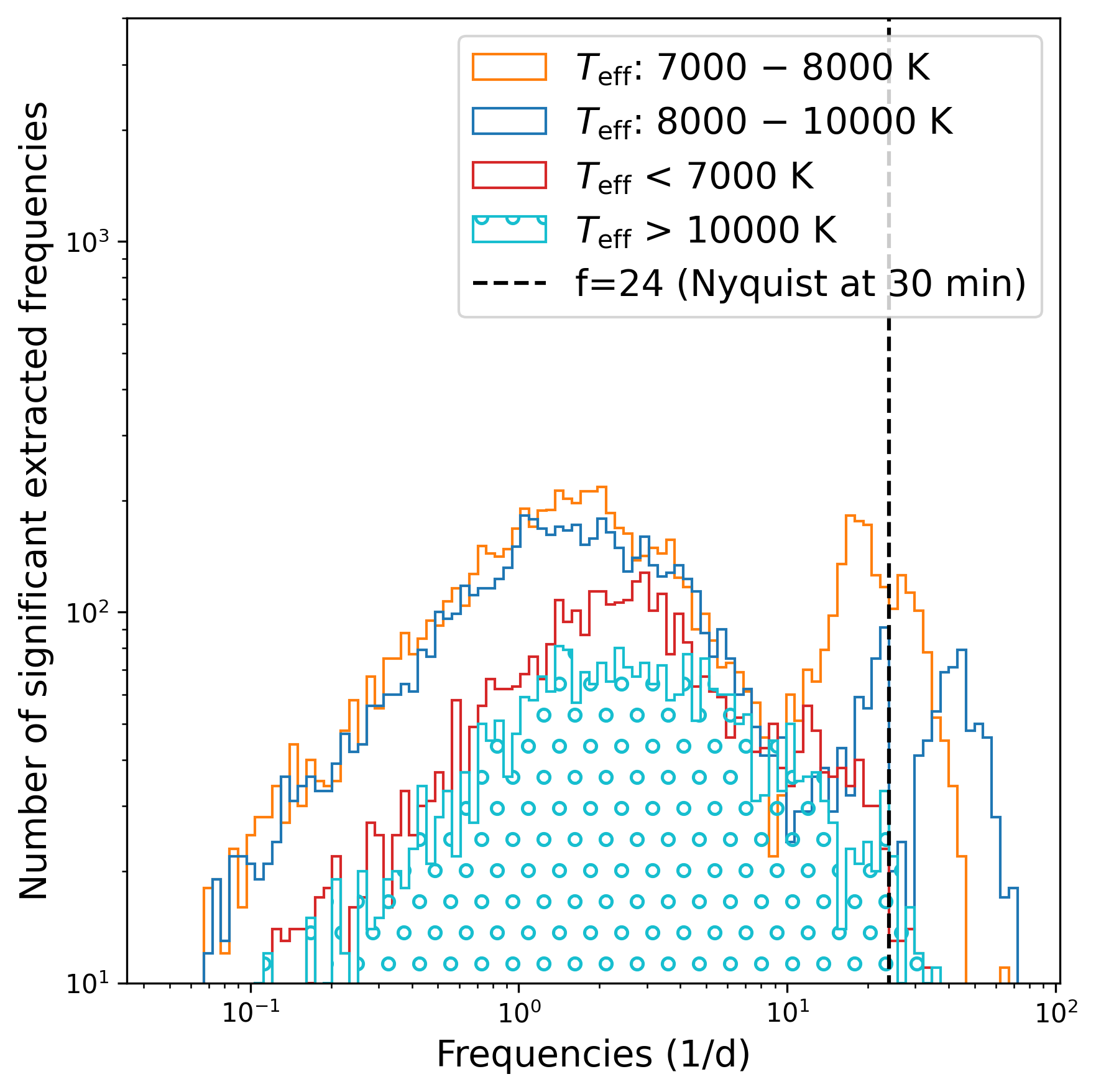}
    \caption{Histogram of the extracted significant frequencies in each temperature range for hybrids. Bins are scaled logarithmically in size to better view the lower densities at high frequency. Included are just those targets in the group of p-mode pulsators that are also in the g-mode pulsators. Same as \ref{fig:hist_freqs_h}, but with logarithmic a frequency axis.}
    \label{fig:hist_freqs_h_log}
\end{figure}

\begin{figure}
\centering
\includegraphics[width=\hsize]{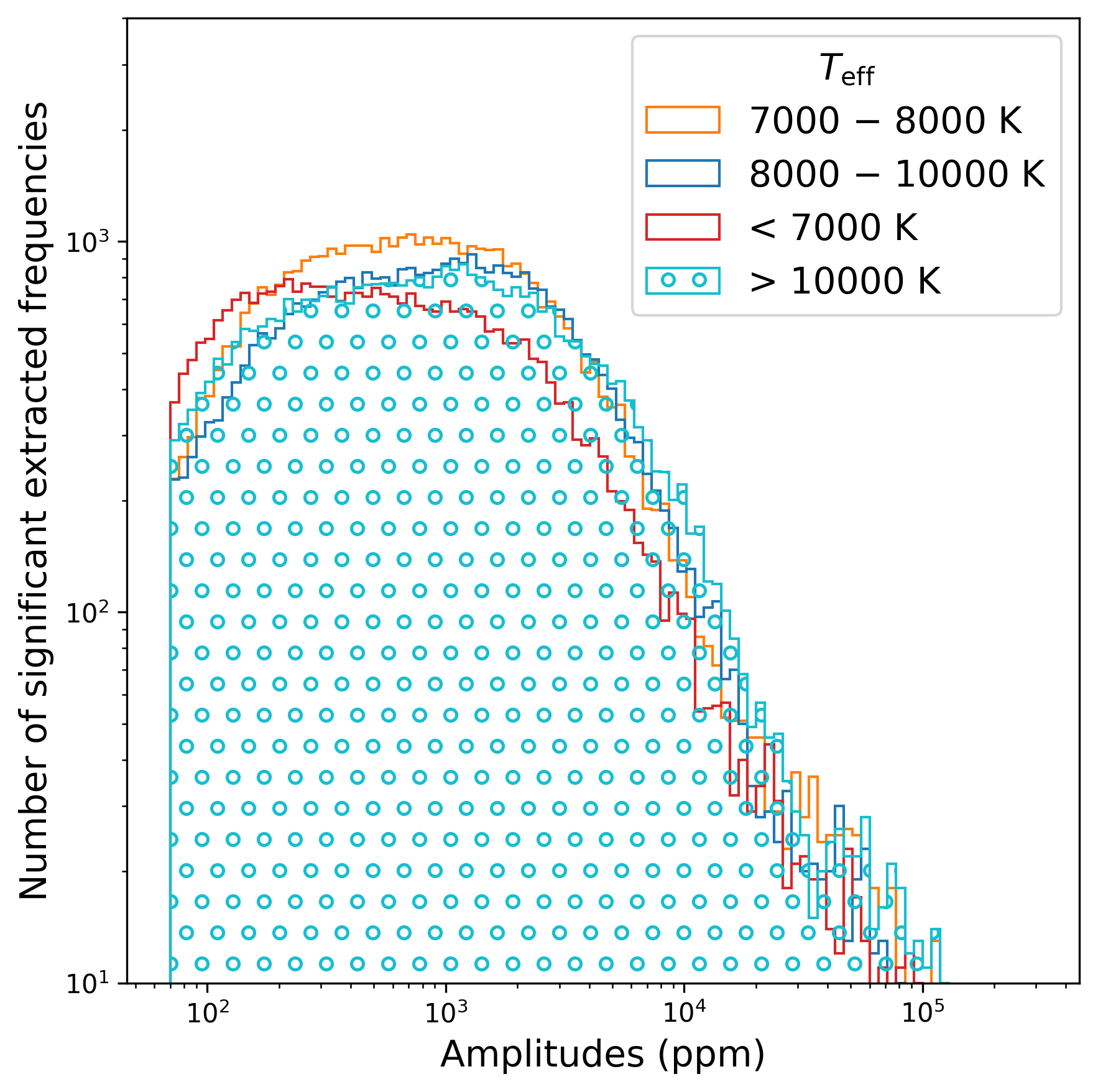}
    \caption{Histogram of the extracted significant amplitudes in each temperature range for all pulsators. Bins are scaled logarithmically in size. Included are just those targets in the group of p-mode pulsators that are not in the g-mode pulsators.}
    \label{fig:hist_ampls_apx}
\end{figure}

\begin{figure}
\centering
\includegraphics[width=\hsize]{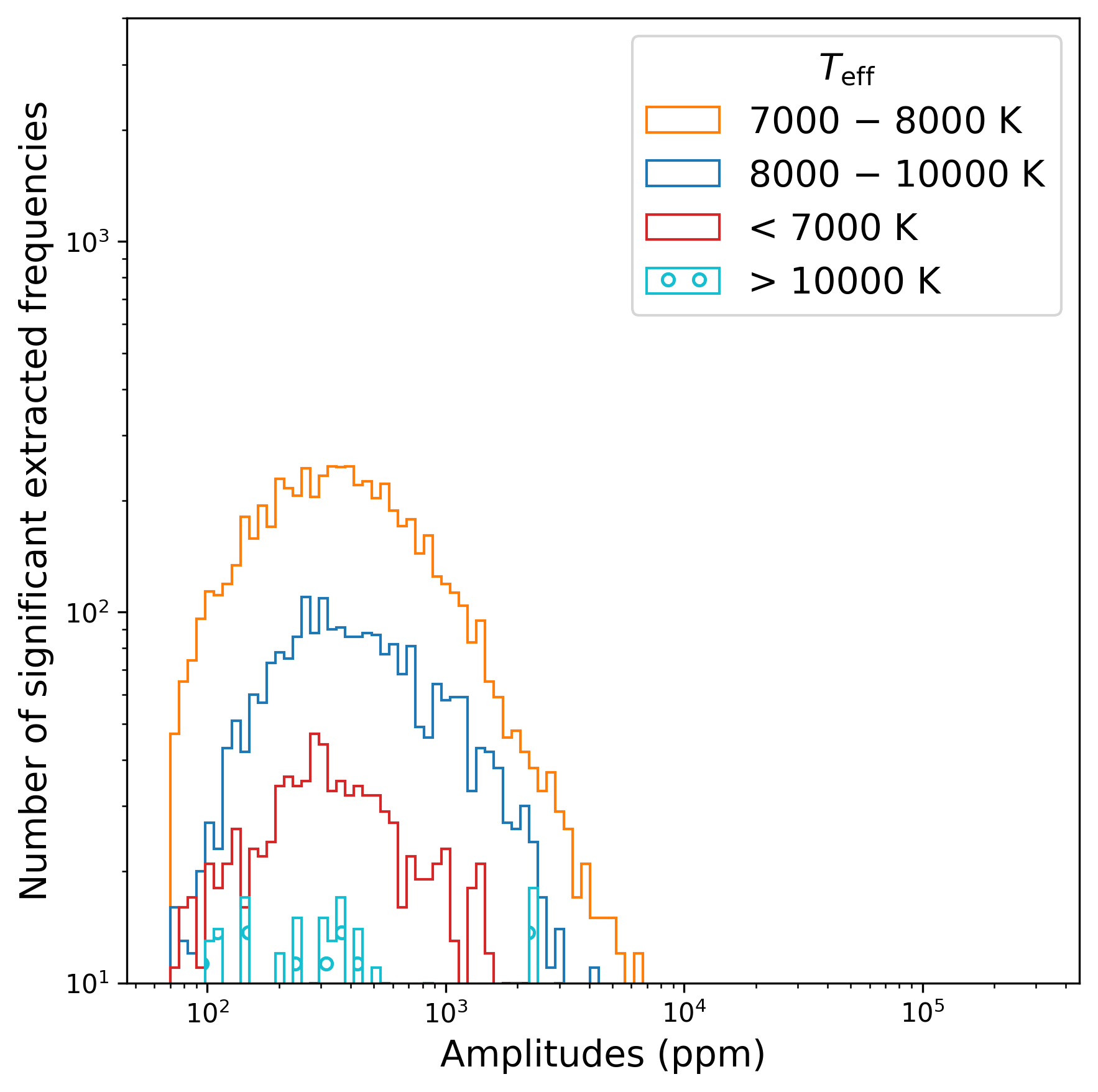}
    \caption{Histogram of the extracted significant amplitudes in each temperature range for p-mode pulsators. Bins are scaled logarithmically in size. Included are just those targets in the group of p-mode pulsators that are not in the g-mode pulsators. The distribution of the lowest $T_{\rm eff}$ is significantly different from all others. The distribution of $T_{\rm eff}$ between 7000 and 8000\,K is significantly different from the two higher ones with p-values of 0.0014 and 0.012, respectively. The two distributions of the highest temperature bins are not distinguishable (p-value 0.809).}
    \label{fig:hist_ampls_p}
\end{figure}

\begin{figure}
\centering
\includegraphics[width=\hsize]{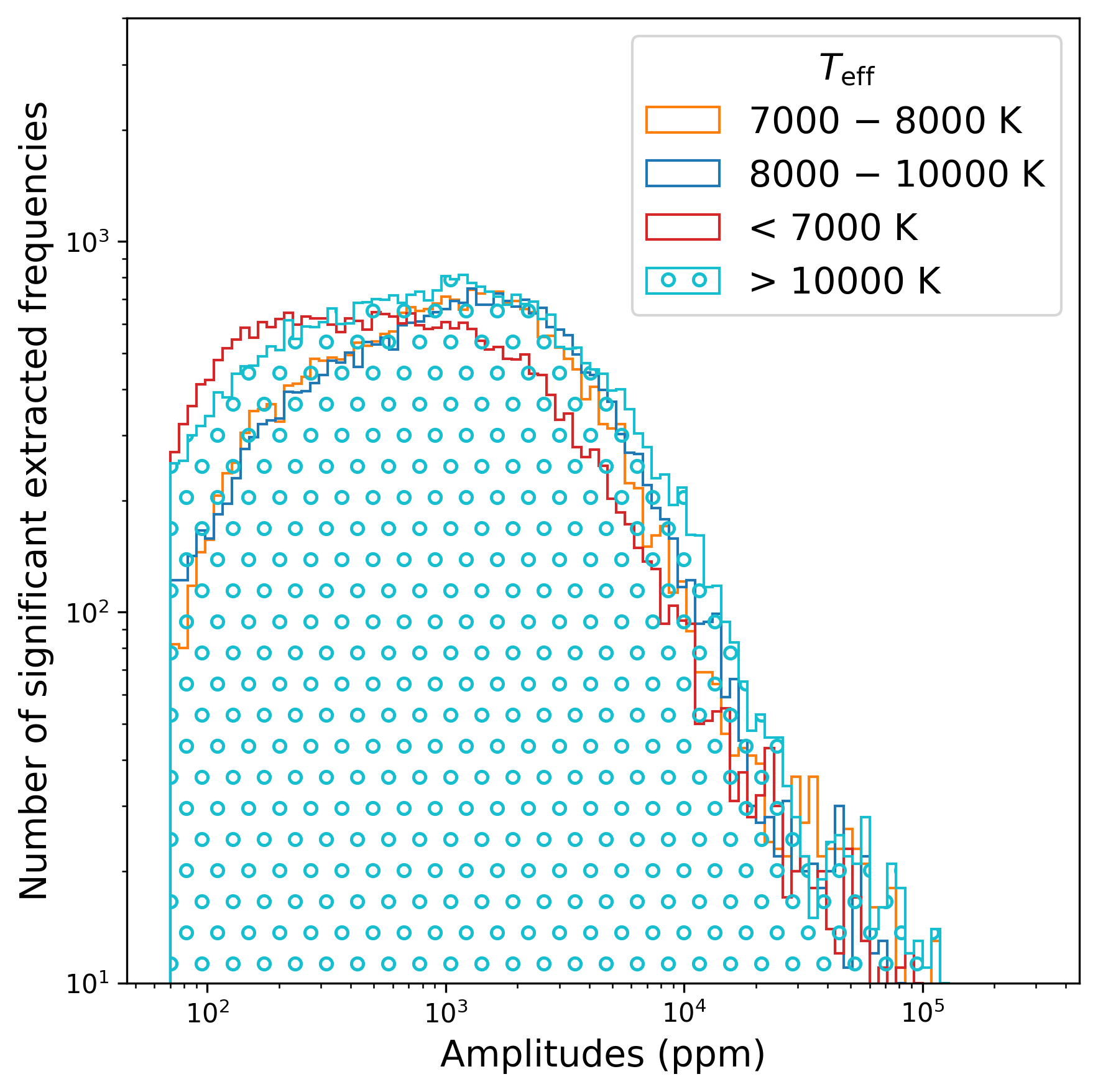}
    \caption{Histogram of the extracted significant amplitudes in each temperature range for g-mode pulsators. Bins are scaled logarithmically in size. Included are just those targets in the group of g-mode pulsators that are not in the p-mode pulsators. The distributions of the two middle $T_{\rm eff}$ bins are statistically identical (p-value of 0.952), all others are significantly different (p-value < $10^{-13}$).}
    \label{fig:hist_ampls_g}
\end{figure}

\begin{figure}
\centering
\includegraphics[width=\hsize]{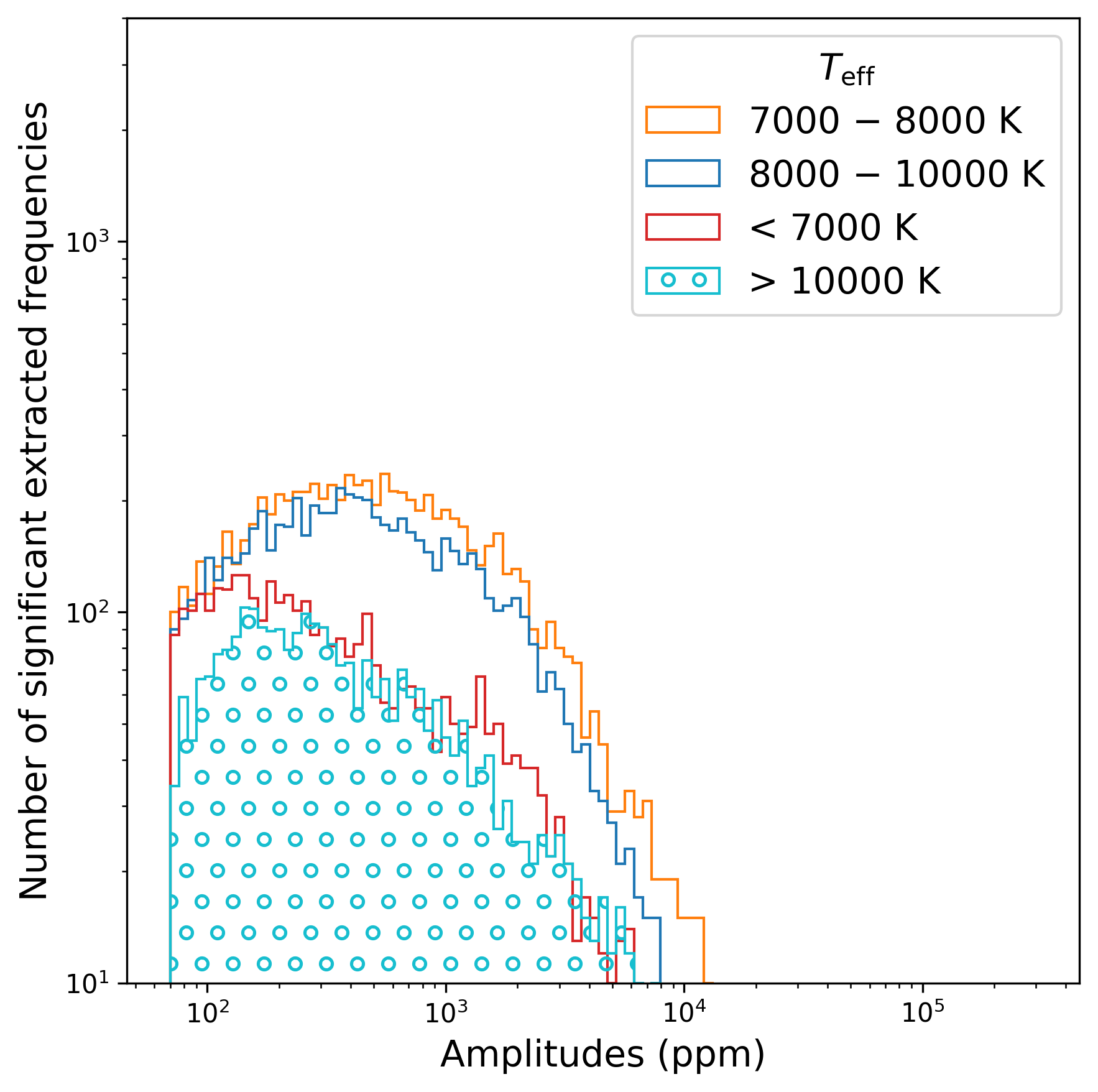}
    \caption{Histogram of the extracted significant amplitudes in each temperature range for hybrids. Bins are scaled logarithmically in size. Included are just those targets in the group of p-mode pulsators that are also in the g-mode pulsators. All distributions are significantly different from each other (p-value < $10^{-5}$).}
    \label{fig:hist_ampls_h}
\end{figure}

\end{appendix}

\end{document}